\documentclass[a4paper,12pt]{article}
\usepackage{graphicx,rotating}
\usepackage{a4p}
\usepackage{mcite}
\graphicspath{.}
\DeclareGraphicsExtensions{.eps,.ps,.eps.gz,.ps.gz}
\newcommand{\gev}{{\rm Ge}\kern-1.pt{\rm V}}
\newcommand{\gevsq}{\mbox{$\mathrm{{\rm Ge}\kern-1.pt{\rm V}}^2$}}
\newcommand{\gamstar}{\mbox{$\gamma^*$}}
\newcommand{\rhoz}{\mbox{$\rho^0$}}
\newcommand{\jpsi}{\mbox{$J/\psi$}}
\newcommand{\xbj}{\mbox{$x$}}
\newcommand{\qsq}{\mbox{$Q^2$}}
\newcommand{\kk}{\mbox{\bf k}}
\newcommand{\kkp}{\mbox{{\bf k}$'$}}
\newcommand{\pp}{\mbox{\bf P}}
\newcommand{\qq}{\mbox{\bf q}}
\newcommand{\vv}{\mbox{\bf V}}
\newcommand{\psih}{\mbox{${\psi}_h$}}
\newcommand{\phih}{\mbox{${\phi}_h$}}
\newcommand{\Phih}{\mbox{${\Phi}_h$}}
\newcommand{\Imj}{\mbox{Im}}
\newcommand{\Rej}{\mbox{Re}}
\newcommand{\thetah}{\mbox{${\theta}_h$}}
\newcommand{\fthree}{\mbox{$F_3$}}
\newcommand{\fsq}{\mbox{$F_2$}}
\newcommand{\fsqem}{\mbox{$F_2^{em}$}}
\newcommand{\fs}{\mbox{$F_1$}}
\newcommand{\fl}{\mbox{$F_L$}}

\newcommand{\ppbar}{\mbox{$\mathrm{p}\bar{\mathrm{p}}$}}

\newcommand{\Pma}{I\!\!P}

\newbox\struttbox
\setbox\struttbox=\hbox{\vrule height1.65ex depth.485ex width0pt}
\def\strutt{\relax\ifmmode\copy\struttbox\else\unhcopy\struttbox\fi}
\def\stru#1#2{\relax\ifmmode\hbox{\vrule height#1 depth#2 width0pt}
\else\vrule height#1 depth#2 width0pt\fi}

\def\lsim{\mathrel{\rlap{\lower4pt\hbox{\hskip1pt$\sim$}}
    \raise2pt\hbox{$<$}}} 
\def\gsim{\mathrel{\rlap{\lower4pt\hbox{\hskip1pt$\sim$}}
    \raise2pt\hbox{$>$}}} 

\begin{document}
\title {
\mbox{ } \hfill {\normalsize DESY 97--068}\\
\vspace*{-3mm}
\mbox{ } \hfill {\normalsize BONN--IR--97--01}\\
\vspace*{-3mm}
\mbox{ } \hfill {\normalsize April 1997}\\[1cm]
{\bf Exclusive Production\\[5mm]
of\\[5mm]
Neutral Vector Mesons\\[5mm] 
at\\[5mm]
the Electron--Proton Collider \mbox{HERA}\\[1cm]}
\author{\rm J.A.~Crittenden\\[5mm]
       {\em Physikalisches Institut, Universit\"at Bonn}\\ {\em
Nu{\ss}allee 12, D--53115 Bonn, Germany}} }
\date{}
\maketitle
\vspace{0.5cm}
\begin{abstract}
The first five years of operation of the multi--purpose experiments
ZEUS and H1 at the electron--proton storage ring facility HERA
have
opened a new era in the study of vector--meson production in
high--energy photon--proton interactions. The high
center--of--mass energy available at this unique accelerator
complex allows investigations in hitherto unexplored kinematic
regions, providing answers to long--standing questions concerning the
energy--dependence of the {\rhoz}, $\omega$, $\phi$, and {\jpsi}
production cross sections. The excellent angular acceptance of these
detectors, combined with that of specialized tagging detectors at small
production angles, has permitted measurements of elastic and inelastic
production processes for both quasi--real photons and those of virtuality
exceeding the squared mass of the vector meson.
This report provides a quantitative picture
of the present status of these studies, comparing them to the extensive
measurements in this field at lower energies and summarizing 
topical developments
in theoretical work motivated by the new data.
\end{abstract}
\thispagestyle{empty}
\cleardoublepage
\thispagestyle{empty}
\tableofcontents
\cleardoublepage
\setcounter{page}{1}
\section{Introduction}
\setcounter{equation}{0}
Experimentation at the accelerator complex HERA
(Hadron--Elektron--Ring--Anlage) operating at the German
national research laboratory DESY (Deutsches Elektronen--Synchrotron)
during the past five years
has revealed an array of Nature's secrets to perusal by the
curiosity--driven human intellect.
The opening of hitherto unexplored kinematic regions at high interaction 
energies is resulting in qualitative
progress in our
understanding of hadronic interactions. A variety of questions both old
and new are being stimulated by the access to this information
on the elementary particles and their interactions. Decades--old conceptual
quandaries are once again being addressed as we witness renewed studies of
electron--proton interactions. The \mbox{studies}
at HERA have fulfilled the expectations of unprecedented breadth 
in the accessible
fields of research. The classic studies of deep inelastic scattering and 
the structure of the proton have been extended to a
kinematic domain in which our microscopes now resolve 
objects orders of magnitude smaller than the proton itself. 
The hadronic nature of the photon enjoys a rejuvenated interest stimulated not
only by the recently 
obtained experimental information on photon--hadron interactions, but also
by exciting developments in theoretical approaches to understanding these
interactions. For the first time, the energy dependence of the hadronic
coupl\-ing of the photon can be studied over a broad range of precisely measured
photon virtualities. In particular, data samples of high statistical precision
have now become available for study of the production of vector mesons in both
exclusive and inclusive reactions. These vector mesons are objects both familiar
and arcane: familiar due to extensive studies of their properties in the 1960's
during an era of discovery in particle physics, as accelerators progressively
achieved the energies necessary to exceed the production energy thresholds,
and arcane because they carry the quantum numbers
of the photon, resulting in quantum--mechanical mixing 
between the mesonic and photonic versions
of this mysterious form of matter/energy, and yet represent the simplest
form of quark bound states. Indeed, studies of deep inelastic
electron--proton scattering stimulated the early development
of the theory of quantum 
chromodynamics. This youngest addition to our field theoretical descriptions
of the forces in nature remains after twenty--five years our most 
convincing conceptual framework for understanding the interactions between
the objects we call quarks: objects which have never been observed as 
free particles, yet
interact as such at high energy. The vector mesons {\rhoz}, $\omega$,
$\phi$, 
{\jpsi} and $\Upsilon$, represent the quark/anti--quark bound states
corresponding to the generational hierarchy employed in the Standard
Model. In this article we investigate the recent 
experimental and phenomenological developments in the understanding of
vector--meson production in photon--proton interactions, profiting from
the broad kinematic domain accessible at HERA where a strict delineation
between ``hard'' and ``soft'' interactions, motivated in part by experimental 
constraints in prior studies, is no longer sufficient, or even useful.

The content of this article must be viewed as an intermediate step in the
development of this field of study for the following two reasons. First,
the vigorous
pursuit of improved experimentation at HERA presently underway 
permits the expectation
of more than 
an order--of--magnitude increase in statistical precision over the coming two
years, an improvement certain to open qualitatively different avenues of
investigation, as well as to sharpen the experimental results presented in this
article. Secondly, the theoretical research addressing the above--mentioned
distinction between hard and soft processes is still in its early stages
and we can certainly allow ourselves the hope that major progress will be
achieved in the coming years. However, the successful operation of the
HERA research facilities during these past five years has clearly delineated
the range of physics issues involved in the field of vector--meson production
in photon--proton interactions, and it is time to  
take stock of the situation and prepare for the future. Such is the intent
of the present work.

We begin by placing the HERA project in historical perspective, briefly 
describing in sections~\ref{sec:history} and~\ref{sec:hera} 
the developments which led to the commissioning of the 
accelerator complex in 1992
and its peformance during the first five years of operation. 
Thereafter follow descriptions in section~\ref{sec:detectors}
of the two multi--purpose
detectors built and operated by the H1 and ZEUS collaborations.
In section~\ref{sec:heraphysics} we provide 
an overview
of the physics topics addressed at HERA to date, beginning with a review
of the kinematics of electron--proton interactions. After a short synopsis
in section~\ref{sec:vmprod} 
of vector--meson production studies since the 1960s, we turn to a review
of contemporary
topical theoretical considerations in section~\ref{sec:theory}. 
Section~\ref{sec:experiment} begins the discussion of experimental results
with a description of the total photon--proton cross section
measurements in section~\ref{sec:gammapxsect} before turning to the
production of single neutral vector mesons in section~\ref{sec:expvm}.
A discussion of general experimental considerations is followed by 
a detailed description of re\-solutions, acceptances, and backgrounds 
associated with each of the decay channels for which results have been
published by the two collaborations. Observations on the energy
and momentum transfer dependences
of the production cross sections and a synopsis of results on production
ratios are followed by a description
of the helicity analyses leading to determinations of the ratio of
longitudinal to transverse photoabsorption cross sections as a function
of photon virtuality. This section on experimental results from HERA ends
with brief descriptions of the measurements of
vector--meson production accompanied by dissociation of the proton.
Section~\ref{sec:future} summarizes the foreseeable improvements in
present measurements, discussing luminosity upgrade plans, additional
detector components, prospects for beam polarization, and those 
for nuclear beams.
The article itself is summarized in section~\ref{sec:summary}
and some general concluding remarks are offered in 
section~\ref{sec:conclusions}.
This article covers the theoretical and experimental work published prior
to the end of 1996. We use natural units ($\hbar=c=1$).

\subsection{Historical Context}
\setcounter{equation}{0}
\label{sec:history}
A strong interest in the study of electron--proton interactions at the high
energies provided by colliding beam accelerators followed the successes of
the deep inelastic scattering experiments at the Stanford Linear
Accelerator (SLAC) in the late 1960s. Both
the PEP (Positron--Electron--Proton) accelerator facility at SLAC
and its counterpart PETRA (Proton--Electron--Tandem--Ring--Accelerator) at 
DESY were originally proposed as 
electron--proton colliding beam 
facilities ~\cite{epseminar73,*dr_72_22}. 
However, the astonishing discoveries at
the e$^+$e$^-$ facility SPEAR at SLAC beginning with that of the {\jpsi}
vector meson in November, 1974, and continuing with the investigations
of charmed mesons and of the $\tau$--lepton turned the tide in favor of 
electron--positron colliding facilities, leading to the exhaustive studies
of $e^+e^-$ interactions 
at the PEP and PETRA accelerators throughout the mid--1980s
and continuing today at the LEP accelerator at CERN 
(European Laboratory for Particle Physics). Among the prime motivating
factors for the effort put into this research was the hope that the higher--mass
quarks and their spectroscopy could be studied as thoroughly at these machines
as were charmed quarks at the SPEAR facility. However, both the bottom and top
quarks have since been discovered via their production in 
hadronic interactions.
The electron--positron colliders CESAR at Cornell University
and DORIS at DESY have indeed proved essential for the spectroscopy of the 
charm and bottom quarks and LEP continues today to add to the wealth of 
information on this rich topic. Concurrently, the continued study of deep
inelastic scattering with neutrino and muon beams at the Fermi National
Accelerator Laboratory (FNAL) in Chicago and CERN  
in Geneva were producing impressive
contributions to the understanding of elementary particles and their 
interactions. The disco\-very of weak neutral currents in 1973 at CERN provided
an important boost to the credibility of the electroweak Standard Model. 
Precision measurements of scaling violations
in the interactions of muons and neutrinos with nucleons at both
CERN and FNAL provided quantitative information necessary to the progress
in the new theory of strong interactions, quantum chromodynamics. Thus
the motivation to continue the early studies in electron--proton interactions
remained strong, especially since advances in accelerator technology
for both proton and electron beams made accessible kinematic regions in which
the Standard Model predicted that the so--called ``weak'' coupling should be
comparable to the electromagnetic coupling, far exceeding the energy range
accessible in fixed--target applications. One crucial question remained:
Could a solution be found for 
the hitherto unaddressed difficulties associated with focussing beams of
vastly different characteristics 
to the minute transverse size necessary to obtain high interaction 
rates?

By the early 1980s the superconducting magnet development studies at FNAL had
made clear the feasibility of 800~{\gev} proton accelerators, and this work was
being applied to the construction of a symmetric {\ppbar} storage
ring accelerator complex. A proposal to add a 30~{\gev} electron--ring to
this facility~\cite{cheer} 
ended up being rejected in favor of FNAL's {\ppbar} and
fixed--target programs.

In 1979 the European Committee for Future Accelerators published a feasibility
study for a high--energy electron--proton accelerator facility~\cite{ecfahera}.
On the basis of this and further design studies at DESY, a proposal for
a colliding beam facility featuring 30~{\gev} electrons 
and 800~{\gev} protons was
submitted to German funding agencies in 1983~\cite{heraprop,*hera1,*hera2}. This proposal
was approved in 1984, contingent on substantial financial and technical
contributions from abroad. The success of the resulting international 
collaboration led to the coining of the term ``HERA model'' as a paradigm
for future particle accelerator projects.

By the end of 1987 the civil construction work on the HERA tunnel and 
experimental halls had been completed, and a huge liquid helium plant
for the superconducting proton ring magnets was operational. The electron
ring was commissioned in mid--1988, as the industrial construction of the 
proton ring magnets was getting underway. The last of these magnets was
installed in the tunnel in September, 1990. The first electron--proton
collisions were achieved in October, 1991, and after the two large 
multi--purpose detectors ZEUS and H1 were commissioned during the Spring
of 1992, the first physics run of the machine took place in June.
\subsection{The Performance of the HERA Accelerator Complex}
\label{sec:hera}
The rapid improvement in HERA 
performance~\cite{arnps_44_413,*ijmp_2a_28,*dr_95_08a} 
during its first five years
of operation is described by the integrated luminosities per year
shown in Table~\ref{table:lumi}. 
\begin{table}[htbp]
  \centerline{
     \begin{tabular}{ccc*{2}{p{2cm}}} \hline \hline
Year & \multicolumn{2}{c}{Energy} & \multicolumn{2}{c}{Integrated Luminosity}\\
     & \multicolumn{2}{c}{(GeV)} & \multicolumn{2}{c}{(pb$^{-1}$)}\\
\cline{2-5}
     & e & p & \multicolumn{1}{c}{e$^-$p} & \multicolumn{1}{c}{e$^+$p}\\
\hline
1992 & 26.7 & 820  & \parbox{1.4cm}{\raggedleft 0.05}  & \multicolumn{1}{c}{--} \\
\hline
1993 & 26.7 & 820  & \multicolumn{1}{c}{1} & \multicolumn{1}{c}{--} \\
\hline
1994 & 27.5 & 820  &\multicolumn{1}{c}{1} & \parbox{1.2cm}{\raggedleft 5 } \\
\hline
1995 & 27.5 & 820  & \multicolumn{1}{c}{--} & \parbox{1.2cm}{\raggedleft 12 } \\
\hline
1996 & 27.5 & 820  & \multicolumn{1}{c}{--} & \parbox{1.2cm}{\raggedleft 17 } \\
\hline
\hline
    \end{tabular}}
  \caption[Integrated Luminosity]{Yearly integrated luminosities during the first five years of HERA operation.}\label{table:lumi}
\end{table} 
Studies of limitations in the electron beam lifetime in 1992 and 1993
concluded that space charge and recombination 
effects involving positively charged ions
in the beam pipe were to blame. Indeed the switch to positrons in early 1994
led to an immediate improvement in the beam lifetime and hence
increased integrated luminosity. As a result, most of the results
we discuss in this article were obtained with a positron beam, though
we often use the term ``electron'' generically in referring to the
scattered lepton. Further
running with an electron beam is presently planned following extensive
improvements to the electron ring vacuum system prior to the 1998 running 
period.
\subsection{The Two Multi--Purpose Detectors H1 and ZEUS}
\label{sec:detectors}
The technical specifications of the HERA accelerators posed unique problems
not only for the machine physicists, but also for those designing the 
detectors. The bunch crossing rate of 10~MHz was a factor of forty greater
than that of the CERN {\ppbar} collider, resulting in new problems
for data acquisition and triggering. The Lorentz boost of the center--of--mass
reference frame in that of the laboratory 
necessitated the design of large--acceptance
detectors more highly segmented and deeper in the proton flight direction. 
Sensitivity to interaction rates of a few events per day (e.g. charged--current
deep inelastic scattering) was required in the
presence of background rates of 100~kHz 
from the interactions of high--energy protons
with residual gas in the beam pipe. The calorimeters
had to be designed for the measurements of
hadronic jets of several hundred {\gev} as well as for those of muonic 
energy depositions of a few hundred MeV. In the following we will discuss
briefly how each of the two multi--purpose detectors addresses these common
problems with complementary methods, emphasizing the detector components
necessary to the triggering and reconstruction of events with single neutral
vector mesons in the final state. We will refer to a right--handed
coordinate system in which the $Z$--axis points in the proton flight direction,
called the forward direction, and the $X$--axis points toward the center of
the HERA ring, with the origin at the nominal interaction point.
\subsubsection{The H1 Detector}

\begin{figure}[htbp]
\begin{center}
\includegraphics[width=\linewidth,bb=19 84 441 574]{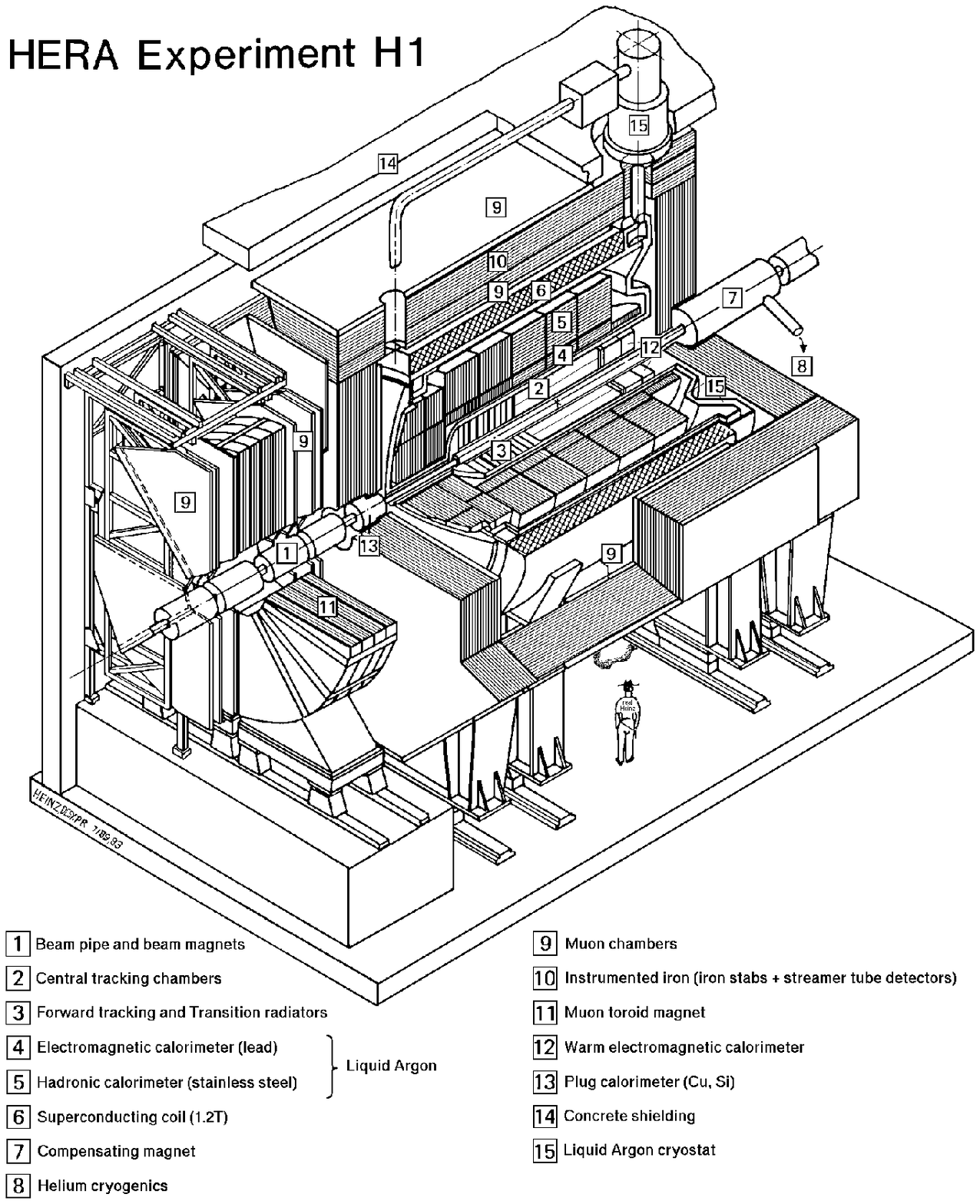}
\end{center}
\caption
{
\label{fig:h1det}
\it The H1 detector
} 
\end{figure}

The design of the 2800 ton 
H1 detector~\cite{dr_96_001}, shown in Fig.~\ref{fig:h1det}, 
emphasizes charged particle tracking
in the central region along with full calorimetric coverage and high
re\-solution in the measurement of electromagnetic energy depositions. The 
primary components of the tracking system are two coaxial cylindrical jet--type
drift chambers~\cite{nim_279_217} covering the polar angle region between 20$^\circ$
and 160$^\circ$. Position re\-solutions per hit of better than 200~$\mu$m in the
radial direction and 2~cm in the longitudinal direction have been attained.
Further inner drift chambers and two proportional chambers
serve to measure the longitudinal track coordinates and to provide triggering
information. The tracking system 
achieves a momentum re\-solution in the
coordinate transverse to the 1.2 Tesla solenoidal field of 
\mbox{$\sigma(p_T)/p_T \, = \, 0.01  \, p_T \, \oplus \, 0.015$}, 
where $p_T$ is expressed
in units of {\gev}. Specific ionization measurements provide 
information used for particle 
identification.
The magnetic field is produced by an impressive 5--m--long 
superconducting solenoid
of sufficient diameter (5.8~m) to enclose the entire calorimeter, storing
100~MJ of energy during operation. The H1 calorimeter~\cite{nim_336_460}
consists of a lead/liquid--argon shower--sampling sandwich section designed to measure electromagnetic
energy depositions, followed by hadronic sections employing steel plates as
absorbers. The depths of the hadronic sections 
range from 6 interaction lengths ($\lambda_I$)
in the forward direction to 4 $\lambda_I$ toward the rear.  The angular coverage
of the calorimeter extends from 4$^\circ$ to 155$^\circ$ in polar angle.
The transverse segmentation of the electromagnetic calorimeter is 
\mbox{$4 \times 4$ cm$^2$},
taking advantage of the flexibility in electrode design of the liquid--argon
calorimetric technique. The electromagnetic calorimeter is further segmented
in the direction of longitudinal shower development into 3 (4) sections in
the barrel (forward) regions. The transverse segmentation of the hadronic
sections is \mbox{$4 \times 4$ cm$^2$} and there are 4 (6) longitudinal sections in 
the barrel (forward) region. This fine segmentation results in 
31,000 electromagnetic and
14,000 hadronic readout channels.
The electromagnetic calorimeter achieves an energy re\-solution of 
\mbox{$\sigma(E)/E \, = \, 0.12/\sqrt{E ({\gev})} \, \oplus \, 0.01$.}  
The high degree of segmentation
permits an ingenious reconstruction algorithm for hadronic energy depositions
which reweights the electromagnetic and hadronic components of the shower
such as to achieve the desired linearity and re\-solution characteristics
despite sampling fractions which differ by 30\% for the two components.
The hadronic energy re\-solution thus obtained is
$\sigma(E)/E \, = \, 0.55/\sqrt{E ({\gev})} \, \oplus \, 0.01$. During the first
three years of operation, the rear region between polar angles of 150$^\circ$
and 175$^\circ$ was covered by a lead/scintillator--plate 
calorimeter with a re\-solution
of 
\mbox{$\sigma(E)/E \, = \, 0.10/\sqrt{E ({\gev})} \, \oplus \, 0.01$.}
In 1995 this calorimeter was
replaced with a lead/scintillating--fiber calorimeter of transverse
cell size \mbox{$4 \times 4$ cm$^2$}, 
energy re\-solution
\mbox{$\sigma(E)/E \, = \, 0.07/\sqrt{E ({\gev})} \, \oplus \, 0.01$,}
and time re\-solution better than 1~ns,
which also included a hadronic compartment 1~$\lambda_I$ thick.
Muon identification is provided by limited--streamer tube instrumentation in
the magnetic field return yoke outside the calorimeter cryostat, comprising
16 layers with a re\-solution of 1~cm per layer.

Our discussion of exclusive vector--meson production will necessitate the
distinction of event classes with and without dissociation of the
proton. The H1 measurements make use of two detectors in the extreme forward
direction to distinguish experimentally these two event types: a toroidal muon 
spectrometer covering the polar angular region between 3$^\circ$ and 17$^\circ$
and a system of scintillators 24~m from the interaction point which cover
the angular region from 0.06$^\circ$ to 0.25$^\circ$. These devices
provide sensitivity to secondary particles produced
in the beam--pipe wall and other beam--line elements 
by interactions of the dissociated--proton system.

The H1 triggering technique relies on track reconstruction in the proportional
chamber and drift chamber systems, which are efficient for tracks with
transverse momenta exceeding a value of 
0.5~{\gev}. Final--state electrons can be identified in the trigger
system via depositions in the electromagnetic calorimeter for energies greater
than 800~MeV and also via depositions in the small--angle luminosity monitor
electron detector. Muons are  identified at the trigger level via hits in
the central muon system.

\begin{figure}[htbp]
\begin{center}
\includegraphics[height=0.4\textheight,angle=270,bb=174 116 538 549,clip=]{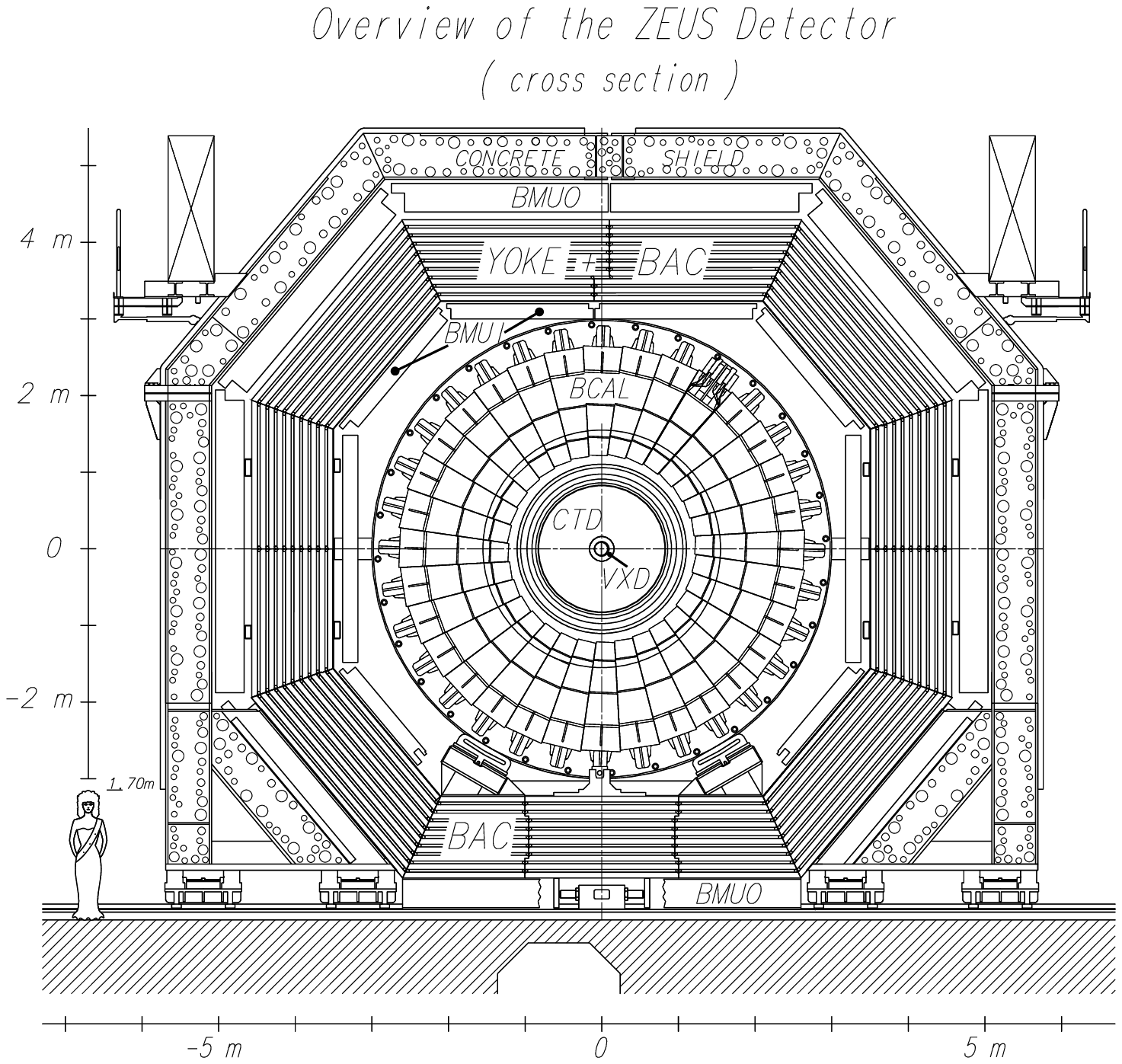}
\includegraphics[width=0.64\textheight,bb=152 255 612 686,clip=]{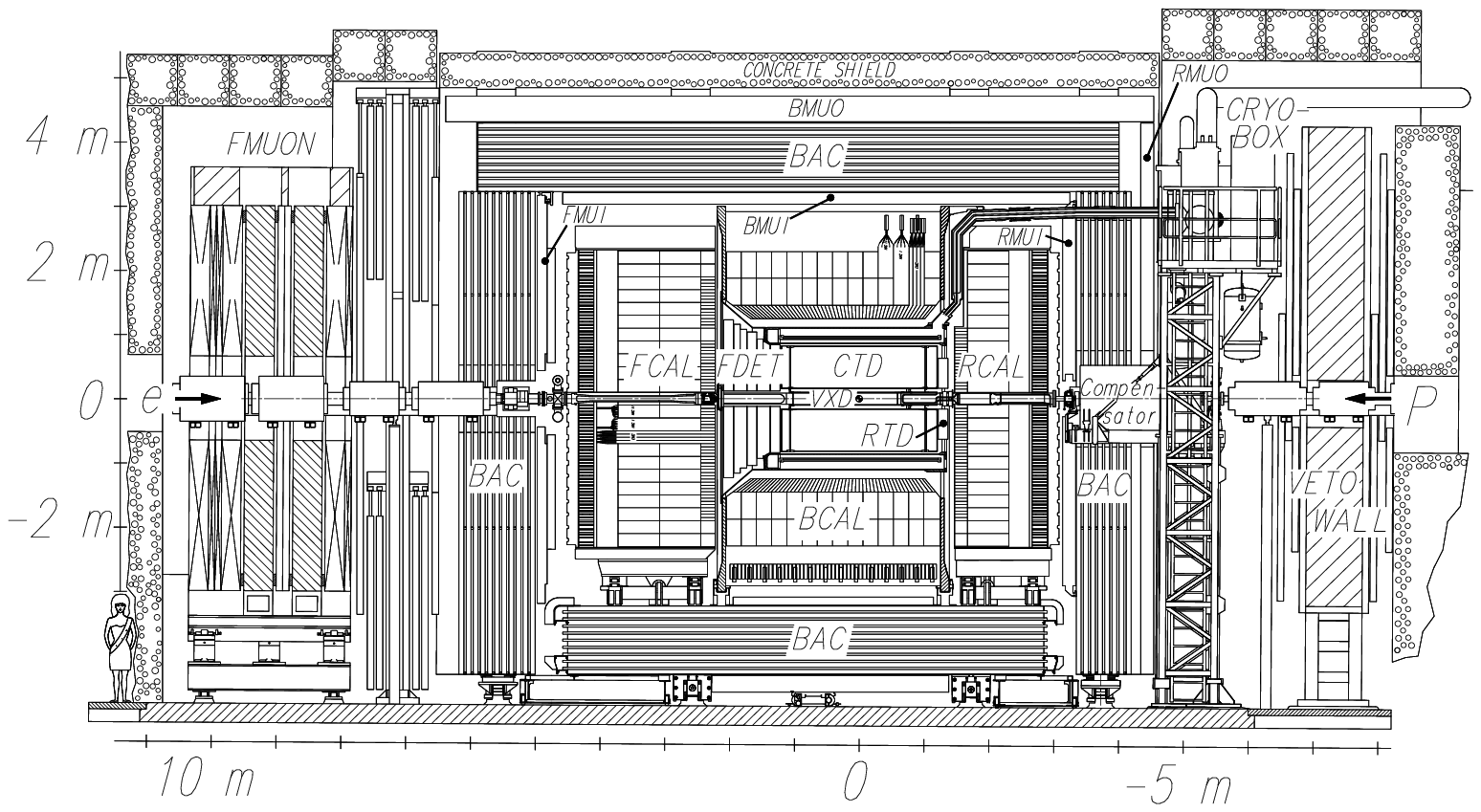}
\end{center}
\vspace*{-7cm}
\begin{center}
\begin{footnotesize}
\begin{it}
\begin{tabular}{rlrlrl}
VXD: & Vertex Detector & FMUON: & Forward Muon & FCAL: & Forward Calorimeter\\
CTD: & Central Tracking Detector & BMUI: & Barrel Muon Inner & BCAL: & Barrel Calorimeter\\
FDET: & Forward Detector& BMUO: & Barrel Muon Outer & RCAL: & Rear Calorimeter\\
RTD: & Rear Tracking Detector & RMUON: & Rear Muon & BAC:  & Backing Calorimeter\\
\end{tabular}
\end{it}
\end{footnotesize}
\end{center}
\caption
{
\label{fig:zeusdet}
\it The ZEUS detector
} 
\end{figure}
\subsubsection{The ZEUS Detector}
The design of the ZEUS detector~\cite{zeusdetector}, a schematic diagram
of which is shown in Fig.~\ref{fig:zeusdet},
profited from innovative long--term
test beam programs at CERN and FNAL 
which led to substantial progress in the understanding
of hadron calorimetry~\cite{nim_265_200,*nim_289_115,*nim_290_95,*nim_336_23}. In particular, the ability to
design a uranium sampling calorimeter with equal sampling fractions
for electromagnetic and hadronic shower components 
(termed a ``compensating'' 
calorimeter~\cite{nim_238_489,*nim_259_389,*nim_263_136}) 
permitted unprecedented specifications in hadronic energy
re\-solution and linearity at the trigger level as well as for the offline 
reconstruction of hadron energies. This leads to one of the remarkable
contrasts between these two HERA detectors: the ZEUS trigger algorithm 
is primarily calorimeter--based, while that of H1 
emphasizes tracking algorithms for reconstruction of the
interaction vertex.
The 700--ton ZEUS calorimeter~\cite{nim_309_77,*nim_309_101} consists of 
forward, central barrel, and rear
sections, covering the polar angle ranges $2^{\circ}-40^{\circ}$,
$37^{\circ}-129^{\circ}$, and  $128^{\circ}-177^{\circ}$. 
Each is made up of layers of 2.6~mm SCSN--38 scintillator and 3.3~mm
stainless--steel--clad depleted-uranium plates. Such a layer
corresponds to 1.0 radiation length ($X_0$)
and $0.04$ interaction lengths. This uniformity in structure 
throughout the entire calorimeter and the distributed nature of the
natural radioactivity of the uranium used to set the gains of the phototubes 
are decisive factors in the
stability and precision of its calibration~\cite{vcalhep}.
The choice of active
and passive thicknesses results in a sampling 
fraction of $4\%$ for electromagnetic and hadronic shower components
(hence compensation) 
and $7\%$ for 
minimum--ionizing particles. 
The compensation results in an hadronic energy re\-solution
of  
\mbox{$\sigma(E)/E \, = \, 0.35/\sqrt{E ({\gev})} \, \oplus \, 0.02$.} 
The re\-solution for electromagnetic showers is  
\mbox{$\sigma(E)/E \, = \, 0.18/\sqrt{E ({\gev})} \, \oplus \, 0.01$.}

The wavelength-shifting
optical readout components subdivide the calorimeters longitudinally
into electromagnetic  and hadronic sections. The electromagnetic sections are
$25 \, X_0$ ($1.1 \, \lambda_I$) deep. The forward (barrel) hadronic 
sections are further
segmented longitudinally into two sections, 
each of $3.0 \, \lambda_I$ ($2.0 \, \lambda_I$) depth. 
The rear calorimeter has a single hadronic section
$3.0 \, \lambda_I$ deep. The optical components subdivide the forward and rear
calorimeters
into towers of transverse dimensions \mbox{$20 \times 20$ cm$^2$}. The 
electromagnetic sections
are further divided vertically into four (two) cells in the forward (rear)
calorimeter. Thus
the transverse segmentation of the electromagnetic 
cells is \mbox{$20 \times 5$ cm$^2$} in the forward region
and  \mbox{$20 \times 10$ cm$^2$} in the rear. The front face of the forward 
(rear) calorimeter is
2.2~m (1.5~m) distant from the nominal electron--proton 
interaction point.
The barrel calorimeter 
consists of 32 wedge-shaped modules situated 1.2~m distant from the beam axis.
Its electromagnetic sections
are divided by the optical 
readout components into 53 cells per module, each of interior 
transverse dimensions
\mbox{$24 \times 5$ cm$^2$}, project\-ive in polar angle. The 14
hadronic towers in each module measure \mbox{$24 \times 28$ cm$^2$} in
the interior transverse dimensions. A $2.5^{\circ}$
rotation of all the barrel calorimeter 
modules relative to the radial direction ensures
that the intermodular regions containing the wavelength shifters do not
point at the beam axis.

The rear calorimeter is further instrumented with a layer of 
\mbox{$3 \times 3$ cm$^2$} silicon
diodes located at a depth of 3.3 $X_0$ in the electromagnetic section, an
element essential to the reconstruction of 
low--energy neutral pions from $\omega$ decays. 

The ZEUS solenoidal coil of diameter 1.9~m and length 2.6~m
provides a 1.43 T magnetic field for the charged--particle
tracking volume. 
The tracking system
consists of a central 
jet--type wire chamber~\cite{nim_279_290,*nim_283_473,*nim_338_181} covering
the polar angular region from 15$^\circ$ to 164$^\circ$ ,
a forward planar tracking detector from 8$^\circ$ to 28$^\circ$ 
and a second planar tracking chamber in the backward direction, covering
the region from 158$^\circ$ to 170$^\circ$. 
The momentum re\-sol\-u\-tion
attained is
\mbox{$\sigma(p_T)/p_T\,=\,0.005 \, p_T \, \oplus \, 0.015$} and
a track is extrapolated to the calorimeter face with a
transverse 
re\-solution of about 3~mm. 
Ionization measurements from the central tracking chamber 
also serve to identify electron--positron pairs from {\jpsi} decays.

The coil
represents about 1~$X_0$ of passive absorber between the barrel calorimeter
and the interaction region, similar to the amount of structural material
in front of the forward and rear calorimeters. A small angle rear tracking
device constructed of 1--cm--wide scintillator strips and a forward and
rear scintillator--tile device with \mbox{$20 \times 20$ cm$^2$} 
segmentation serve as
shower presampling detectors, improving the energy measurement for the
scattered electron or positron~\cite{nim_382_419}.

The muon system~\cite{nim_333_342} 
is constructed of limited streamer tubes inside and outside
of the magnetic return yoke, covering the region in polar angle from
34$^\circ$ to 171$^\circ$. Hits in the inner chambers 
provide muon triggers for {\jpsi} decays.

The shaping, sampling, and pipelining algorithm of the readout
developed
for the calorimeter~\cite{nim_321_356,*nim_277_217,*dr_92_130,*hervasthesis,*ieee_36_465} and used in modified form for the silicon and presampler
systems permits the reconstruction of shower times with respect to the
bunch crossings with a re\-solution of better than 1~ns, providing essential
rejection against upstream beam--gas interactions, as well as allowing
5~$\mu$s for the calculations of the 
calorimeter trigger processor~\cite{nim_355_278}.
For the triggering of the simple final states associated with 
vector--meson 
decays in the exclusive production channels we discuss in this article
a trigger threshold of 0.5~{\gev} was used in the calorimeter, along with
track candidates in the central drift chamber, or a threshold of 0.5~{\gev}
with hits in the muon chambers.

In 1994 the Leading Proton Spectrometer 
(LPS)~\cite{dr_96_183,sacchithesis} 
began operation. This large--scale system of six stations of 
silicon--strip detectors located between 24 and 90~m from the interaction
point employs ``Roman Pot'' technology to place active detector planes
within a few millimeters 
of the proton beam. Beamline magnets provide a resolving
power of 0.4\% in the longitudinal coordinate of proton momentum and 5~MeV
in the momentum component transverse to the nominal proton beam axis.
The beam transverse momentum distribution 
has r.m.s. values of 40~MeV in the horizontal
plane and 90~MeV in the vertical plane, which dominate the 
LPS re\-solution. The acceptance
in longitudinal momentum 
extends from 40\% to 100\% of the proton beam momentum. When
used to identify exclusive vector--meson production, the LPS reduces 
contamination by processes in which the proton dissociates into a 
multiparticle final state to under 0.5\%, compared
to 10--20\% achieved with the other detectors in the forward region. This
measurement of the final--state proton also provides a direct measurement
of the momentum transfer to the proton, a measurement more accurate than can
be derived from the other detectors.

The Proton Remnant Tagger (PRT) consists of a 
set of seven pairs of scintillators situated 5.2~m from the interaction
point in the forward direction. They cover polar angles between 6 and 26~mrad
and serve to discriminate between exclusive vector--meson production processes
and those in which the proton dissociates into a multiparticle final state
producing depositions in the PRT.

Prior to the 1995 data--taking period the ZEUS col\-lab\-ora\-tion in\-stalled
a small tung\-sten/scin\-tillator
electromagnetic sampling calorimeter around the beam pipe
in the rear direction 3~m from the
nominal interaction point. Covering the region of small electron scattering
angle between 17 and 35~mrad, this beam--pipe calorimeter (BPC)
allows measurements for values of photon virtuality extending from 
0.1 to 0.9~{\gevsq}. At the time of this writing
analysis results are becoming available and promise to play an
important r\^ole in 
understanding the transition region between photopro\-duction
processes and deep inelastic scattering.

\subsection{Investigations of Photon--Proton Interactions at HERA}
\label{sec:heraphysics}
This chapter provides a brief general overview of the issues in
photon--proton interactions
addressed by investigations at HERA. After an introduction to the notation
and a discussion of the accessible ranges in the various kinematic variables,
three broad categories of physics topics are treated. These categories
will be seen to have some overlap and serve primarily as a rhetorical tool
of organization rather than as indicators of physics--based distinction.
The choice of these categories is furthermore conditional on the present--day
state of HERA operation. As the instantaneous luminosity increases and
data with other beam energies become available, this categorization of
physics topics studied at HERA will require modification.
\subsubsection{Kinematics}
\label{sec:kinematics}
Much of the notation and choice of variables in this section is motivated
by the investigations of deep inelastic electron--nucleon scattering
performed at SLAC in the late 1960s
which led to the discovery of Bjorken
scaling~\cite{prl_23_935}  and the further development of 
the infinite momentum frame 
formalism~\cite{pr_179_1547,*photonhadron,*prl_23_1415,*pr_1_2901,*prl_22_744,pr_3_1382}. 
For a detailed general treatment of
particle kinematics, extending beyond that of deep inelastic scattering,
we refer to the book by Byckling and Kajantie~\cite{byckling}.

We consider the neutral--current process $e^{\pm} + p \rightarrow e^{\pm} + X$, and
choose the notations {\kk}, {\kkp}, and {\pp} for the four--momenta
of the initial--state electron, 
the final--state electron, and the initial--state
proton respectively. The charged--current process with its final--state
neutrino has also been measured
at HERA; we restrict ourselves in this discussion to photon--exchange, 
neglecting the contributions from exchange of the heavy intermediate
vector bosons W$^{\pm}$ and Z$^0$. These contributions 
become important at virtualities
comparable to squared vector--boson masses, whereas the present--day 
event statistics for the processes we consider in this article are limited
to virtualities less than about 30~{\gevsq}. The Mandelstam variable
$s$ for the two--body scatter represents the square of
the electron--proton center--of--mass energy
\begin{eqnarray}
s &\equiv& (\kk + \pp)^2 \, = \, m_e^2 + m_p^2 + 4 \, E_e \, E_p,
\end{eqnarray}
where $E_e$ and $E_p$ are the beam energies. The nominal beam energies
\begin{eqnarray*}
E_e &=& 27.5 \; {\gev}, \hspace*{1cm} E_p = 820 \; {\gev} 
\end{eqnarray*}
result in a center--of--mass energy
\begin{eqnarray*}
\sqrt{s} &\approx& 300 \; \gev.
\end{eqnarray*}
Defining {\qq} to be the
photon four--momentum,
\begin{eqnarray}
\qq &\equiv& \kk - \kkp,
\end{eqnarray}
any two of the following three Lorentz--invariant
quantities suffice to describe the kinematics:
\begin{itemize}
\item
the photon virtuality
\begin{eqnarray}
\label{eq:q2}
\qsq &\equiv& -{\qq}^2 \, = \, Q^2_{min} + 4 p_ep'_e \sin^2{{\theta_e}/2},
\end{eqnarray}
with
\begin{eqnarray}
Q^2_{min} &=& 2(E_eE'_e -  p_ep'_e - m^2_e),
\end{eqnarray}
where $\theta_e$ is the electron scattering angle with respect to
its initial direction, $E'_e$ is the energy of the 
final--state electron, and the magnitudes of the electron
initial-- and final--state momenta are designated as $p_e$ and $p'_e$,
\item
the Bjorken scaling variable {\xbj}
\begin{eqnarray}
\xbj &\equiv& \frac{\qsq}{2 \, \pp \cdot \qq},
\end{eqnarray}
\item
and the Bjorken scaling variable $y$
\begin{eqnarray}
y &\equiv& \frac{\pp \cdot \qq}{\pp \cdot \kk}\,\approx\,1-\frac{E'_e}{E_e}\cos^2{{\theta_e}/2}.
\end{eqnarray}
\end{itemize}
These three variables are simply related in the approximation that the
masses are negligible compared to the momenta:
\begin{eqnarray}
\qsq &\approx& s{\xbj}y.
\end{eqnarray}
The minimum kinematically allowed value for {\qsq}, $Q^2_{min}$, can be
written in the approximation that the electron mass be neglected with
respect to its momentum as follows:
\begin{eqnarray}
\label{eq:q2min}
Q_{min}^2 &\approx& m_e^2 \frac{y^2}{(1-y)}
\end{eqnarray}
and does not exceed $10^{-8}$ \gevsq\ in the kinematic region of the
measurements considered in this article. We will discuss data samples
for exclusive vector--meson production
char\-acter\-ized by three {\qsq}--ranges:\\[5mm]
\begin{minipage}[t]{0.32\textwidth}
\begin{center}
{$Q^2_{min} < Q^2 < 4 \;$ GeV$^2$\\[2mm]
                            $\langle Q^2 \rangle \approx 0.01-0.1 \;$ GeV$^2$}
\end{center}
\end{minipage}
\hfill
\begin{minipage}[t]{0.32\textwidth}
\begin{center}
{$0.2 < Q^2 < 0.9 \;$ GeV$^2$\\[2mm]
                            $\langle Q^2 \rangle \approx 0.3-0.6 \;$ GeV$^2$}
\end{center}
\end{minipage}
\hfill
\begin{minipage}[t]{0.32\textwidth}
\begin{center}
{$Q^2 > 1 \;$ GeV$^2$\\[2mm]
                            $\langle Q^2 \rangle \approx 2-20 \;$ GeV$^2$},
\end{center}
\end{minipage}\\[5mm]
corresponding to photopro\-duction studies, where the final--state 
electron is not detected, studies at intermediate values of {\qsq}, where
the electron is detected at small scattering angles in 
a special--purpose calorimeter (the ZEUS BPC), and studies
where the electron is detected in the main detector and the upper limit
on {\qsq} is due simply to luminosity limitations. The values for the
average {\qsq} cited for 
the untagged photoproduction studies are rough estimates
based on the {\qsq}--dependence
of the photon flux. The corresponding {\em median} {\qsq} value is
approximately 10$^{-4}$ {\gevsq}.

Two further Lorentz--invariant variables which will turn out to be 
useful in our discussions of the HERA measurements are
\begin{eqnarray}
\nu &\equiv& \frac{\pp \cdot \qq}{m_p} \, = \, \frac{\qsq}{2\,x\,m_p},
\end{eqnarray}
which is the value of the photon energy in the proton rest frame,
and
\begin{eqnarray}
W^2 &\equiv& (\qq + \pp)^2 \, = \, -\qsq + 2\nu{m_p} + m_p^2 \, \approx \, sy-\qsq \, \approx \, \qsq \, \frac{1-x}{x},
\end{eqnarray}
the squared center--of--mass energy of the photon--proton system.
The kinematic limit for the value of $\nu$ is the electron energy
in the proton rest frame:
\begin{eqnarray}
\nu_{max} \, \approx \, \frac{s}{2m_p} \, \approx \, 48 \; \mbox{TeV},
\end{eqnarray}
and the Bjorken--$y$ variable can be written
\begin{eqnarray}
y &=& \frac{\nu}{\nu_{max}}.
\end{eqnarray}
In the context of the diffractive production of vector mesons, 
the negative squared 
momentum transfer at the proton vertex, the Mandelstam
variable $t$, is essential to our discussions of interaction size 
(see section~\ref{sec:general}). It is defined as follows:
\begin{eqnarray}
t & \equiv & (\pp - {\pp}')^2 \, = \, (\vv - \qq)^2,
\end{eqnarray}
where ${\pp}'$ is the four--momentum of the final--state proton and {\vv}
that of the vector meson. Neglecting the proton mass with respect to 
its momentum we can write
\begin{eqnarray}
\label{eq:trec}
- t &\approx& p_T^2(P')=  (\vec{p}_T(V)-\vec{p}_T(\gamma))^2,
\end{eqnarray}
where $p_T^2(P')$ is the transverse momentum of the final--state proton;
$\vec{p}_T(V)$ and $\vec{p}_T(\gamma)$ are the transverse momenta of the
vector meson and photon. The value of $|t|$ deviates from the squared magnitude
of the vector--meson transverse momentum by less than {\qsq}.
The minimum 
value of $|t|$ kinematically necessary to the elastic process is
\begin{eqnarray}
|t|_{min} &=& m_p^2 \, \frac{(M_V^2 + \qsq)^2}{W^4} \approx \frac{(M_V^2 + \qsq)^2}{4\,{\nu}^2},
\end{eqnarray}
where $M_V$ is the mass of the vector meson,
and does not exceed 10$^{-4}$~{\gevsq} for the data sets considered in
this article.
 
\begin{figure}[htbp]
\begin{center}
\includegraphics[width=0.8\linewidth]{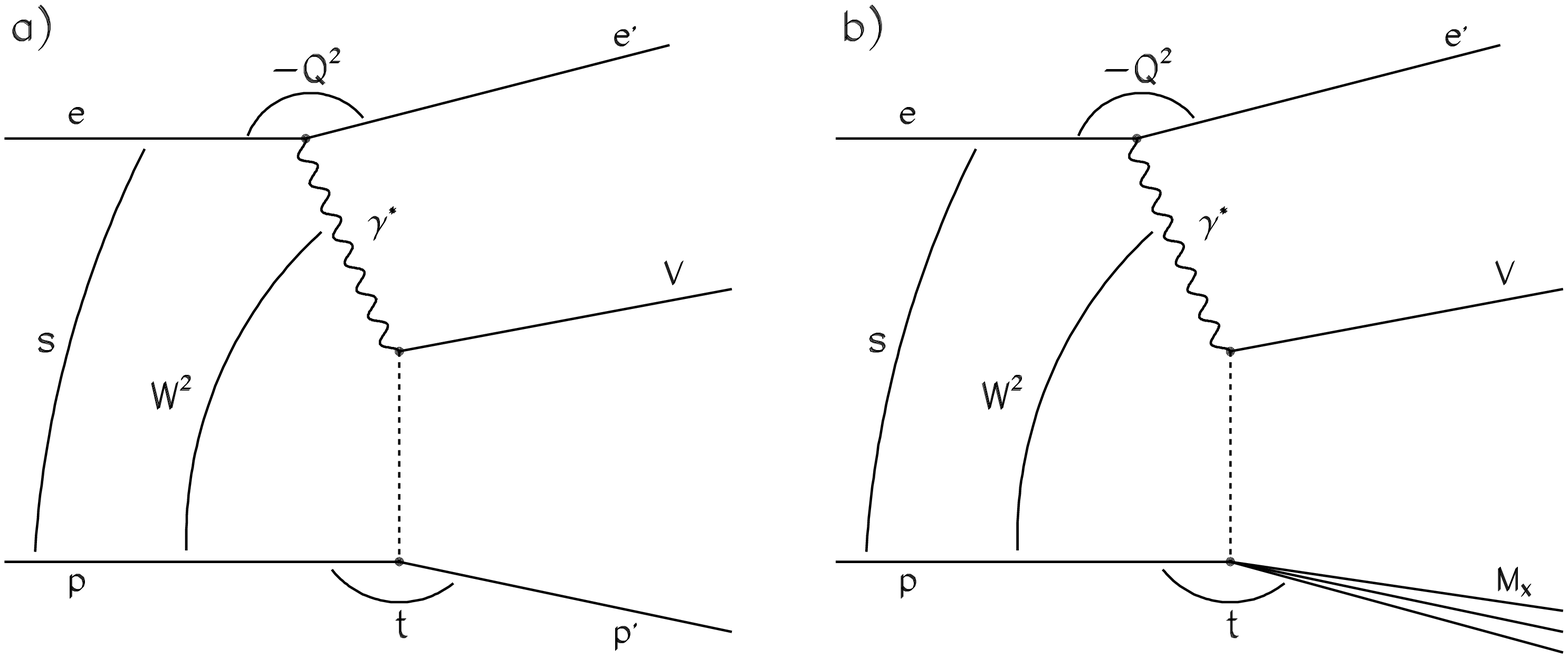}
\end{center}
\caption
{
\label{fig:newvector}
\it Schematic diagrams for the diffractive electroproduction of single vector
mesons: a)~elastic production, the principal topic of this article, and 
b)~production with proton dissociation into a hadronic state of
invariant mass M$_X$, one of the primary backgrounds.
The dashed line indicates the Pomeron exchange associated with the 
diffractive process. 
The definitions of the kinematic variables shown are given in the text.
}
\end{figure}

In concluding our discussion of the kinematics, we introduce 
the elasticity $z$, 
a variable important to the definition of the physical process central
to this article. The elasticity is defined as
\begin{eqnarray}
z &=& \frac{\pp \cdot \vv}{\pp \cdot {\qq}}.
\end{eqnarray}
Using the fact that the sum of the squared particle masses is equal to the
sum of the Mandelstam variables in the reaction
$\gamstar + p \rightarrow V + p$ one can show
\begin{eqnarray}
z&=&1+\frac{t}{2m_p\nu}.
\end{eqnarray}
Thus for values of $|t|$ negligible with respect to \mbox{$2 m_p \nu$},
the process we refer to in this article as ``elastic'' or
``exclusive'', shown in Fig.~\ref{fig:newvector}~a), 
restricts the value of the elasticity to unity. 
Figure~\ref{fig:newvector}~b) shows the process accompanied by dissociation
of the proton, which results in elasticity $z \lsim 1$. This process 
is the primary source of background in the investigations of the elastic
process at HERA, except for those in which the final--state proton is
detected. In this article we will mention
only briefly inelastic vector meson production, which is
characterized by substantially lower values for the elasticity.

\vspace*{3mm}
The high photon energy for which data at HERA have been obtained, 
spanning two orders of magnitude from
400 to 40000 GeV in the proton rest frame is precisely
the feature of HERA which makes it attractive as a laboratory for
photopro\-duction studies. The high center--of--mass energies in
the photon--proton system (40 to 200~{\gev}) allow an unprecedented
sensitivity to the energy dependence of $\gamma^* p$--interactions.
Equivalently, formerly unmeasured regions of high \qsq\ and of low \xbj\
have now been made accessible, which brings us to our first physics topic:
the proton structure functions.
\subsubsection{Proton Structure Functions}
\label{sec:f2}
\begin{figure}[htbp]
\begin{center}
\includegraphics[width=0.8\linewidth,bb=25 158 525 650,clip=]{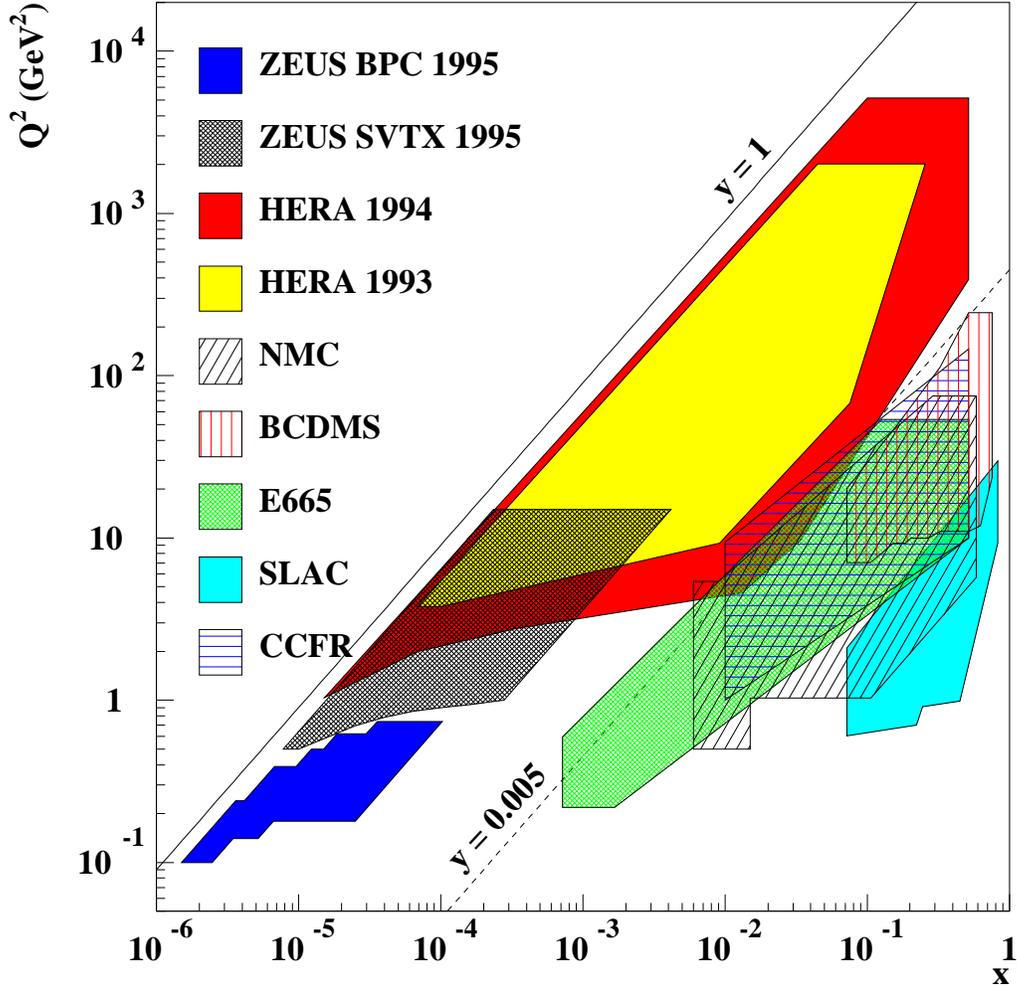}
\end{center}
\caption
{
\label{fig:xq2plane}
\it The kinematic region covered by the measurements of the proton structure
function F$_2$({\xbj},{\qsq}) by the H1 and ZEUS experiments at HERA in 1993
and 1994, compared to a region covered by ZEUS studies using data from
runs in 1995 where the $e^+p$ interaction vertex was shifted by 67~cm in
the forward direction (SVTX), and to a region covered by data where
the scattered electron was detected in a beam--pipe calorimeter (BPC)
in the ZEUS experiment. Also shown are the kinematic domains
in which the proton structure function was measured in the fixed--target 
muon experiments NMC and BCDMS at CERN, and E665 at FNAL. The region covered
by the fixed--target neutrino experiment CCFR at FNAL is shown, as well
as that covered in the early SLAC fixed--target studies with electron beams.} 
\end{figure}

One of the primary contributions from investigations at HERA to
date has been the observation of a strikingly strong {\xbj}--dependence in
the proton structure function {\fsq}({\xbj},{\qsq}) 
in the hitherto unmeasured kinematic region \mbox{$\xbj < 10^{-3}$.} 
Figure~\ref{fig:xq2plane} shows the kinematic region covered in
measurements performed by the H1 and ZEUS collaborations at HERA in 1993
and 1994, compared to a region covered by ZEUS studies using data from
runs in 1995 where the $e^+p$ interaction vertex was shifted by 67~cm in
the forward direction (SVTX), and to a region covered by data where
the scattered electron was detected in the beam--pipe calorimeter
in the ZEUS experiment. Also shown are the kinematic domains
in which the proton structure function was measured in fixed--target 
neutrino, electron, and 
muon experiments.

For the kinematic domain covered by the HERA measurements, the
differential neutral--current cross section is related to the structure
functions {\fs}({\xbj},{\qsq}), {\fsq}({\xbj},{\qsq}), and {\fthree}({\xbj},{\qsq}) as follows:
\begin{eqnarray}
\label{eq:sigf2}
 \frac{d^2\sigma^{e^{\pm}p}}{d{\xbj}d\qsq} 
&=& \frac{2{\pi}\alpha^2}{{\xbj}Q^4}
   \left[ 2(1-y) \,  {\fsq} + {y^2}\;2{\xbj}{\fs} \mp {(2y-y^2)}\;{\xbj}{\fthree} \right]\; \left[ 1 + \delta_r \right] \\[8mm]
&=& \frac{2{\pi}\alpha^2}{{\xbj}Q^4}
   \left[ (2-2y+y^2)\, {\fsq} - y^2{\fl} \mp {(2y-y^2)}\;{\xbj}{\fthree} \right]\; \left[ 1 + \delta_r \right]
\end{eqnarray}
with
\begin{eqnarray}
\fl({\xbj},{\qsq})&=&\fsq({\xbj},{\qsq}) - 2\xbj\fs({\xbj},{\qsq}),
\end{eqnarray}
and where
$\delta_r({\xbj},{\qsq})$ represents a small correction for electroweak 
radiative effects.
The contribution of the parity--violating part from Z$^0$--exchange, 
{\fthree}({\xbj},{\qsq}),
does not exceed 1\% for \mbox{${\qsq} < 10^3\;{\gevsq}$,} and
can be reliably estimated using
the expectations of the electroweak Standard Model. The Z$^0$ contribution
to {\fsq} is similarly small and can be corrected for in the same manner.
Thus the differential cross section is primarily determined by 
the value of the electromagnetic part of {\fsq}, which we call {\fsqem}
\begin{eqnarray}
 \frac{d^2\sigma^{e^{\pm}p}}{d{\xbj}d\qsq} = \frac{2{\pi}\alpha^2}{{\xbj}Q^4}
   \left[ 2(1-y) + \frac{y^2}{1+R} \right] \; {\fsqem}
   \; \left[ 1 + \delta_Z \right]\; \left[ 1 \mp \delta_3 \right]\; \left[ 1 + \delta_r \right],
\end{eqnarray}
with $R$ defined as
\begin{eqnarray}
R({\xbj},{\qsq})&=&\frac{\fl({\xbj},{\qsq})}{2x{\fs}({\xbj},{\qsq})} \, = \, \frac{\fsq({\xbj},{\qsq})}{2{\xbj}\fs({\xbj},{\qsq})} \, - \, 1,
\end{eqnarray}
and the small positive values of $\delta_3({\xbj},{\qsq})$ and $\delta_Z({\xbj},{\qsq})$ are the 
aforementioned corrections. The value of $R$ contributes corrections
of order 10\% at values of $y \lsim 1$ and can be calculated in the context
of perturbative quantum chromodynamics (pQCD)~\cite{pl_76_89}.

\begin{figure}[htbp]
\begin{center}
\includegraphics[width=0.8\linewidth,bb=88 117 473 667,clip=]{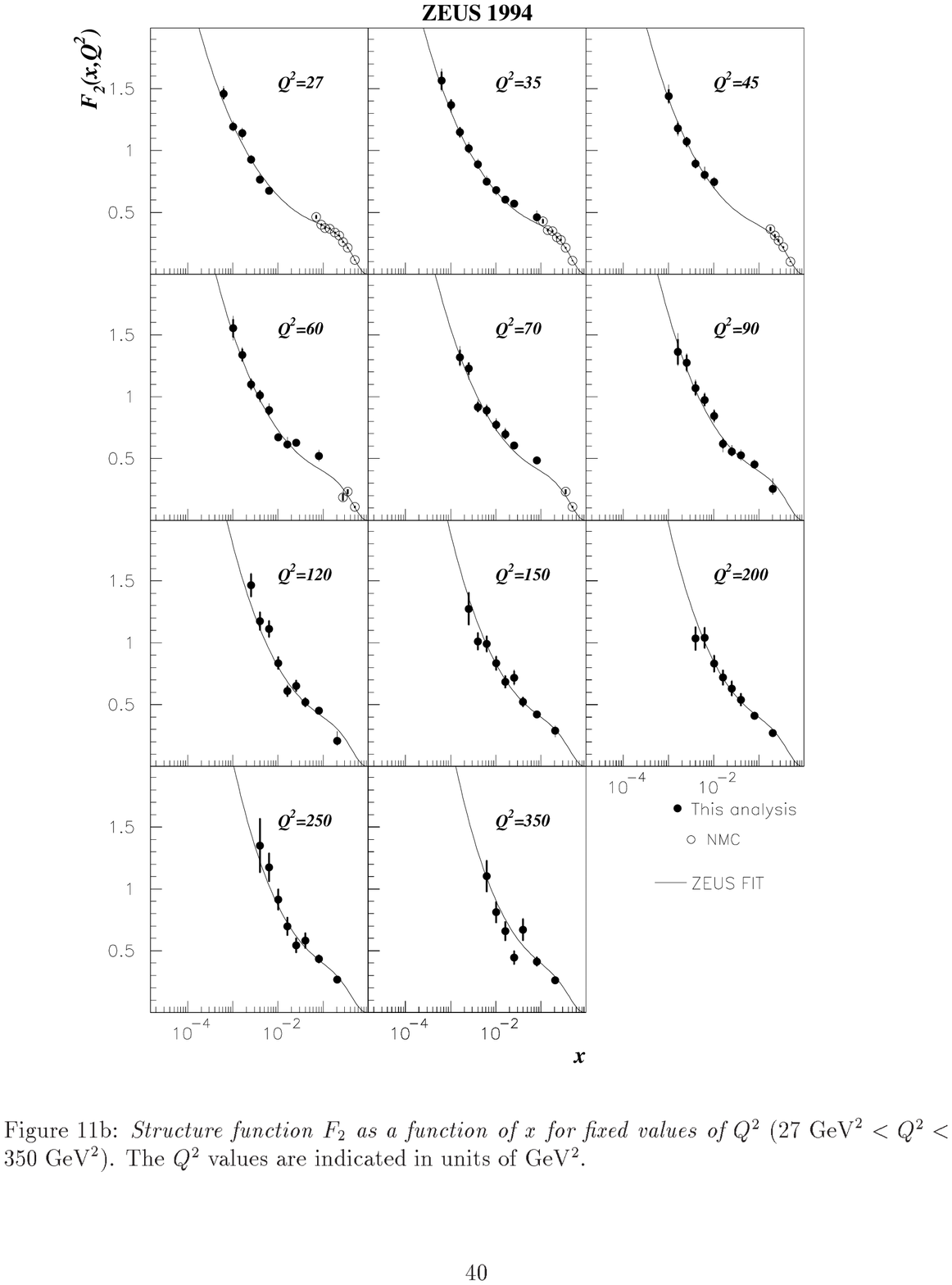}
\end{center}
\caption
{
\label{fig:f2}
\it An example of measurements by the ZEUS
collaboration of the structure function {\fsq}({\xbj},{\qsq}), shown 
as the solid circles, and the results of a QCD NLO fit
to the data~\protect\cite{zfp_72_399}. The {\qsq}--values are given in units
of {\gevsq}. 
The error bars represent the sum in quadrature of the systematic 
and statistical uncertainties. The open circles show 
measurements obtained by the NMC collaboration
in fixed--target studies with a muon beam~\protect\cite{pl_364_107}. 
} 
\end{figure}

The H1 and ZEUS collaborations have 
published a series of
measurements of {\fsq}({\xbj},{\qsq}) 
in the range \mbox{$3.5 \cdot 10^{-5} < \xbj < 3.2 \cdot 10^{-1}$} and 
\mbox{$1.5 < \qsq < 5.0 \cdot 10^{3} \; \gevsq$}
%
~\cite{pl_299_385,*np_407_515,*pl_321_161,*pl_303_183,np_439_471,
*pl_316_412,zfp_65_379,zfp_72_399}. 
Figure~\ref{fig:f2} shows an example of these measurements featuring 
the strong {\xbj}--dependence and compares them to the measurements
with muon beams at higher {\xbj} performed by the NMC 
collaboration~\cite{pl_364_107}. Also shown is the result of 
a fit~\cite{pl_332_393}
to the data based on a next--to--leading--order (NLO) approximation
to the Dokshitzer--Gribov--Lipatov--Altarelli--Parisi 
(DGLAP)~\cite{sjnp_15_438,*sjnp_15_675,*sjnp_20_95,*jetp_46_641,*np_126_298}
evolution equations which describe the dependence on {\qsq}.
Such a dependence of {\fsq} on {\xbj} has 
long been the subject of theoretical interest.
In 1974 De~R\'ujula {\em et al.}~\cite{pr_10_1649} motivated a dependence growing
with 
${\xbj^{-1}}$ more weakly than a power law but more rapidly than any power of
its logarithm. L\'opez {\em et al.}~\cite{prl_44_1118,*dr_96_87} 
have discussed the possibility of
a power--law dependence. In section~\ref{sec:gammapxsect} we will see how this {\xbj}--dependence is related to the
energy dependence in the total photon--proton cross section.

\begin{figure}[htbp]
\begin{center}
\includegraphics[width=0.7\linewidth,bb=12 143 563 704,clip=]{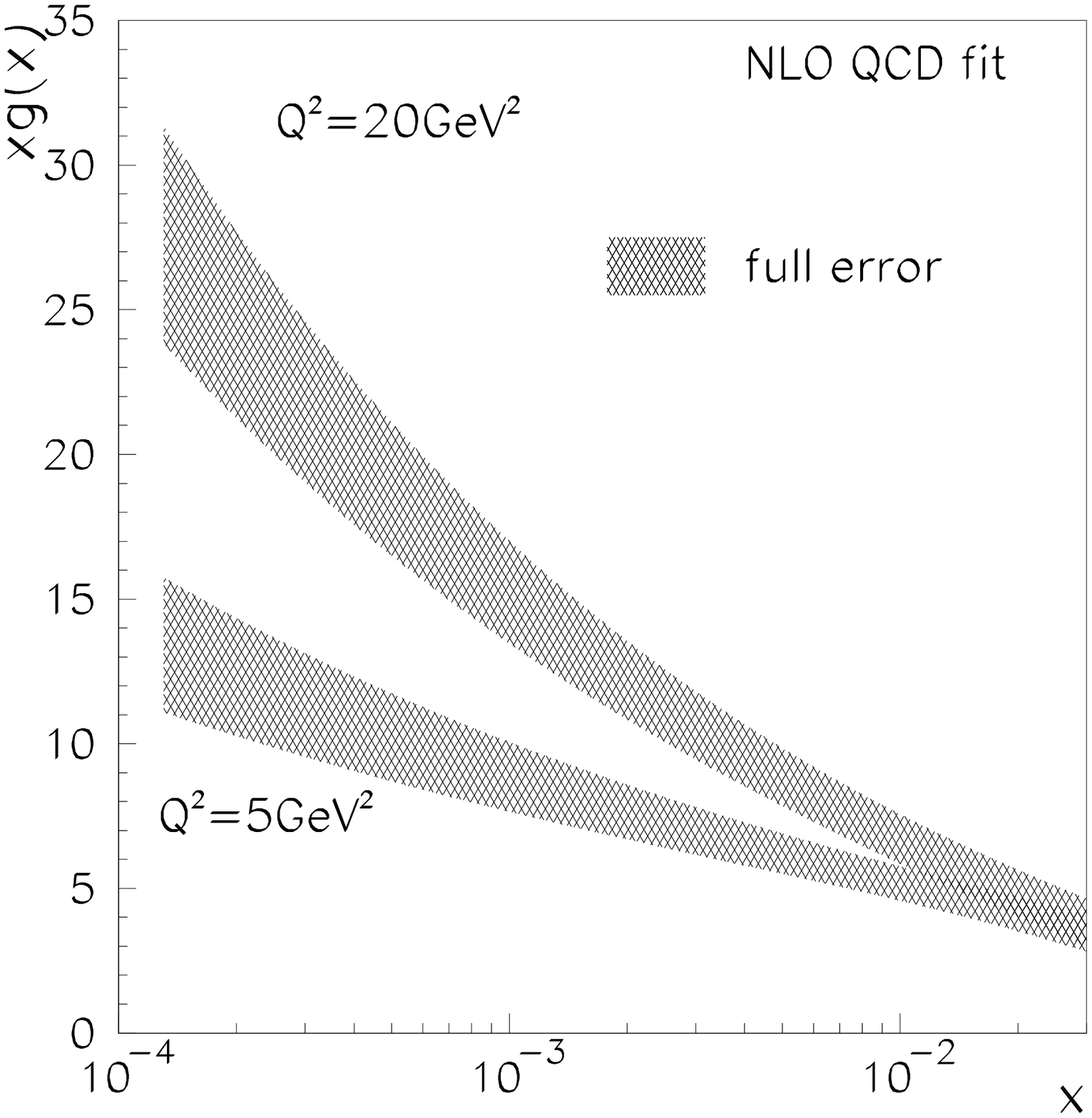}
\end{center}
\caption
{
\label{fig:xg}
\it Results for the gluon momentum density ${\xbj}G({\xbj,{\qsq}})$ 
determined by the H1 collaboration 
using NLO QCD fits to 
their 1994 measurements of the proton structure 
function {\fsq}~\protect\cite{np_470_3}. 
The shaded bands indicate the region of uncertainty in the result for
the gluon density, obtained by quadratic addition of systematic uncertainties
with those due to the uncertainties
in the fit parameters
determined via the minimization procedure.
} 
\end{figure}

While the photon does not couple to the gluons at the Born level, its
interactions with the quarks and antiquarks at low {\xbj} 
result in sensitivity to the gluon 
momentum distribution.
Both the H1 and ZEUS collaborations have derived gluon densities from
their measurements of the proton structure 
function~\cite{np_470_3,pl_354_494,*np_449_3,*pl_345_576}.  
The structure function {\fsq} can be decomposed into singlet and nonsinglet
parts. The latter is insensitive to the gluon content of the proton and
evolves with {\qsq} differently than the singlet part. The singlet part
depends on the gluon content
and its {\qsq}--dependence
allows the extraction of the gluon momentum density ${\xbj}G({\xbj},{\qsq})$.
Figure~\ref{fig:xg} shows the results for this density
obtained from a global NLO QCD fit by the H1 
collaboration
to their 1994 data. A steep decrease with {\xbj} is evident within the
margin of uncertainty, as the density function decreases by a factor
of approximately 
four over the measured region of {\xbj} for ${\qsq}=20\;{\gevsq}$.
We will see that the gluon density function plays a crucial
r\^ole in the interpretations 
of elastic vector--meson
production at HERA  based on perturbative QCD.

We close this brief review of HERA studies of the structure of the proton
by citing the first presentation of a result for 
the longitudinal structure function
{\fl}({\xbj},{\qsq})
~\cite{dr_96_236}.
The H1 collaboration has used the {\fsq} measurements at low $y$, where
the contribution from {\fl} is small, to extract
parton densities, then extrapolated these results to estimate {\fsq} at 
high $y$ and to isolate the contribution of {\fl}, obtaining the result
\mbox{$\fl \, = \, 0.52 \pm 0.03 \;(stat)^{+0.25}_{-0.22} \;(sys)$} at 
\mbox{$\xbj \, = \, 2.43 \cdot 10^{-4}$} and 
\mbox{$\qsq \, = \, 15.4 \; \gevsq$}. The extrapolation algorithm
presumes the validity of the DGLAP evolution procedure. The result is consistent
with a QCD calculation which employs the gluon density function determined
from the measurement of {\fsq}. The experimental
challenge of accurately measuring the energy of the scattered electron
in the range \mbox{6$\, < \, E'_e \, <$ 11~{\gev}} at high $y$ is evinced by
the systematic uncertainties in the result.

\subsubsection{Hadronic Final States in Deep Inelastic Scattering}
The electron--proton center--of--mass system at HERA moves in the laboratory
reference frame in the proton flight direction with a value for the Lorentz
factor $\gamma$ equal to~2.8. 
The unique laboratory reference frame provided by the advent of these 
asymmetric storage rings
in the study of deep inelastic scattering results in good acceptance for
the scattered proton constituent and thus opens a vast field of study
of hadronic final states. The high center--of--mass energy furthermore
extends the accessible rapidity range and results in more collimated, and
hence more easily identified, jets in the final state. A wide range of
investigations have been performed, including dijet production, the measurement
of $\alpha_s$ from multijet rates, energy flows, 
inclusive charged particle distributions, strangeness production, charm
production, and much more
~\cite{pl_298_469,*zfp_61_59,*zfp_63_377,*pl_346_415,*np_445_3,*pl_356_118,*pl_358_412,*zfp_72_573,*dr_96_215,
*pl_306_158,*zfp_59_231,*zfp_67_81,*zfp_67_93,*pl_363_201,
*np_480_3,
*zfp_68_29,
*zfp_72_593,
*pl_338_483,zfp_70_1}.
The early investigations of event topologies in the deep inelastic
scattering revealed a class of events with no energy flow in the forward
region, amounting to approximately 10\% of the total observed rate. Thus a
clean signal for diffractive physics at high photon virtuality was obtained,
opening the way to a variety of studies ranging from inclusive
transverse momentum distributions and energy flows to studies of Pomeron
structure
~\cite{
zfp_70_1,
np_429_477,*pl_348_681,*zfp_70_609,
*zfp_68_569,*zfp_70_391,*pl_315_481,pl_384_388}. 

The investigations of exclusive final states in deep inelastic scattering
have now become of topical theoretical interest, as diffractive vector--meson
production at high {\qsq}
has been recognized to yield information on the relative contributions of hard
and soft processes. A survey of the current status of these theoretical
considerations is presented in section~\ref{sec:theory}.
A great deal of data has become available and these 
studies will be described in detail in section~\ref{sec:expvm}.
\subsubsection{Photopro\-duction}
The ability of the HERA experiments
to trigger on the hadronic final state independently of the scattered
electron results in sensitivity to photoproduced final states
provided by 
the quasi--real photons radiated at small angles by the electron beam.
In our discussions of the HERA data
we will refer as ``photopro\-duction'' 
to studies of final states where the scattered electron was
either undetected or detected in special--purpose detectors at very small
scattering angle, comparing the results to those
obtained in photon beams, basing the validity of the comparison on the
region of photon virtuality covered 
($\langle \qsq \rangle \approx 0.01-0.1 \, \gevsq$) and on the 
very small contribution by longitudinal photons to
the photoabsorption cross section compared to that
by transverse photons at such low values of {\qsq}. The HERA data
extend the range of accessible photon energies by two orders of magnitude
compared to that achieved in fixed--target experiments.

The large photopro\-duction cross section at HERA, three orders of magnitude
higher than that for deep inelastic processes where the scattered electron
is observed in the central detectors, has resulted in the availability
of high statistics investigations of photon interactions and hadronic final
states characterized by transverse energies far exceeding the QCD scale
parameter $\Lambda_{\mathrm{QCD}}$. An overview of photopro\-duction
physics topics addressed by
the first three years of HERA operation can be found in ref.~\cite{dr_94_215}.
A general review of theoretical issues in photopro\-duction is given in
ref.~\cite{np_407_539}. The H1 and ZEUS collaborations have published
numerous studies on final--state jet topologies, inclusive charged particle
rapidity distributions and energy flows, quantified the occurence of
rapidity gaps between jets and distinguished contributions from direct and
resolved photon interactions
%
~\cite{pl_297_205,*pl_314_436,*pl_328_176,*np_445_195,*pl_358_412,*zfp_70_17,
*pl_297_404,*pl_322_287,*pl_342_417,*pl_354_163,*pl_384_401,*pl_348_665,np_435_3,*zfp_67_227}.
Diffractive processes have been identified via the presence of a rapidity
gap with respect to the proton direction, resulting in
a number of studies of diffractive 
hadronic final states, including the observation
of hard scattering
~\cite{np_435_3,*zfp_67_227,
pl_346_399,*pl_356_129,*pl_369_55}.
Cross sections for the photopro\-duction of charmed mesons have been determined
~\cite{
np_472_32,
*pl_349_225}.
The measurements of total photon--proton cross sections and 
vector--meson photopro\-duction are essential to the topic of this article
and will be discussed in detail in section~\ref{sec:PhysHERA}.
\subsection{Vector--Meson Production in Photon--Proton Interactions}
\label{sec:vmprod}
\subsubsection{Experimental Techniques Prior to HERA Operation}
\label{sec:earlystudies}
A comprehensive description 
of vector--meson photopro\-duction and electroproduction
experiments prior to 1978 is available in the review by
Bauer {\em et al.}~\cite{rmp_50_261}. A wide variety of photopro\-duction 
studies on hydrogen targets were performed
at the Cambridge Electron Accelerator, the Wilson Synchrotron Laboratory
at Cornell, SLAC, and DESY. Photon beams
of energies between 2 and 18~{\gev}
were obtained via electron--beam bremsstrahlung and via backscattered laser 
beams. Hybrid bubble chamber studies at SLAC~\cite{pr_5_545} and 
streamer chamber studies at DESY~\cite{np_113_53}
permitted complete reconstruction of the final state in many cases, while
wire chamber spectrometers of relatively limited acceptance detected either
the proton alone, reconstructing the vector meson via missing--mass techniques,
or the vector--meson decay products, or both, in two--arm setups. Studies of
{\rhoz} mesons dominate the statistics, and asymmetries in the dipion
mass spectrum generated much theoretical interest, as did the helicity
analyses of the decay angle distributions. Evidence for the 
interference of the {\rhoz} and $\omega$ production amplitudes was 
observed~\cite{sjnp_9_69,*np_29_349,*pl_39_289,*np_256_365}
in the dipion invariant mass spectra, which also exhibited
further asymmetries in the {\rhoz} resonance region attributed 
to interference with nonresonant backgrounds.
Photopro\-duction on protons of the 
$\omega$, $\phi$, ${\rho}'$ was 
studied~\cite{prl_43_657,*prl_43_1545,*np_209_56}. 
Electro-- and muoproduction of the {\rhoz}, $\omega$, and $\phi$ 
mesons up to {\qsq}
values of about 2~{\gevsq} was
investigated at DESY, SLAC, and
Cornell~\cite{np_113_53,np_36_404,*pr_8_687,*pr_10_765,*pr_19_1303,*pr_24_2787,*pr_25_634}.
High--energy muoproduction of the {\rhoz} meson was studied by
the CHIO collaboration at FNAL~\cite{prl_38_633,*pr_26_1} and 
by the EMC~\cite{pl_161_203,zfp_39_169} and NMC~\cite{zfp_54_239,np_429_503}
collaborations at CERN.
Recently 
results for {\rhoz} muoproduction from the FNAL experiment E665 have also
become available~\cite{mpi_97_3,*pa03_09}, 
reaching a virtual photon energy of 420 GeV.
Diffractive {\rhoz} production has also been investigated in bubble chamber studies
of neutrino and antineutrino interactions with 
protons~\cite{zfp_58_375,*ijmp_9_513}.
Soon after their discovery, the {\jpsi} and ${\psi}(2S)$ mesons 
were observed in photoproduction at Cornell~\cite{prl_35_1616}, 
FNAL~\cite{prl_34_1040,*prl_36_1233}, and SLAC~\cite{prl_35_483}, 
the FNAL measurements extending the photon
energy range to 100~{\gev}.
Photopro\-duction of charm states prior to 1985 is covered in the review
by Holmes {\em et al.}~\cite{arnps_35_397}. The production of {\jpsi} and ${\psi}(2S)$
mesons was studied using fixed--target spectrometers
in tagged and wideband photon beams for photon energies
up to 225~{\gev} at FNAL~\cite{prl_48_73,*prl_52_795}. 
Comparable energies were reached in muoproduction
Berkeley/Princeton/FNAL collaboration~\cite{prl_43_187} at FNAL and 
by the EMC~\cite{np_213_1} and NMC~\cite{pl_332_195} collaborations
at CERN. The NA14 collaboration at CERN also measured {\jpsi}
photopro\-duction in a CERN photon beam at such energies~\cite{zfp_33_505}.
In more recent years the fixed--target experiment E687 
in the photon beam at FNAL has 
published results for {\jpsi} photopro\-duction~\cite{pl_316_197}.
\subsubsection{The Contribution from Investigations at HERA}
\label{sec:PhysHERA}
This section briefly summarizes the contributions to this field of research
presented by the two collaborations H1 and ZEUS prior to the end of 1996. 
The range of forthcoming \mbox{studies} 
at HERA to be expected in the coming years
will be covered in section~\ref{sec:future}. The section \mbox{entitled} ``Suggestions
for Future Work'' in the review of Bauer {\em et al.}~\cite{rmp_50_261} presents
an instructive comparison to these issues.

The measurements at HERA we discuss in this article
concern a range of photon energies in the proton rest frame
extending from 400 {\gev} to 40 TeV. 
Of immediate consequence are the small lower kinematic bounds on
photon virtuality, $\qsq_{min}$, and 
momentum transfer at the 
hadronic vertex (see section~\ref{sec:kinematics}), 
rendering their influence on
the determination of kinematical variables negligible for most studies. 
The variety of
diffractive processes which contribute coherently to the interaction
with the proton increases with decreasing $|t|_{min}$~\cite{dr_95_47}. 
Thus the kinematic
region covered at HERA allows an unprecedented sensitivity to the partonic
structure of the photon in its diffractive interactions.
In contrast to the fixed--target studies, the HERA experiments study
both photopro\-duction and electroproduction processes simultaneously, 
distinguishing them via selection algorithms requiring
either a small energy deposit in the main calorimeters and at least one
charged particle track
or a large electromagnetic energy deposit signalling the detection
in the central detector of an electron scattered with large momentum transfer.

The present HERA results also extend the range of accessible photon virtualities
by more than two orders of magnitude, 
reaching {\qsq} values of $5 \cdot 10^3$~{\gevsq}. Thus the energy dependence
of the total photon--proton cross section for highly virtual photons has been
studied at high energy for the first time, and comparisons to the studies
at low energy provide high sensitivity even to the 
very weak energy dependences expected in diffractive processes. Particular
attention has been given to the intermediate region of photon virtuality
\mbox{0.1 $< \qsq <$ 1.0~{\gevsq}}, where the transition from
vector dominance to the region of applicability of perturbative
quantum chromodynamical calculations can be studied.

Investigations of photoproduction of {\rhoz}, $\omega$, $\phi$, $\rho'$,
{\jpsi}, and ${\psi}(2S)$ mesons have been published.
In addition results on 
the diffractive production of the {\rhoz}, $\phi$, and $\jpsi$
mesons at high {\qsq} have been presented.
Each meson has been studied only in its dominant two--charged--particle
decay mode, but for the $\omega (\pi^0\pi^+\pi^-)$ and the 
$\rho' (\pi^+\pi^-\pi^+\pi^-)$. The energy dependence of the forward
($t = 0$) cross sections has been studied via comparison with
measurements at low energy. Production ratios have been quantified 
in this high--energy region where, for example, the rates of $\omega$ and
$\phi$ production are expected to become comparable since the contributions to
$\omega$ production other than Pomeron exchange become negligible.
The dependence on the momentum transfer at
the proton vertex $t$ has been studied up to values of $|t|$
of about 4 $\gev^{2}$. Helicity analyses of the vector--meson decay
products have been presented, leading to estimations for
the ratio of longitudinal to transverse photon cross sections.
This variety of measurements now avail\-able 
has promp\-ted a great deal of new theoretical work, which we
summarize in the following section.
\cleardoublepage
\section{Contemporary Theoretical Approaches}
\setcounter{equation}{0}
\label{sec:theory}
A broad array of theoretical considerations has been brought to bear on the 
problem of exclusive vector--meson production at HERA over the past few years.
The high level of interest has arisen in part due to 
the wide kinematic range covered
by the data, with photon virtualities ranging from 10$^{-10}$ to 30~{\gevsq}
and photon--proton center--of--mass energies from 40 to 200~{\gev}.
The range of vector--meson masses for which measurements have become
available, extending from that of the {\rhoz} to that of the {\jpsi}, provides a
further parameter to be used in testing theoretical ideas. The photopro\-duction
data show features characterizing diffractive processes and measurements
at high {\qsq} have yielded strong evidence for the applicability of
approaches based on perturbative QCD calculations. Thus the HERA measurements
are demanding the reconciliation of theoretical ideas which have hitherto
been tested only in distinct kinematic domains. The purpose of this
chapter is to summarize the current status of these theoretical approaches.
Following a brief discussion of model--independent issues, we summarize
a variety of topical nonperturbative calculations and end with a discussion
of the status of calculations based on perturbative quantum chromodynamics.
\subsection{General Considerations}
\label{sec:general}
The early measurements at HERA of total photon--proton cross sections and
of {\rhoz} meson production exhibited characteristics typical of soft
diffractive processes. The magnitudes of the cross sections were comparable
to those measured at an order of magnitude lower energy, and low momentum
transfers $t$ were strongly favored. In the brief outline of the general
characteristics of such processes we follow the discussion of 
ref.~\cite{perl}. A beautifully detailed and comprehensive discussion
of the Regge theory which successfully describes these phenomena can be found
in ref.~\cite{collins}.

The strong forward peaking of the diffractive processes mentioned above
refers to an exponential $t$--dependence in the region of small $|t|$:
\begin{eqnarray}
\label{bdep}
\frac{d\sigma}{d{|t|}} &=& A e^{-b |t|} 
\end{eqnarray}
In terms of a simple optical model interpretation, the exponential slope
$b$ is related to the radii of two Gaussian--shaped 
objects participating in the elastic scatter
\begin{eqnarray}
b &\propto& R_1^2 + R_2^2.
\end{eqnarray}
This article discusses the HERA results on elastic scattering from the proton,
for which the relevant squared 
radius has been determined to be near 4 ${\gev}^{-2}$
in its hadronic interactions, so the slopes discussed in this article yield
information on the size of the object scattering on the proton, usually
interpreted as a $q \bar{q}$--wave packet.

The optical theorem relating the imaginary part of the forward scattering
amplitude to the total cross section is useful here, since for purely
diffractive elastic processes the real part is zero and measurements 
show the imaginary part exceeding the real part
by at least a factor of
three even
at low energies. At HERA energies the real part is less than a few percent
and we can relate the above differential cross section to the total cross
section as follows:
\begin{eqnarray}
\left. \frac{d\sigma}{d{|t|}} \right|_{t=0} \, =  \, A \, = \, \frac{\sigma_{tot}^2}{16 \pi}.
\end{eqnarray}
If we now integrate the differential cross section to get the total
elastic cross section and express the cross sections in millibarns 
and the exponential slope $b$ in units
of {\gev}$^{-2}$ we obtain,
\begin{eqnarray}
\label{eq:optical}
\sigma_{el} &=& \frac{0.051 \, \sigma_{tot}^2}{b}. 
\end{eqnarray}
In this manner it becomes obvious that the energy dependences of the
elastic and total cross sections are closely related by that
of the exponential slope $b$. Such an energy dependence, termed ``shrinkage''
since the width of the forward diffractive peak in the $t$--distribution
decreases with energy,
has been accurately measured in low--energy hadronic interactions
and found to vary starkly
with the process considered. 
Of uncertain dynamical origin, this effect
remains topical today in the investigations of the HERA data.

The energy dependence of of the forward diffractive 
cross section derived from Regge theory
follows a
power law, where the power is determined by the exchanged trajectory:
\begin{eqnarray}
\frac{d\sigma}{d{|t|}} \propto e^{-b_0 |t|} \left( \frac{s}{s_0} \right)^{2({\alpha}(t)-1)}. 
\end{eqnarray}
The energy scale factor $s_0$ is determined empirically to be comparable to
the hadronic mass scale of about 1~{\gevsq}. The $t$--dependence of the 
linear trajectory
\begin{eqnarray}
\alpha (t) &=& \alpha_0 + {\alpha}' t
\end{eqnarray}
leads directly to the aforementioned shrinkage effects 
\begin{eqnarray}
b &=& b_0\,+\, 2\,{\alpha}' \ln{\frac{s}{s_0}}
\end{eqnarray}
and an energy dependence of the total cross section 
determined by the $t$=0 intercept of the trajectory (eq.~\ref{eq:optical})
\begin{eqnarray}
\label{eq:powerdep}
\sigma_{tot} \propto  \left( \frac{s}{s_0} \right)^{(\alpha_0-1)}. 
\end{eqnarray}
The weak energy dependence of the total scattering cross sections measured
at low energy motivated the introduction of the Pomeron trajectory with
intercept near unity, and these considerations were again brought to the
fore when the early HERA results for the total photopro\-duction cross
sections gave results quantitatively comparable to the measurements at low
energy.

We conclude this general discussion of diffractive processes with a remark
clarifying a convention of terminology. Since the exclusive reactions
$\gamstar + p \rightarrow V + p$ we discuss in the article involve 
more final--state rest mass than in the initial state, these reactions are not,
strictly speaking, ``elastic'' interactions. However, the recent literature
has adoped the term ``elastic'' for these processes. The reason is based
on kinematics: the minimum momentum transfer required by the
generation of the final--state rest mass is negligibly small compared
to typical values of the momentum transfer measured at the high
center--of--mass energies of the HERA regime. The reactions we consider
are called ``diffractive inelastic'' in ref.~\cite{perl}, for example, while
today's theoretical and experimental articles reserve the term ``inelastic''
for processes in which the photon or proton
dissociates into several final--state 
particles. Thus an essential 
difference between diffractive elastic and diffractive inelastic processes
discussed in ref.~\cite{perl}, namely the conservation of  intrinsic spin
and parity in the elastic interactions, 
becomes moot in the context of the present article, since in
the helicity analyses of our ``elastic'' reactions we investigate the possibility
of spin and
parity changes in order to gain information on the dynamical origin of
the interaction.

We continue our discussion of general theoretical considerations by turning
to the issue of vector--meson production ratios. The quark--flavor 
structure of the vector mesons, together with the assumption of a
flavor--independent production mechanism governed by the electromagnetic
quark--photon coupling in the quark--parton model result in 
production rates in the relationship
\begin{eqnarray}
{\rhoz}\,:\,\omega\,:\,\phi\,:\,\jpsi &=& 9\,:\,1\,:\,2\,:\,8.
\end{eqnarray}
This flavor symmetry can be expected to be broken by effects related
to the vector meson masses. Indeed, a suppression factor of about a factor
of three was observed in elastic {$\phi$} photoproduction at low 
energies~\cite{rmp_50_261} and for {\jpsi} photoproduction~\cite{arnps_35_397}
this suppression was shown to be more than two orders of magnitude.
On a model--independent basis it is reasonable to expect
the symmetry to be restored for \mbox{$\qsq \gg M_V^2$}. Further discussion
of expectations for the production ratios at high {\qsq} can be found
in ref.~\cite{dr_95_47}.

General arguments favoring the dominance of vector--meson photoproduction
by longitudinal photons over that by transverse photons at high {\qsq} 
have been made based on the point--like nature of the longitudinal 
photon. The resulting factorization theorem and dimensional counting
considerations~\cite{hep96_11_433} 
result in a  prediction for the {\qsq}--dependence of
the \mbox{$\gamma^* p \rightarrow V p$} cross section to vary as $Q^{-6}$.
However, there are no model--independent arguments for the {\qsq}--dependence
in the ratio of the two cross sections (aside from the fact that the ratio
must vanish at ${\qsq}=0$ due to gauge invariance), complicating the comparison
with the HERA data, which comprise approximately equal contributions
from transverse and longitudinal photons. 

If helicity were conserved in the photon/vector meson transition, then
the ratio of the longitudinal to transverse cross sections could be
determined by measuring the vector meson polarization via its decay
angle distributions. This transition poses a fundamental quandary,
however, and questions concerning the structure of the vector--meson
spin--density matrix have played a prominent r\^ole thoughout the history
of investigations into exclusive vector--meson photo-- and leptoproduction.
The quandary arises from the rest mass of the vector meson, which 
results in an ambiguity in the choice of Lorentz frame appropriate to
a description of the underlying dynamics. Early studies at low energy
and low {\qsq} employed comprehensive 
helicity analyses to show that a frame could indeed
be defined in which
the spin--density matrix elements were independent of the kinematics and
in which helicity was conserved. It is important to bear in mind while
reading this article, however, that no such analysis has yet been possible
with the HERA data, and conclusions concerning the ratio of longitudinal
to transverse cross sections are based on an assumption motivated by
measurements in a kinematic region not covered by the HERA data. 
A detailed 
description of the spin--density matrix, which describes the helicity
structure of the interaction, will be presented in section~\ref{sec:helana}.

\vspace*{3mm}
\begin{center}
---------------------------------------------
\end{center}
The interpretation of the HERA photopro\-duction results for the lower mass
vector mesons as being of soft diffractive origin has been
quite successful. However, the measurements of {\jpsi} photopro\-duction
at high energy as well as the exclusive production of vector mesons
at photon virtualities greater than 7~{\gevsq} show starkly different
energy dependences. Model--independent theoretical 
considerations point to the
need for explanations based on perturbative 
quantum chromodynamics. The mass of the {\jpsi} introduces a hard scale
which casts doubt on the applicability of calculations based on single soft
Pomeron
exchange. And on general grounds one expects the longitudinal component
of the virtual photon to dominate the transverse component at high {\qsq},
so it is plausible to expect considerations based
on the optical model to lose
their validity. Indeed the experimental results exhibit energy
dependences incompatible with those derived from calculations based on
soft diffraction. These considerations have made the exclusive production
of vector mesons in high--en\-ergy $\gamma^* p$ interactions
the subject of lively theoretical interest from experts in both 
nonperturbative and perturbative approaches to the description of hadrons and
their interactions. In the following two sections we briefly summarize
the recent theoretical work in this field.
\subsection{Nonperturbative Approaches}
\label{sec:nonpert}
{\bf Donnachie and Landshoff}~\cite{pl_348_213} have employed
the highly successful parametrizations of the 
hadron--hadron total cross sections~\cite{pl_296_227,*rmp_68_611},
based on the exchange of an object
we will refer to as the phenomenological Pomeron, along with an
assumption of Vector--Meson Dominance 
(VMD)~\cite{rmp_50_261,np_14_543,prl_22_981,*sakurai} and the additive quark model
to describe vector--meson photopro\-duction. They used the $e^+e^-$ decay width of the {\rhoz} to estimate the {\gamstar}--{\rhoz}--coupling.
The Regge--theory--inspired fits
to the hadronic cross sections yield values for the Pomeron trajectory of
\begin{eqnarray}
\label{eq:alpha}
\alpha_{\Pma}(t=0) &\approx& 1.08 \hspace*{2cm} \alpha'_{\Pma} \, \approx \, 0.25 \; \gev^{-2},
\end{eqnarray}
resulting in the expectation of a power--law increase of the
elastic cross 
section with energy, the value of the
power less than $4(\alpha_{\Pma}(t=0)-1) \approx 0.32$, depending
on the $t$--range covered by the data 
(eqs.~\ref{eq:optical} and~\ref{eq:powerdep}), 
and in the expectation of shrinkage.
They furthermore suggested an extension of these considerations to high
{\qsq}, modeling the Pomeron as a pair of nonperturbative gluons, predicting
a $Q^{-6}$ dependence at high {\qsq}.

\vspace*{2mm}
The model of {\bf Haakman {\em et al.}}~\cite{hep95_07_394} derived an effective
intercept based on the hypothesis of multiple Pomeron exchanges 
at high energies. This intercept leads to a stronger energy dependence and 
successfully matches the HERA results for the energy dependence in the
total ${\gamma^*p}$ cross section as well as in exclusive {\rhoz} and 
{\jpsi}
photopro\-duction and in {\rhoz} production at high {\qsq}.

\vspace*{2mm}
{\bf Jenkovszky {\em et al.}}~\cite{hep96_08_384} have addressed the energy dependence
of {\jpsi} photopro\-duction.
They succeed in explaining the steep energy dependence within the framework
of single Pomeron exchange (indeed, the ``pure'' Pomeron of unity intercept)
attributing the steep energy dependence to preasymptotic effects
arising from the high {\jpsi} mass. This high mass alters the 
relationship between the elastic and inelastic contributions to the total
cross section, moving the contribution from the 
elastic cross section to higher energy compared to that from 
the elastic production of the low--mass vector mesons.

\vspace*{2mm}
{\bf Pichowsky and Lee}~\cite{pl_379_1} have discussed electroproduction of {\rhoz} mesons in terms
of a Pomeron--exchange model where the {\rhoz}--Pomeron coupling is
represented by a nonperturbative quark--loop, predicting the {\qsq}--dependence
of the elastic cross section in the region of transition to the perturbative
regime \mbox{0.1$< {\qsq} <$ 10~{\gevsq},} and successfully matching 
results from the fixed--target muon experiments. The energy dependence of
the elastic cross section in this model is weak, similar to that for the
phenomenological Pomeron exchange.

\vspace*{2mm}
The nonperturbative calculations of {\bf Dosch {\em et al.}}~\cite{hep96_08_203}
used a stochastical model of the QCD vacuum yielding linear confinement
to derive high--energy vector--meson leptoproduction cross sections
for each of the vector mesons,
based on hadron sizes. They found $t$--dependences which are less steep
than exponential and {\qsq}--dependences which do not reach the
asymptotic behavior of $Q^{-6}$ even for {\qsq} values as high as 10~{\gevsq}.
Of particular interest are their predictions for the {\qsq}--dependence
of the ratio of longitudinal to transverse cross sections,  
which show a much
weaker dependence than the proportionality to {\qsq}/${M_V^2}$ 
expected on the basis
of vector--meson dominance, and which match the currently available HERA
measurements of this ratio. However, this model has no prediction for the energy
dependence of the elastic cross sections.

\vspace*{2mm}
{\bf Niesler {\em et al.}}~\cite{pl_389_157} have published a calculation
of {\rhoz} leptoproduction which derives the $M_{\pi^+\pi^-}$ spectrum from
the two--pion contribution to the photon spectral function, which is given
by the pion form factor. The coupling of the {\rhoz} meson to the nonresonant
continuum was understood in terms of an effective field theory approximating
QCD for composite hadrons. The calculation yielded good agreement to 
a measurement of the mass spectrum at low energy where a cusp arising from
{\rhoz}-{$\omega$} mixing is evident. Consistency with the 
ZEUS photopro\-duction
result~\cite{zfp_69_39} was also achieved, 
though the statistical accuracy of this early
measurement at high energy is much weaker, preventing a clear observation
of such an interference cusp, if it is present.

\subsection{Calculations Based on Perturbative Quantum Chromodynamics}
\label{sec:pqcd}
Shortly before data--taking began at HERA in 1992, {\bf Ryskin}~\cite{zfp_57_89}
suggested that diffractive {\jpsi} electroproduction could be calculated
in perturbative QCD (pQCD) even at low {\qsq}, since the mass
of the {\jpsi} provided the necessary scale large compared to the
confinement scale $\Lambda_{\mathrm{QCD}}$.
His calculation, based on the exchange of a momentum--symmetric
pair of perturbative gluons, indicated that the cross section is sensitive
to the square of the gluon density evaluated at effective values
of {\qsq} and {\xbj}:
\begin{eqnarray}
\bar{Q}^2 &=& \frac{\qsq+M_{J/{\psi}}^2}{4} \hspace*{5mm} and 
\hspace*{5mm} \bar{\xbj} \, = \, \frac{4\bar{Q}^2}{W^2}.
\end{eqnarray}
Interest in this approach was further stimulated when the early experimental
results showed both that the proton structure function rises sharply
as {\xbj} decreases and that the energy dependence of the diffractive
{\jpsi} photopro\-duction cross section is much steeper than expected
from a soft diffractive process. The Ryskin calculation has since been
implemented in a Monte Carlo simulation package~\cite{cpc_100_195}.
Some theoretical uncertainties, including the need for  the unmeasured
two--gluon form factor, mitigate the quantitative use of this process as
a measure of parton densities, but hope remains strong that this quadratic
dependence will allow high sensitivity to the gluon density. A first attempt
to discriminate among parton density parametrizations has been 
made by {\bf Ryskin {\em et al.}}~\cite{hep95_11_228}, where the authors 
extended the calculation beyond the leading--log approximation
and indeed
concluded 
that the energy dependence of the diffractive {\jpsi} photopro\-duction
cross section provides a more sensitive probe of the gluon density than
the proton structure function measurements.

\vspace*{2mm}
{\bf Brodsky {\em et al.}}~\cite{pr_50_3134} have generalized this type of calculation
to the diffractive production of low--mass vector--meson states (indeed
any hadronic states with the quantum numbers of the photon), making several
distinctive predictions for measurements in the kinematic range
of low~{\xbj}, $s \gg \qsq + M_V^2$, and $\qsq \gg \Lambda_{\mathrm{QCD}}^2$ 
addressed by measurements at HERA.
These included the predominance of the amplitude for longitudinally polarized
photons producing longitudinally polarized vector mesons, $t$--slopes 
independent of the produced vector--meson wave function and flavor content,
a slow increase of these $t$--slopes with energy, ratios of the 
vector--meson 
production cross sections closely related to standard SU(3) wave function
predictions, as well as the $Q^{-6}$ dependence and the dependence on the
square of the gluon density function in the proton. The calculation
employed the momentum distribution of constituents in 
the virtual photon, the two--gluon
form factor of the proton depending on the gluon density 
function $G({\xbj},{\bar{Q^2}})$ (which
alone determines the $t$--dependence at high {\qsq}) 
and the nonperturbative wave function 
of the final--state vector meson, for which they
used the minimal Fock--state wave functions of Brodsky and 
Lepage~\cite{pr_22_2157}. 
Their results reduce to those
of Ryskin when a simple nonrelativistic symmetric pair of quarks
is assumed for the vector meson.
They showed in
a double--leading--log approximation that the longitudinal part of the
forward cross section can be written
\begin{eqnarray}
\left. \frac{d\sigma_{{\gamma^*p} \rightarrow \mbox{V} \, p}}{d{|t|}} \right|_{t=0} \, =  \, A_{V} \, \frac{\alpha_s^2({\bar{Q^2}})}{{\alpha}\bar{Q^6}} \left| \left[ 1+\frac{i\pi}{2}\frac{d}{d(\ln{\xbj})} \right] {\xbj} G({\xbj},{\bar{Q^2}}) \right|^2,
\end{eqnarray}
where $\alpha_s$ is the strong coupling constant and $\alpha$ the fine structure
constant. The additional term proportional to the logarithmic derivative of the
gluon momentum density is a perturbation arising from the real part of
the amplitude calculated via dispersion relations, and results in 
an effective QCD Pomeron intercept exceeding 
unity at small {\xbj} and high {\qsq}.
The normalization factor A$_{V}$ 
is a constant derived from the vector--meson wave function
which leads to an uncertainty of approximately a factor of
three in the normalization
of the predicted cross section for the {\rhoz}, for example. 
For the interactions of the longitudinal photons 
the ratio of the production
of longitudinally polarized vector mesons to that of transversely
polarized vector mesons was predicted to rise even faster than 
$\qsq/{M^2_{V}}$. Particularly interesting predictions
were also made for diffractive leptoproduction of vector mesons off nuclei,
including decisive tests of color transparency, but these will not be
tested at HERA until nuclear beams are introduced.

\vspace*{2mm}
{\bf Frankfurt {\em et al.}}~\cite{pr_54_3194} have applied the above ideas in
quantitative calculations of the magnitudes, energy dependences, and
cross section ratios for vector--meson leptoproduction and compared
them to the NMC and HERA measurements. They discussed pQCD predictions
for the vector--meson production ratios at high {\qsq}, as well as for
the production of excited states of vector mesons and for the diffractive
production off nuclei. The kinematic limits of validity for these perturbative
calculations were explicitly quantified. They concluded that the study
of exclusive
diffractive production of vector mesons presents an effective method
for determining the vector--meson 
wave functions and may therefore provide useful tests
of QCD lattice gauge calculations at high energy.

\vspace*{2mm}
{\bf Hoodbhoy}~\cite{hep96_11_207}
has addressed the magnitude of two contributions to diffractive
{\jpsi} electroproduction
neglected
in the early calculations: that of the Fermi motion of the $c$ quarks in the
{\jpsi} and that of asymmetric gluon pairs in the proton, finding them to
be small.

\vspace*{2mm}
{\bf Collins {\em et al.}}~\cite{hep96_11_433} have also succeeded in generalizing the
calculations beyond the leading log approximation. The calculation
was restricted to longitudinally polarized photons, though the contributions 
by transversely polarized photons were extensively discussed.
One interesting conclusion is that the ratio of longitudinally to transversely
polarized vector mesons is sensitive to the polarized quark densities in 
the proton, so that information on these densities can be obtained from
diffractive vector--meson production without requiring polarized proton beams.
This remark holds only at high {\xbj}, however, since the production 
of transversely polarized vector mesons
is predicted to vanish at small {\xbj}.
The predicted ratio of longitudinal to transverse vector--meson 
polarization is much larger than the value measured at HERA,
leading the authors to surmise that the HERA results
have a substantial contribution from transversely polarized photons.

\vspace*{2mm}
{\bf Martin {\em et al.}}~\cite{hep96_09_448} have addressed this issue of the {\qsq}--dependence of the vector meson polarization,
showing that not only the above perturbative calculations but
also nonperturbative calculations of vector--meson production by transverse
photons predict contributions to the production of transversely polarized
{\rhoz} mesons 
which decrease more rapidly with {\qsq} than the early experimental
results from HERA indicate. They proposed a model of open production of
light $q\bar{q}$--pairs and used 
an assumption of parton--hadron duality to calculate the ratio
of the production cross section for 
longitudinally polarized {\rhoz} mesons to that
for transverse polarization. Their results
yielded a {\qsq}--dependence for this ratio which is weaker than linear and is
consistent with the present measurements within the rather large 
experimental uncertainties.

\vspace*{2mm}
{\bf Nemchik {\em et al.}}~\cite{pl_374_199,*pl_341_228,hep96_05_231} 
have shown that
diffractive leptoproduction of vector mesons provides a sensitive
probe of Pomeron dynamics at high energies and that a measurement of
the Balitsky--Fadin--Kuraev--Lipatov 
(BFKL) 
Pomeron~\cite{hep96_10_276,*sjnp_23_642,*pl_60_50,*jetp_44_443,*jetp_45_199,*sjnp_28_822} intercept will be possible when {\rhoz} and {\jpsi}
production cross sections are available for {\qsq} values greater than
100~{\gevsq}. The generalized BFKL dynamics were used to make specific
predictions for the interaction of the target nucleon
with the color dipole representing
the virtual--photon state. The dipole size
was determined by the sum \mbox{$M_V^2+{\qsq}$}, 
allowing a scanning of the dynamics by observing energy dependences
while varying $M_V$ and {\qsq}. Particularly striking effects were 
predicted for the energy dependence of the radially excited vector--meson
states.

\vspace*{2mm}
{\bf Amundson {\em et al.}}~\cite{hep96_01_298} 
have used the nonrelativistic QCD factorization
formalism of Bodwin {\em et al.}~\cite{pr_51_1125} to calculate 
{\jpsi} forward photopro\-duction. This calculation results in a linear
dependence on the gluon density function. Good agreement
with the fixed--target and HERA measurements of the energy dependence
of the {\jpsi} photopro\-duction cross sections was found at the level
of present experimental uncertainties.

\vspace*{2mm}
{\bf Ginzburg and Ivanov}~\cite{pr_54_5523} have
discussed the hard diffractive
photoproduction of light mesons and photons for both real and virtual
initial--state photons in the approximation that the transverse momentum
of the final--state meson or photon exceed the QCD confinement scale. 
They quantify the region of pQCD validity for the production
of {\rhoz} and {$\phi$} mesons and for the hard Compton effect.
They emphasize that a signature for this region of validity
is the exclusively longitudinal polarization of a final--state vector meson,
independent of the initial--state photon polarization, 
which is a consequence of 
the chiral nature of perturbative couplings. This article also contains a
useful survey of contemporary theoretical work in this field.

\vspace*{2mm}
{\bf Bartels {\em et al.}}~\cite{pl_375_301} and 
{\bf Forshaw and Ryskin}~\cite{zfp_68_137}
have pointed out that the hard scale necessary to the applicability of
perturbative calculations can be given by the momentum transfer at the
proton vertex $t$ as well as by the photon virtuality and the vector--meson
mass. They show
that the BFKL equation can be used to predict
the $t$--dependence in diffractive
vector--meson production in the approximation 
$|t| \gg \Lambda_{\mathrm{QCD}}^2$.
At present the HERA measurements at high $t$ are statistically
too weak to permit a quantitative comparison, but sufficient information
should become available in coming years to test the calculations.
\cleardoublepage
\section{Experimental Results from Investigations at HERA}
\setcounter{equation}{0}
\label{sec:experiment}
Our discussion of the present experimental status of investigations of
vector--meson electroproduction at HERA begins with
a brief description of the total inclusive cross sections for the interaction
of real and virtual photons with the proton. This description will contrast
the various physical processes contributing to these cross sections, 
resulting in a model--independent perception of their dependence on energy
and photon virtuality in the kinematic domains covered by the available
measurements. Our goal is to maintain enough generality to permit a 
unified understanding of these interactions, bridging the gaps between
the various kinematic regions presently described independently in their
corresponding kinematic approximations. We are motivated to such an
approach by the broad kinematic range covered by HERA measurements, which
beautifully complements earlier investigations of photopro\-duction processes
and of deep inelastic lepton scattering.
\subsection{Total Photon--Proton Cross Sections}
\label{sec:gammapxsect}
The use of a high--energy electron beam for investigations of proton structure
at HERA has as a direct consequence the dominating contribution of real--photon--proton interactions, referred to as photopro\-duction processes. The flux
of radiated photons in the energy range used for the HERA measurements 
from such a beam is a factor of 2--3 higher than that from
a muon beam of similar energy, for example, and the total cross section for 
photon--proton interactions is a mere factor of $\alpha^{-1}$ less than the
hadron--hadron total cross section, where $\alpha$ is the fine structure
constant. This ${\approx}100$ $\mu$b cross section
for photopro\-duction is three orders of magnitude greater than the part of 
the deep inelastic cross section for which the central H1 and ZEUS 
detectors accept the
scattered electron (${\qsq}  \gsim 1\;{\gevsq}$). While for photopro\-duction
processes the scattered electron exits the central detector via the
rear beam pipe, any part of the
rest of the final state which mimicks an electron causes a serious
background to studies of deep inelastic scattering and limits
the kinematic region accessible for its accurate measurement. This
high rate for photopro\-duction processes led directly to the first publications
of physics results from the first HERA running in June of 1992~\cite{pl_299_374,
pl_293_465}. 
Each experiment repeated the measurement with the higher statistics
available in later data runs, further 
attempting to quantify the various elastic and dissociative subcomponents
of the total cross section~\cite{zfp_69_27,zfp_63_391}.

The photon--proton total cross section is related to the
observed electron--proton cross section via the relation
\begin{eqnarray}
\label{eq:photot}
 \frac{d^2\sigma_{ep}}{dy\,d{\qsq}} = \frac{\alpha}{2\pi \qsq}
   \left( \frac{1+(1-y)^2}{y} - \frac{2(1-y)}{y} 
   \cdot \frac{Q_{min}^2}{\qsq}\right) \cdot
   \sigma_{{\gamma^*}p}^{tot}(W_{\gamma^* p}),
\end{eqnarray}
where $y$ is the fraction of
the electron beam energy carried by the interacting quasi--real photon. 
The parameter $Q_{min}^2$ is the minimum virtuality kinema\-tical\-ly 
allowed by the photon energy and the electron
mass (see eq.~\ref{eq:q2min}). Reconstruction
algorithms 
for
the photon virtuality {\qsq} and the center--of--mass energy $W_{{\gamma^*}p}$ 
employed by both the H1 and the ZEUS collaboration
exploit the information from 
their luminosity--monitoring small--angle electron detectors, 
which measure the final--state
angle $\theta$ and energy $E'_e$ of the scattered electron:
\begin{eqnarray}
y &\approx& \frac{E_\gamma}{E_e} \, = \, 1 - \frac{E'_e}{E_e},\\[5mm]
\qsq &\approx& 4 E_e E'_e \sin^2{{\theta}_e/2},\\[5mm]
W_{{\gamma^*}p} &\approx& \sqrt{4 E_\gamma E_p}.
\end{eqnarray}
The photon and electron energies are expressed in the laboratory reference
system and a small--scattering--angle approximation has been made.
The acceptance of the electron detectors is limited to
final--state electron angles which deviate from the electron beam flight
direction by less than 5~mrad, resulting in measured {\qsq} ranges
\mbox{$Q_{min}^2 \, < \,$ {\qsq} $\, < \,$ 0.01~{\gevsq}}, 
with an average
of about $5\cdot10^{-4}\;{\gevsq}$. The H1 experimental result employed photon
energies ranging between 8.3~{\gev} and 19.3~{\gev} 
during dedicated running in 1994, covering
a much wider range than that of the ZEUS analysis, which used data taken
in 1992 and limited the photon energy to a region where the efficiencies
were well understood: \mbox{8.5 $< \, E_\gamma \, <$ 11.5~{\gev}.} 
The results for the total cross section were\\[5mm]
\hspace*{1.5cm} $\sigma_{{\gamma^*}p}^{tot} = 165 \pm 2 \;(stat) \pm 11 \;(sys) \, {\mu}$b  
for $\langle W_{{\gamma^*}p}\rangle$ = 200  GeV   (H1)\\[5mm]
\hspace*{1.5cm} $\sigma_{{\gamma^*}p}^{tot} = 143 \pm 4 \;(stat) \pm 17 \;(sys) \, {\mu}$b 
for $\langle W_{{\gamma^*}p}\rangle$ = 180  GeV  (ZEUS).\\[5mm]
Comparisons with earlier photopro\-duction measurements confirmed
that the weak energy dependence (a mere 30\% increase as $W_{{\gamma^*}p}$ 
increases from
20 to 200~{\gev}), characteristic of the soft diffractive processes
observed for hadronic cross sections, applies to photopro\-duction as well.\\

We turn now to measurements of the total photon--proton cross section with
virtual photons at HERA. Such a cross section is well defined in a  kinematic
region where the 
lifetime of the virtual photon exceeds the
interaction time associated with the
longitudinal extent of the proton in its rest frame. 
This region is characterized by squared photon
momenta which greatly exceed their virtuality, precisely the region measured
at HERA. If we designate $k_\gamma$ as the momentum of the virtual
photon in the rest frame of the proton, 
then we can estimate its lifetime by applying the Heisenberg uncertainty
relation to the amount of energy by which the photon is off--shell
as follows~\cite{dr_95_47,pl_30_123,hardproc}:
\begin{eqnarray}
{\Delta{t}} \geq \frac{1}{2 \Delta{E}},
\end{eqnarray}
with
\begin{eqnarray}
 \Delta{E} = k_\gamma - \sqrt{k_\gamma^2 - \qsq}.
\end{eqnarray}
The HERA results we discuss cover a kinematic range
\begin{eqnarray}
 10^{-1} < {\qsq} < 5 \cdot 10^3 \; {\gevsq}, \hspace{5mm} 4 \cdot 10^2 < k_\gamma < 4 \cdot 10^4 \; {\gev}.
\end{eqnarray}
For the kinematic region in which $k_\gamma^2$ dominates {\qsq}, we can 
approximate $\Delta{E}$ as follows:
\begin{eqnarray}
 \Delta{E} \approx \frac{1}{2} \, \frac{\qsq}{k_\gamma}.
\end{eqnarray}
Since the Lorentz--invariant variable $\nu$ is the photon
energy in the proton rest frame we can write to a good approximation
\begin{eqnarray}
 \Delta{E} \approx \frac{1}{2} \, \frac{\qsq}{\nu} = m_p {\xbj}.
\end{eqnarray}
The speed of the virtual photon in the proton rest frame, $v_\gamma$,
is only slightly greater than the vacuum speed of light
\begin{eqnarray}
  v_\gamma = \frac{k_\gamma}{\sqrt{k_\gamma^2 - \qsq}} \approx 1 + \frac{1}{2} \frac{\qsq}{k_\gamma^2},
\end{eqnarray}
so the condition for a well defined ${\gamma^*p}$ cross section becomes
\begin{eqnarray}
  \frac{1}{2 m_p {\xbj}} >  2 R_p,
\end{eqnarray}
where $R_p$ is the radius of the proton. Setting $R_p = 0.8$ fm, we obtain
the condition
\begin{eqnarray}
\label{eq:xlimit}
   \fbox{${\xbj} < 0.06.$}
\end{eqnarray}
In order to relate this condition to the center--of--mass energy we recall
\begin{eqnarray}
\label{eq:wq2}
   W^2 = \qsq \, \Biggl[ \frac{1}{\xbj}-1 \Biggr] \, + \, m_p^2 \; \approx \; \frac{\qsq}{\xbj}
\end{eqnarray}
and observe that our condition becomes
\begin{eqnarray}
\label{e:xseccond}
\frac{W^2}{Q^2} > 17.
\end{eqnarray}

In order to relate the total ${\gamma^*p}$ cross section to
the differential deep inelastic electron--proton cross section
we require a convention for the virtual photon flux $K$.\footnote{A flux of virtual particles is an ill--defined
concept in quantum field theory; only the product of the flux with
the cross section is unambiguously defined. The conventions for defining
this flux share a common limiting value as $\qsq \rightarrow 0$ with
fixed photon energy, which is the
flux as defined for real photons.}
The convention of Hand~\cite{pr_120_1834},
\begin{eqnarray}
K &=& \nu \, - \, \frac{\qsq}{2\,m_p} \, = \, \frac{\qsq (1-x)}{2 m_p x},
\end{eqnarray}
is defined such that the photon--proton center--of--mass energy,
\begin{eqnarray}
W^2 &=& m_p^2 \, + \, 2 m_p K,
\end{eqnarray}
corresponds to a specific value for the flux independent of {\qsq},
which is convenient for the comparison of photoproduction with real and
virtual photons. We can now relate the differential electron--proton
cross section to the photoproduction cross section~\cite{pl_30_123,pr_167_1365,perl}:
\begin{eqnarray}
 \frac{d^2\sigma^{ep}}{d{\xbj}d\qsq} &=& \Gamma_T(\xbj,\qsq)\; (\sigma_{{\gamma^*p}}^T \, + \, \epsilon\, \sigma_{{\gamma^*p}}^L),
\end{eqnarray}
with the polarization parameter $\epsilon$ defined as
\begin{eqnarray}
\label{eq:epsilon}
\epsilon &=& \frac{2(1-y)}{1+(1-y)^2-2(1-y){{\qsq}_{min}}/{\qsq}},
\end{eqnarray}
and where
\begin{eqnarray}
\Gamma_T(\xbj,\qsq) &=& \frac{\alpha}{\pi} \, \frac{m_p K}{Q^4}\,(1+(1-y)^2-2(1-y){{\qsq}_{min}}/{\qsq)}.
\end{eqnarray}
In the approximation $\qsq \gg \qsq_{min}$ the polarization parameter
reduces to
\begin{eqnarray}
\epsilon &=& \frac{1-y}{1-y+y^2/2},
\end{eqnarray}
which deviates little from unity in the kinematic region covered by
the HERA data, resulting in equal sensitivity to the longitudinal and
transverse cross sections.

Following eq.~\ref{eq:sigf2} we can write the photon--proton cross sections
in terms of the structure functions {\fs} and {\fsq}, neglecting
the contribution from {\fthree}, which is significant only
at higher values of {\qsq}, but including terms which contribute
at low {\qsq} and high {\xbj}:
\begin{eqnarray}
\sigma_{{\gamma^*p}}^T\,&=&\,\frac{4\pi^2\alpha}{\qsq (1-\xbj)} \,2 \xbj \fs({\xbj},\qsq),\\[5mm]
\sigma_{{\gamma^*p}}^L\,&=&\,\frac{4\pi^2\alpha}{\qsq (1-\xbj)} \, \left[ \frac{\qsq + 4\,m_p^2{\xbj}^2}{\qsq} \; {\fsq}({\xbj},\qsq) \,-\,2 \xbj \fs({\xbj},\qsq) \right].
\end{eqnarray}
In the approximation $\xbj^2m_p^2 \ll \qsq$, 
\begin{eqnarray}
\sigma_{{\gamma^*p}}^L  \, \approx \frac{4\pi^2\alpha}{\qsq} {\fl}({\xbj},\qsq),
\end{eqnarray}
and the sum of the longitudinal and transverse cross sections is simply
\begin{eqnarray}
\label{eq:virtot}
\sigma_{{\gamma^*p}}^T + \sigma_{{\gamma^*p}}^L &\approx& \frac{4\pi^2\alpha}{\qsq (1-\xbj)} {\fsq}({\xbj},\qsq).
\end{eqnarray}

\begin{figure}[htbp]
\begin{center}
\includegraphics[width=0.6\linewidth]{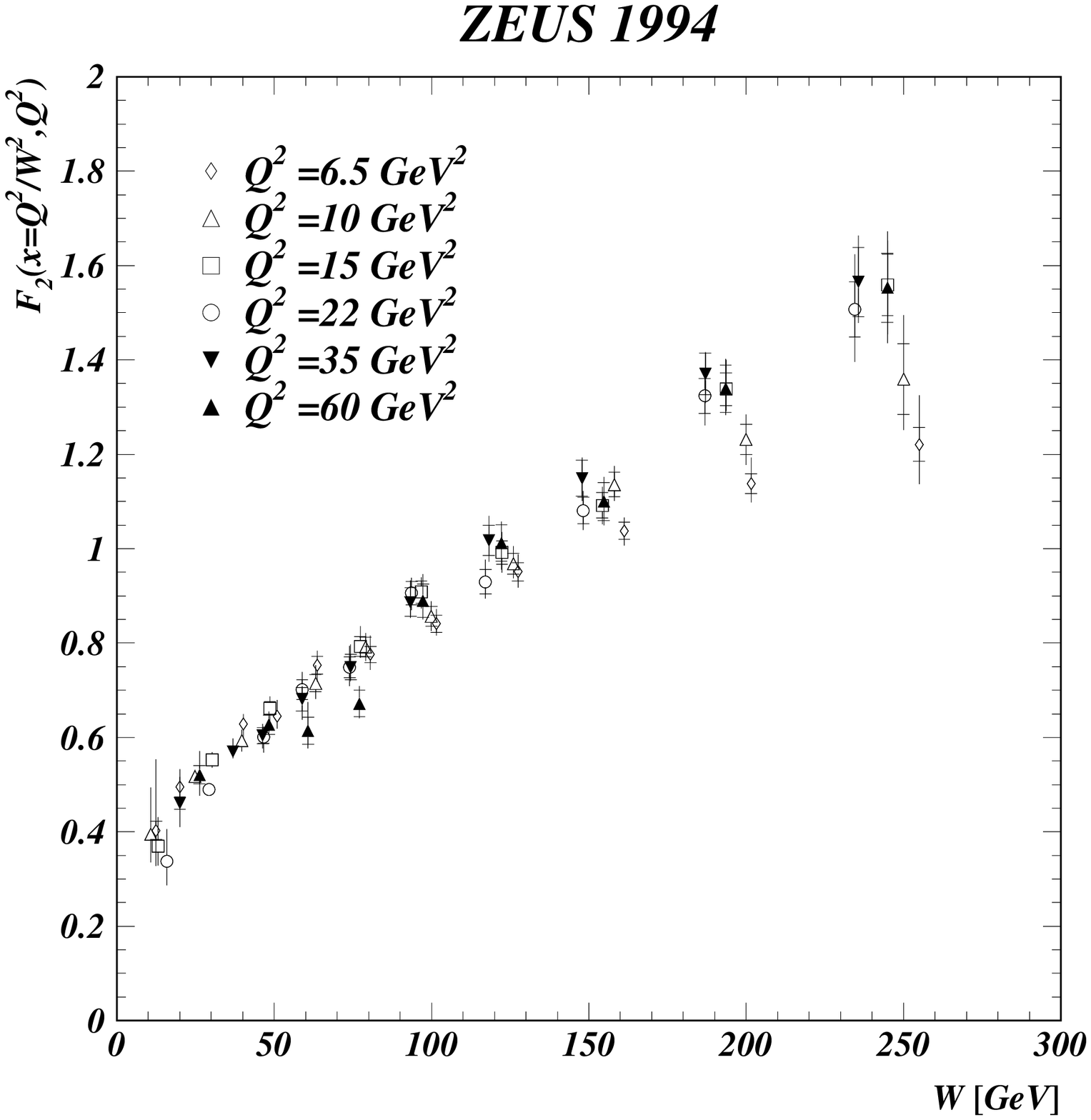}
\end{center}
%
\begin{center}
\includegraphics[height=0.6\linewidth]{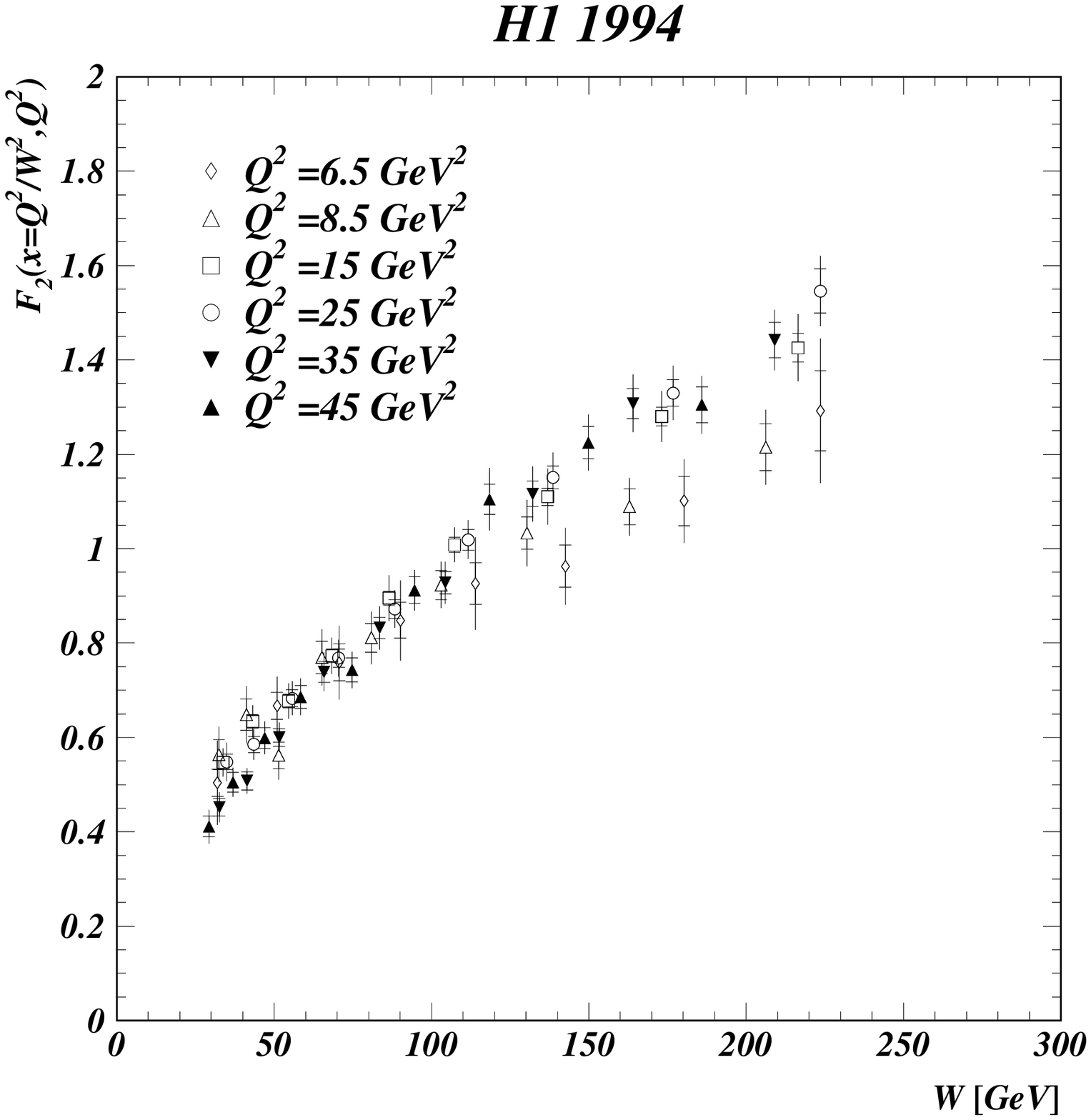}
\end{center}
\caption
{
\label{fig:q2dep}
\it The value of $\fsq({\xbj}={\qsq}/W^2_{\gamma^* p},{\qsq})$ as a function of
the $\gamma^* p$ center--of--mass energy $W_{\gamma^* p}$ for various values of {\qsq} 
as measured by the ZEUS~\protect\cite{zfp_72_399}
and H1~\protect\cite{np_470_3} collaborations with data recorded in 
1994. The inner error bars represent
the statistical uncertainties and the outer error bars show the 
quadratic sum of statistical and systematic uncertainties.
} 
\end{figure}

This {\qsq}--dependence is shown in Fig.~\ref{fig:q2dep}, which includes
measurements by the ZEUS~\cite{zfp_72_399} and H1~\cite{np_470_3} 
collaborations obtained with the 
data recorded in 1994.
The magnitude of the cross section indeed scales with $Q^{-2}$ in this region,
as the {\qsq}--dependence of {\fsq}({\xbj},{\qsq}) is weak, and the
energy dependence is also independent of {\qsq} over a wide range in {\qsq}.
The energy--dependence is seen to be strong, 
the cross section rising by more than
a factor of two as the energy increases from 50 to 200 GeV. Given the
simple relationship between $W_{\gamma^* p}$, {\xbj}, and {\qsq} in this kinematic region (see eq.~\ref{eq:wq2}), 
this rise 
is easily identified as 
equivalent to the strong {\xbj}--dependence of the structure function {\fsq}
discussed in section~\ref{sec:f2}.

\begin{figure}[htbp]
\begin{center}
\includegraphics[width=0.8\linewidth]{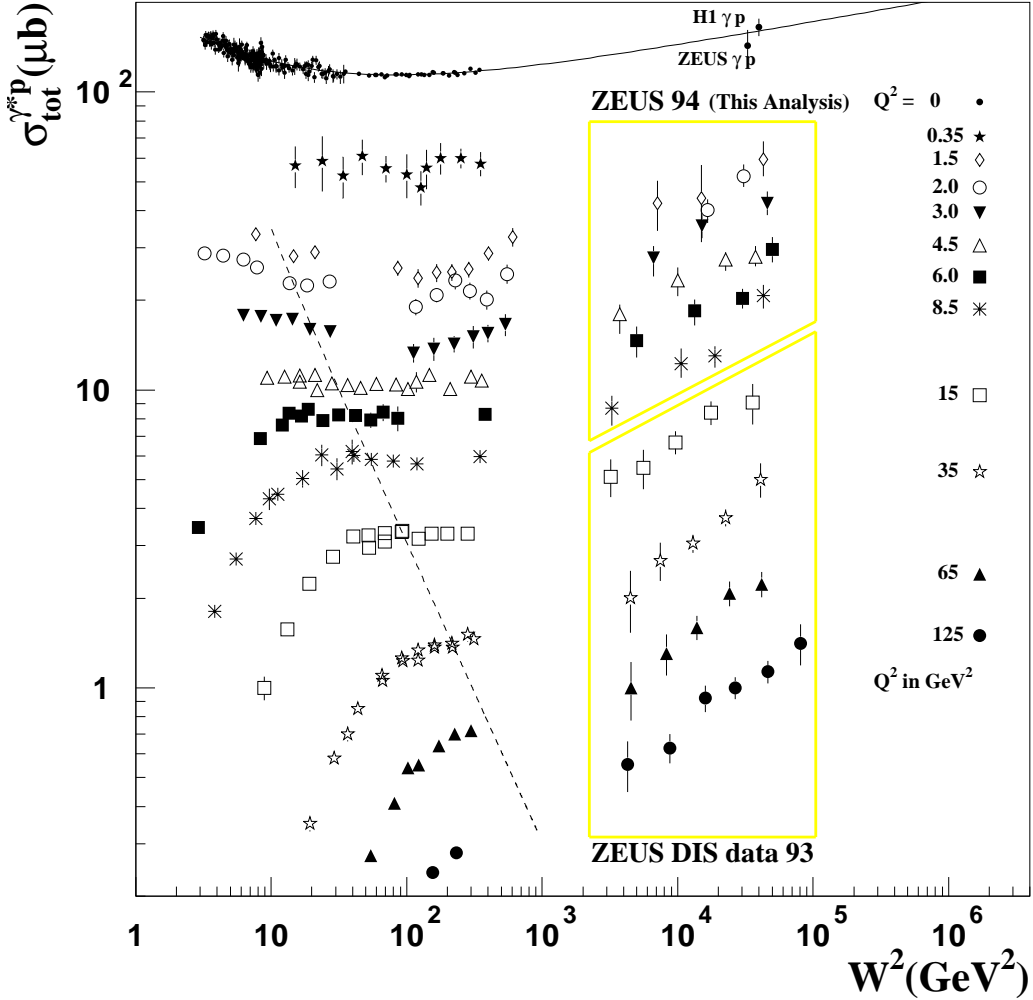}
\end{center}
\caption
{
\label{fig:zloq2}
\it The value of the total photon--proton cross section 
$\sigma_{{\gamma^*p}}^{tot}$  as a function of
the squared center--of--mass energy for various values of {\qsq} 
as measured by the ZEUS collaboration with
subsamples (see text for details) 
of the 1994 data~\protect\cite{zfp_69_607}, compared to measurements at higher
{\qsq} using the 1993 data~\protect\cite{zfp_65_379}, the HERA photopro\-duction cross section
measurements~\protect\cite{zfp_69_27,zfp_63_391}, and the low--energy
deep inelastic~\protect\cite{pl_364_107,pl_223_485,pr_54_3006}, and 
photopro\-duction~\protect\cite{prl_40_1222,*ch_87_1} measurements. 
The curve represents a parametrization of 
the photopro\-duction measurements based on phenomenological
Regge theory~\protect\cite{zfp_61_139}.
The dashed line connects points where {\xbj}=0.1.
}
\end{figure}

Having observed the strong energy dependence of the proton
interaction cross section for
virtual photons when compared to that of the cross section
for real photons, it becomes interesting
to study the transition region. 
In a first attempt to investigate this region, the ZEUS collaboration
used two experimental tricks in an early analysis of the data recorded in 1994.
The first was to take advantage of the initial--state radiation from the
electron to measure at effectively lower electron energy and hence lower 
\qsq. The second was to take advantage of HERA accelerator physicists' 
ability to manipulate the
electron RF structure such that the interaction vertex be moved 67~cm in
the forward direction, increasing the acceptance for electrons scattered
at small angles. In this manner it became possible to measure the
virtual--photon--proton cross section down to 
{\qsq}--values of 1.5~{\gevsq}~\cite{zfp_69_607}, as shown in Fig.~\ref{fig:zloq2}.
In this plot we compare as well the very different energy dependences
of the photopro\-duction and deep inelastic cross sections. The solid line
indicates the expectations based on phenomenological 
Regge theory~\cite{zfp_61_139}, which
describe the photopro\-duction measurements well but 
are inconsistent with the virtual--photon--proton measurements. The dashed line
corresponds to the limit of validity of the cross section definition
specified by eq.~\ref{eq:xlimit}, the cross section being well defined
at high $W$. The measurements at higher values of 
{\xbj}~\cite{pl_364_107,pl_223_485,pr_54_3006}
exhibit a threshold--type behavior resulting from the short photon lifetime
compared to the interaction time associated with the
longitudinal size of the proton in its rest frame. The conclusion
from these measurements at low {\xbj} and low {\qsq} is that the strong
energy dependence holds down to {\qsq} values of 1.5~{\gevsq}.

\begin{figure}[htbp]
\begin{center}
\includegraphics[width=0.8\linewidth]{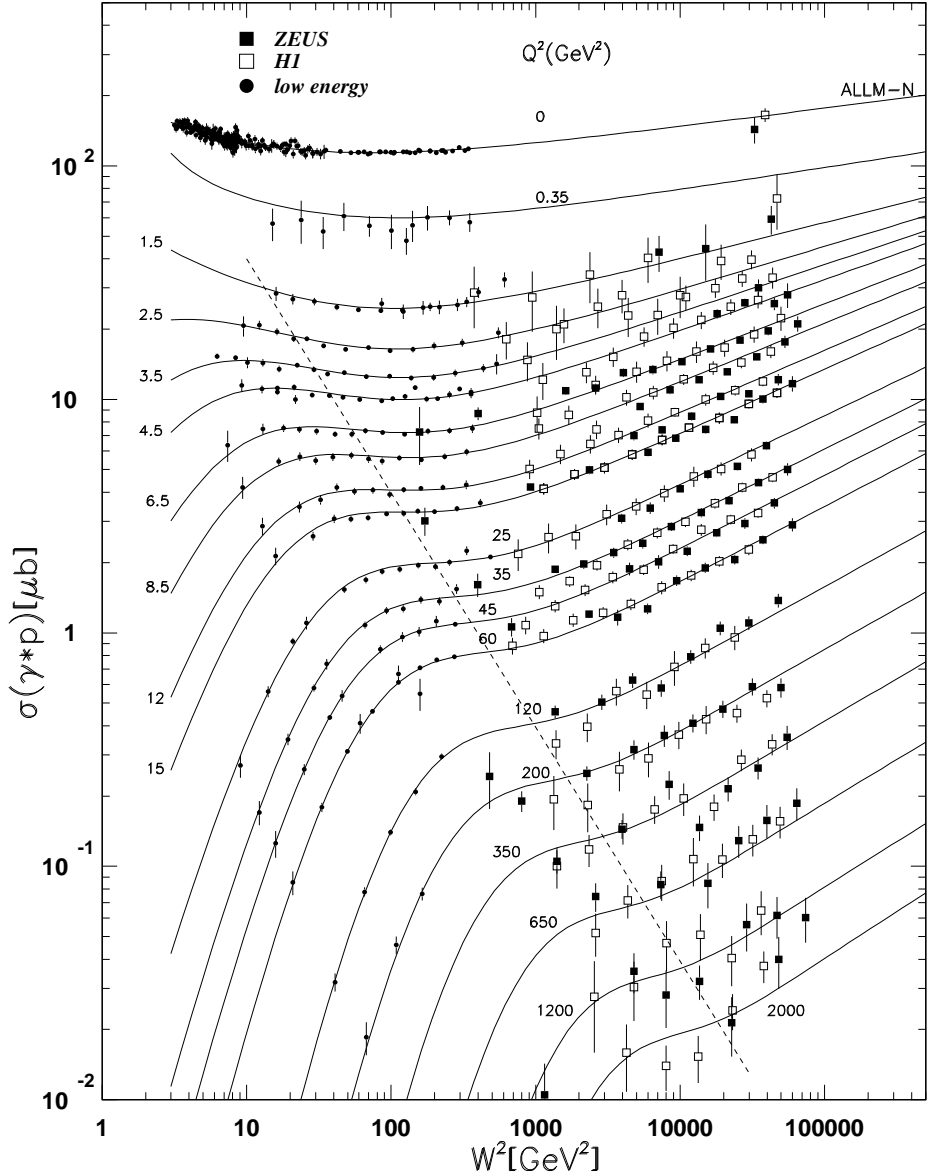}
\end{center}
\caption
{
\label{fig:sigtot94}
\it The value of the total
photon--proton cross section $\sigma_{{\gamma^*p}}^{tot}$ as a function of
the squared center--of--mass energy for various values of {\qsq} 
as measured by the ZEUS~\protect\cite{zfp_72_399} and H1~\protect\cite{np_470_3} collaborations with
the 1994 data and by 
the NMC collaboration~\protect\cite{pl_364_107} at CERN. The curves represent
calculations using the ALLM proton structure 
function parametrizations~\protect\cite{pl_269_465,*marcusthesis}.
The dashed line connects points where {\xbj}=0.1.
}
\end{figure}

The HERA measurements employing the full data set from the 1994 data--taking
period~\cite{zfp_72_399,np_470_3} are shown in Fig.~\ref{fig:sigtot94}
and compared to calculations using the ALLM proton structure 
function parametrizations~\cite{pl_269_465,*marcusthesis}, 
which exhibit consistency
with the
data over the entire range in {\qsq}.
Substantial gains in the kinematic range were achieved, and
the region covered by these measurements now reaches that of the lower energy
deep inelastic fixed--target 
experiments. The transition region from the threshold behavior
at high {\xbj} to the high $W$ region covered
by the HERA data is now precisely measured.
The slope in the energy dependence is seen to decrease with decreasing {\qsq},
but these measurements also do not permit a
judgement as to the {\qsq}--value for which 
the transition to the Regge behavior occurs.

\begin{figure}[htbp]
\begin{center}
\includegraphics[width=0.8\linewidth]{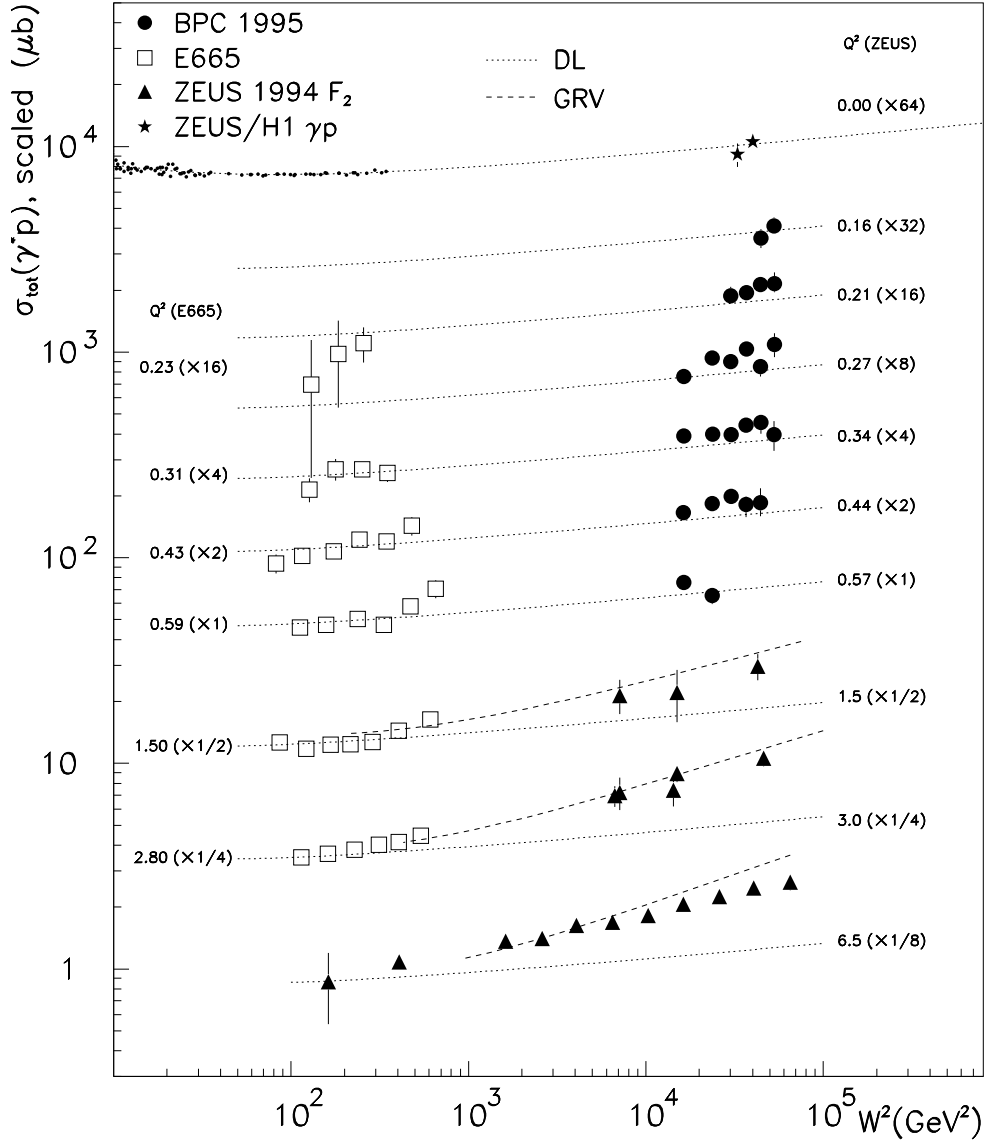}
\end{center}
\caption
{
\label{fig:zbpc}
\it The value of the total 
photon--proton cross section $\sigma_{{\gamma^*p}}^{tot}$ 
as a function of
the squared center--of--mass energy for various values of {\qsq} 
as measured by the ZEUS 
collaboration~\protect\cite{pa02_25} with the 1994 and 1995 data 
compared to results from the E665 collaboration~\protect\cite{pr_54_3006} 
at FNAL. The low--energy photopro\-duction 
measurements~\protect\cite{prl_40_1222,*ch_87_1}
are also shown. Calculations based on 
phenomenological Regge theory~\protect\cite{zfp_61_139}
are indicated as dotted lines and perturbative QCD 
calculations~\protect\cite{zfp_48_471,*zfp_53_127,*zfp_67_433} 
are shown as dashed lines. This comparison
of the HERA data with the lepton scattering measurements at higher values of
{\xbj} indicates that the transition region in {\qsq} between the two types of
energy dependence lies in the 
region \mbox{0.16 $< \qsq <$ 3.0~{\gevsq}}.
} 
\end{figure}

The beam--pipe calorimeter
installed by the ZEUS collaboration prior to the
1995 data--taking period has permitted a more detailed
investigation of the low--{\qsq}
region. Preliminary measure\-ments~\cite{pa02_25} 
of the to\-tal vir\-tual--pho\-ton--pro\-ton cross sections for
\mbox{0.16 $< \qsq <$ 0.57~{\gevsq}}
and
\mbox{130 $< W <$ 230~{\gev}}
are shown in Fig.~\ref{fig:zbpc}, along with the ZEUS results from
the 1994 data at low {\qsq}. The comparison to data from the E665 
collaboration~\cite{pr_54_3006} at higher {\xbj}
using
extra\-polations based on Regge-type energy dependence~\cite{zfp_61_139} 
and that expected from
QCD calculations~\cite{zfp_48_471,*zfp_53_127,*zfp_67_433} 
indicates that the transition occurs between {\qsq} values
of 0.16 and 3.0~{\gevsq}. While this conclusion is mitigated by the necessity
of comparing results from two different experiments and by the 5\% normalization
uncertainty in the ZEUS results not shown in Fig.~\ref{fig:zbpc}, it appears
that future measurements in this kinematic region will provide the information
necessary to test theoretical hypotheses concerning the
transition from perturbative to nonperturbative strong interactions. An
excellent discussion of the topical 
issues addressed by these measurements can be
found in ref.~\cite{hep96_08_9}.
%
\cleardoublepage
\subsection{Exclusive Vector--Meson Production}
\label{sec:expvm}
Since the first published measurements of exclusive vector--meson production
appeared following the 1993 HERA run, ameliorations of detector components,
addition of new detector components, and an increasing number
of analysis strategies have
led to further 
quantitative investigation of a variety of final--state topologies
and kinematic ranges. Table~\ref{table:vmdata} shows a compilation of
results published prior to the end of 1996. 
The designation ``Inel'' refers to processes
which result in a value for the elasticity variable $z$ less than 0.8.
Results which include
quasi--elastic scattering accompanied by proton dissociation ($z \lsim 1$) 
are referred to
as ``El/pDiss''. A comparison of results for {\rhoz}, $\phi$, and 
{\jpsi} at high {\qsq}
can also
be found in ref.~\cite{pa02_28}. 
Preliminary H1 results from the 1995 data run for {\jpsi} production
are described in ref.~\cite{pa02_85}.
The review of HERA results on diffractive
processes in ref.~\cite{hep97_03_245} also includes an instructive chapter
on vector meson production.

We now discuss the experimental issues common to the
measurements for each type of vector meson investigated. Thereafter follows
a discussion of the considerations particular to each of these
vector mesons and to each
decay mode studied.

\begin{table}[htbp]
 \begin{sideways}
  \begin{minipage}[b]{\textheight}
   \begin{center}
    \begin{tabular}{clcc@{--}ccr@{--}lcr@{--}lcr@{$\pm$}lc} \hline \hline
 VM & El/Inel& Decay & \multicolumn{2}{c}{{\qsq} Range} & $\langle \qsq \rangle$&\multicolumn{2}{c}{$W$ Range} & $\langle $W$ \rangle$ &\multicolumn{2}{c}{$p_T^2$ Range} & Events  &\multicolumn{2}{c}{Bckgd}& Ref \\
& & Channel & \multicolumn{2}{c}{(\gevsq)} & (\gevsq) &  \multicolumn{2}{c}{(\gev)} & (\gev) & \multicolumn{2}{c}{(\gevsq)} & & \multicolumn{2}{c}{(\%)} & \\
\hline
{\rhoz} & \hspace*{3mm} El & $\pi^+\pi^-$ &
 $10^{-10}$&$4$&$10^{-1}$&
 $60$&$80$&$70$&
 $0$&$0.5$&$6381$&15&$6$&\cite{zfp_69_39} \\
\hline
{\rhoz} & \hspace*{3mm} El & $\pi^+\pi^-$ &
 $10^{-10}$&$0.5$&$10^{-1}$&
 $40$&$80$&$55$&
 $0$&$0.5$&$358$&33&$10$&\cite{np_463_3} \\
\hline
{\rhoz} & \hspace*{3mm} El & $\pi^+\pi^-$ &
 $10^{-8}$&$0.01$&$10^{-3}$&
 $164$&$212$&$187$&
 $0$&$0.5$&$1000$&$31$&$8$&\cite{np_463_3} \\
\hline
{\rhoz} & \hspace*{3mm} El & $\pi^+\pi^-$ &
 $7$&$25$&$11$&
 $42$&$134$&$79$&
 0&0.6&$82$&$26$&$18$&\cite{pl_356_601} \\
\hline
{\rhoz} & \hspace*{3mm} El & $\pi^+\pi^-$ &
 $8$&$50$&$13$&
 $40$&$140$&$81$&
 0&0.5&$160$&$10$&$8$&\cite{np_468_3} \\
\hline
{\rhoz} & \hspace*{3mm} El & $\pi^+\pi^-$ &
 $10^{-10}$&$1$&$10^{-1}$&
 $50$&$100$&$73$&
 0.07&0.40&$1653$&$7$&$1$&\cite{dr_96_183} \\
\hline
{\rhoz} & \hspace*{3mm} El & $\pi^+\pi^-$ &
 $7$&$36$&$11$&
 $60$&$180$&$120$&
 0&0.8&$260$&$11$&$8$&\cite{pa02_65} \\
\hline
{\rhoz} & \hspace*{3mm} El/pDiss & $\pi^+\pi^-$ &
 $10^{-10}$&$4$&$10^{-1}$&
 $50$&$100$&$70$&
 $0$&$0.5$&$80000$&22&$3$&\cite{pa02_50} \\
\hline
{\rhoz} & \hspace*{3mm} El & $\pi^+\pi^-$ &
 $0.25$&$0.85$&$0.5$&
 $20$&$90$&$55$&
 0&0.6&$1500$&$27$&$9$&\cite{pa02_53} \\
\hline
{\rhoz} & \hspace*{3mm} El/pDiss & $\pi^+\pi^-$ &
 $7$&$36$&$11$&
 $60$&$180$&$120$&
 0&0.8&$90$&$18$&$6$&\cite{pa02_65} \\
\hline
{\rhoz} & \hspace*{3mm} El/pDiss & $\pi^+\pi^-$ &
 $10^{-5}$&$0.01$&$10^{-3}$&
 $85$&$105$&$95$&
 0.5&4.0&$3000$&\multicolumn{2}{c}{{\em n.d.}}&\cite{pa02_51} \\
\hline
${\rho}'$ & \hspace*{3mm} El & $\pi^+\pi^-\pi^+\pi^-$ &
 $4$&$50$&$7$&
 $40$&$140$&$85$&
 0&0.6&$70$&$9$&$6$&\cite{pa01_88} \\
\hline
$\omega$ & \hspace*{3mm} El & $\pi^+\pi^-\pi^0$ &
 $10^{-9}$&$4$&$10^{-1}$&
 $70$&$90$&$80$&
 0&0.6&$170$&$16$&$9$&\cite{zfp_73_73} \\
\hline
$\phi$ & \hspace*{3mm} El & $K^+K^-$ &
 $10^{-10}$&$4$&$10^{-1}$&
 $60$&$80$&$70$&
 0.1&0.5&$570$&$25$&$10$&\cite{pl_377_259} \\
\hline
$\phi$ & \hspace*{3mm} El & $K^+K^-$ &
 $7$&$25$&$11$&
 $42$&$134$&$95$&
 0&0.6&$37$&$22$&$17$&\cite{pl_380_220} \\
\hline
$\phi$ & \hspace*{3mm} El & $K^+K^-$ &
 $6$&$20$&$10$&
 $42$&$134$&$88$&
 0&1.0&$32$&$9$&$8$&\cite{pa02_64} \\
\hline
{\jpsi} & \hspace*{3mm} El & $e^+e^-,\mu^+,\mu^-$ &
 $10^{-10}$&$4$&$10^{-1}$&
 $40$&$140$&$95$&
 0&1.0&$104$&$17$&$13$&\cite{pl_350_120} \\
\hline
{\jpsi} & \hspace*{3mm} El & $e^+e^-,\mu^+\mu^-$ &
 $10^{-10}$&$4$&$10^{-1}$&
 $40$&$140$&$90$&
 0&1.0&$620$&$15$&$7$&\cite{pa02_47} \\
\hline
{\jpsi} & \hspace*{3mm} Inel & $e^+e^-$ &
 $10^{-9}$&$4$&$10^{-1}$&
 $110$&$160$&$135$&
 0&1.0&$25$&\multicolumn{2}{c}{{\em n.d.}}&\cite{pa02_47} \\
\hline
{\jpsi} & \hspace*{3mm} Inel & $\mu^+\mu^-$ &
 $10^{-10}$&$4$&$10^{-1}$&
 $60$&$130$&$95$&
 0&1.0&$65$&\multicolumn{2}{c}{{\em n.d.}}&\cite{pa02_47} \\
\hline
{\jpsi} & \hspace*{3mm} El & $e^+e^-,\mu^+\mu^-$ &
 $10^{-10}$&$4$&$10^{-1}$&
 $30$&$150$&$95$&
 0&2.0&$430$&$12$&$11$&\cite{np_472_3} \\
\hline
{\jpsi} & \hspace*{3mm} El/pDiss & $e^+e^-,\mu^+\mu^-$ &
 $10^{-10}$&$4$&$10^{-1}$&
 $30$&$150$&$105$&
 0&2.0&$320$&\multicolumn{2}{c}{{\em n.d.}}&\cite{np_472_3} \\
\hline
{\jpsi} & \hspace*{3mm} Inel & $\mu^+\mu^-$ &
 $10^{-10}$&$4$&$10^{-1}$&
 $30$&$150$&$105$&
 0&5.0&$85$&\multicolumn{2}{c}{3}&\cite{np_472_3} \\
\hline
{\jpsi} & \hspace*{3mm} El & $e+e^-,\mu^+\mu^-$ &
 $8$&$40$&$16$&
 $30$&$150$&$95$&
 \multicolumn{2}{c}{{\em n.d.}}&$25$&$25$&$11$&\cite{np_468_3} \\
\hline
{\jpsi} & \hspace*{3mm} El & $e+e^-,\mu^+\mu^-$ &
 $7$&$25$&$11$&
 $42$&$134$&$80$&
 \multicolumn{2}{c}{{\em n.d.}}&$14$&$15$&$7$&\cite{pa02_28} \\
\hline
${\psi(2S)}$ & \hspace*{3mm} El & ${\jpsi} \, \pi^+\pi^-$ &
 $10^{-10}$&$4$&$10^{-1}$&
 $40$&$160$&$80$&
 \multicolumn{2}{c}{{\em n.d.}}&$25$&\multicolumn{2}{c}{{\em n.d.}}&\cite{pa02_86} \\
\hline
\hline
    \end{tabular}
  \caption[HERA Vector Meson Data]{
\begin{small}
Measurements of single neutral
vector--meson production at HERA published by the H1 and ZEUS collaborations
prior to the end of 1996. 
The event statistics presented are either quoted directly
or estimated from information provided in the respective 
reference. They are corrected
for nonresonant backgrounds which appear in the invariant mass distributions,
but not for the backgrounds from 
other sources, which are shown in the column 'Bckgd'. The transverse momentum
ranges shown refer to the vector momentum sum of particles observed in the
central detectors and approximate the ranges of momentum transfer at the 
proton vertex. 
Information which was unavailable 
is designated by ``{\em n.d.}''
\end{small}
}
\label{table:vmdata}
\end{center}
\end{minipage}
\end{sideways}
\end{table} 

\subsubsection{General Experimental Considerations}
\begin{itemize}
\item
\underline {The Photon Lifetime}\\*[2mm]
The discussion of the validity of the photon--proton cross section definition
follows closely that in section~\ref{sec:gammapxsect}, with the replacement
\begin{eqnarray}
\qsq \; \rightarrow \; \qsq + M_V^2,
\end{eqnarray}
where $M_V$ is the mass of the final--state vector meson. Thus the limiting
value of {\xbj} is lower than the value of 0.06 found in 
eq.~\ref{eq:xlimit}, but all of the measurements we discuss in the following
cover kinematic ranges in which the lifetime of the photon in the 
proton rest frame far exceeds the interaction time associated with the
longitudinal extent of the proton in its rest frame.
\item
\underline{The Photon Flux}\\*[2mm]
The effective photon flux is obtained
from the re\-la\-tion\-ship be\-tween the dif\-fer\-en\-tial 
el\-ec\-tron--pro\-ton cross sec\-tion
and the longitudinal and transverse photon--proton cross sections
$\sigma^L_{{{\gamma^*p}} \rightarrow Vp}$ and $\sigma^T_{{{\gamma^*p}} \rightarrow Vp}$ as follows:
\begin{eqnarray}
\frac{d^2\sigma_{ep \rightarrow eVp}}{dy\,d{\qsq}}=\frac{\alpha}{2\pi y \qsq}
    \left[\left( 1+(1-y)^2 - 2(1-y)
    \frac{Q_{min}^2}{\qsq}\right) 
    \sigma^T_{{{\gamma^*p}} \rightarrow Vp} +
    2(1-y) 
    \sigma^L_{{{\gamma^*p}} \rightarrow Vp} \right], \nonumber\\
\end{eqnarray}
where $Q_{min}^2 \, = \, m_e^2 \frac{y^2}{(1-y)}$. For photopro\-duction
studies the ratio of the two partial 
cross sections can be motivated by arguments using
VMD assumptions~\cite{sjnp_38_736}:
\begin{eqnarray}
\frac{\sigma^L_{{{\gamma^*p}} \rightarrow Vp}(W_{{\gamma^*p}},\qsq)}{\sigma^T_{{{\gamma^*p}} \rightarrow Vp}(W_{{\gamma^*p}},\qsq)} &=& \xi \, \frac{\qsq}{M_V^2},
\end{eqnarray}
with $\xi$ of order unity, 
and the transverse cross section is expected to scale with {\qsq} as
\begin{eqnarray}
   \sigma^T_{{{\gamma^*p}} \rightarrow Vp}(W_{{\gamma^*p}},\qsq) &=& 
\left( 1 + \frac{\qsq}{M_V^2} \right)^{-2} \, \sigma^T_{{{\gamma^*p}} \rightarrow Vp}(W_{{\gamma^*p}},{\qsq}=0).
\end{eqnarray}
The flux falls rapidly with $y$ and {\qsq}, while the $\gamma^* p$ cross section
itself depends only weakly on these variables. Thus we can write to a
good approximation\footnote{The designation ``{\em tot}'' refers here to
the sum of longitudinal and transverse contributions
to the elastic cross section, in contrast to
its use in eqs.~\protect\ref{eq:photot} and~\protect\ref{eq:virtot}, 
where it referred to the
$\gamma^* p$ interaction cross section
summed over all final states.}
\begin{eqnarray}
\label{eq:sigtot}
   \sigma^{tot}_{{{\gamma^*p}} \rightarrow Vp} \equiv \sigma^T_{{{\gamma^*p}} \rightarrow Vp} + \epsilon \, \sigma^L_{{{\gamma^*p}} \rightarrow Vp} &=& \frac{\sigma_{ep \rightarrow eVp}}{\Phi_{{\gamma^*}/e}}
\end{eqnarray}
where $\epsilon$ is the virtual
photon polarization parameter, defined as in eq.~\ref{eq:epsilon},
and the flux of photons is calculated as
\begin{eqnarray}
\Phi_{{\gamma^*}/e} &=& \int_{y_{min}}^{y_{max}} dy \int_{Q_{min}^2}^{Q_{max}^2}
        d\qsq \; f_{{\gamma^*}/e}(y,\qsq),
\end{eqnarray}
\begin{eqnarray}
f_{{\gamma^*}/e}(y,\qsq)=\frac{\alpha}{2\pi y \qsq}
   \left( 1+\frac{\qsq}{M_V^2} \right) ^{-2}
   \left[{1+(1-y)^2} 
   - 2(1-y) \left( \frac{Q_{min}^2}{\qsq} - \xi \frac{\qsq}{M_V^2}  \right)
    \right],
\end{eqnarray}
with $Q_{min}^2$ defined as above and the other 
limits defined by the acceptance and event selection cuts. The integrand
falls by an order of magnitude over the $y$--range covered by the measurements.
The dependence on the vector--meson mass arises primarily from the longitudinal
component and influences the flux calculation at a level of less than a 
few percent for the photoproduction studies.
We note further that in the kinematic range of the HERA measurements the
polarization parameter exceeds 0.97, resulting in roughly equal sensitivity to the transverse 
and longitudinal cross sections. The latter is negligibly small in
the photopro\-duction studies, while at high {\qsq} the
two cross sections are of comparable magnitude. 
For \mbox{$M_V^2 \gg {\qsq} \gg {m_e^2}$}  the relationship between
the photon--proton
cross section and the differential electron--proton cross section 
reduces to
\begin{eqnarray}
\sigma^{tot}_{{{\gamma^*p}} \rightarrow Vp} &=& \frac{1}{f^T_{{\gamma^*}/e}(y,\qsq)} \frac{d^2\sigma_{ep \rightarrow eVp}}{dy\,d{\qsq}},
\end{eqnarray}
where 
\begin{eqnarray}
   f^T_{{\gamma^*}/e}(y,\qsq) &=&  \frac{\alpha (1+(1-y)^2)}{2\pi y \qsq}.
\end{eqnarray}
The assumption of weak energy dependence applied in the total cross section
calculations as described above is invalid for the studies at high
{\qsq} and for the studies of 
{\jpsi} photoproduction. For such investigations the flux factor was
calculated on an event-by-event basis.
\item
{\underline {Kinematic Reconstruction}}\\*[2mm]
The vector--meson event samples all required the reconstruction of
the invariant mass of the final--state meson. With the sole exception 
of the $\pi^0$ in 
$\omega \rightarrow \pi^+ \pi^- \pi^0$ decay all decay products
were identified as such 
via charged--particle tracking, which provided high--accuracy
determination of their momentum three--vectors at the production vertex.
Using ad hoc mass assumptions for these tracks, 
the vector 
mesons were identified via peaks in invariant mass spectra, and their
momentum four--vectors were determined. 
In the photopro\-duction samples no other information was available, and the $\gamma^*$p center--of--mass energy
was determined via the relation
\begin{eqnarray}
\label{eq:wrec}
W_{{\gamma^* p}} &=& 2 E_p (E - P_z)_{V},
\end{eqnarray}
where $E_p$ is the initial--state proton energy and $ (E - P_z)_{V}$ is
the difference between the energy and the longitudinal momentum component 
of the vector meson. The accuracy provided by the tracking allowed a
re\-solution of about 2~{\gev} in $W_{{\gamma^* p}}$. 
For the investigations at high
{\qsq} additional information on the scattered electron was available via
tracking or calorimetric measurements. In this case the reconstruction
strategies of the two experiments differed, as H1 employed the double--angle
method~\cite{damethod} used in the structure function analyses and ZEUS
used a method which exploited 
the assumption of exclusivity. Under this
assumption, one can determine the energy of the scattered electron from
its scattering angle ${\theta_{e}}$ and the vector--meson four--vector 
as follows:
\begin{eqnarray}
   E'_{e} &=& \frac{2 E_e - (E - P_z)_{V}}{1 + \cos{\theta_{e}}}.
\end{eqnarray}
The re\-solution in the electron energy obtained in this manner was more
than a factor of five better than that obtained from the direct calorimetric
measurement.

A reasonable approximation to the momentum transfer $t$ at the proton
vertex (see eq.~\ref{eq:trec}) is
the 
negative squared total transverse momentum observed in the detector, i.e.~that
of the vector meson for photopro\-duction samples and 
\begin{eqnarray}
t \, = \, - p_T^2 \, = \, - \left[ (p^e_x + p^{V}_x)^2 + (p^e_y + p^{V}_y)^2 \right] 
\end{eqnarray}
for the high--{\qsq} samples, using either the double--angle
electron momentum or the
constrained electron momentum. For photopro\-duction studies in which the
final--state electron was detected in one of the small--angle electron
taggers, the mo\-men\-tum trans\-fer was well de\-ter\-mined by the trans\-verse 
mo\-men\-tum
of the vec\-tor me\-son alone \mbox{(${\qsq} < 0.01 \, {\gevsq}$),}
while in the untagged photopro\-duction studies
the smearing due to the unknown electron transverse momentum 
was
corrected for via unfolding algorithms, and 
clean samples of high--$t$ events
were not obtained. The sole exception to these means of
extracting the momentum transfer was the analysis exploiting data from the 
ZEUS Leading Proton Spectrometer, where a direct measurement of the
scattered proton allowed a re\-solution in $t$ of 0.01~{\gevsq}, dominated
by the proton beam divergence.
\item
{\underline {Luminosity Measurement}}\\*[2mm]
The H1 and ZEUS experiments each base their determination of the
luminosity used in the extraction of the electron--proton cross sections
on the rate from the Bethe--Heitler bremsstrahlung process 
\mbox{$ep \rightarrow e'{\gamma p}$}. This rate can be calculated to an
accuracy of 0.2\% in the context of quantum electrodynamics, i.e. in 
an approximation ignoring proton structure and recoil~\cite{kpthesis}.
The signature for the process is simple and clean, since the energies
of the final--state
electron and photon are measured in small--angle calorimeters and their
sum equals the energy of the initial--state electron. Primary contributions
to the uncertainty in the luminosity measurement are the re\-solution and
acceptances of the calorimeters, and the corrections for contributions
from bremsstrahlung on the residual gas in the beam pipe. This uncertainty
improved from approximately 4\% in the 1993 data to 2\% in 1994. In none
of the investigations described in this article does the uncertainty
in the luminosity dominate the normalization uncertainty in the cross section.
The primary contributions to the normalization uncertainty 
are discussed for each of the studies individually 
in section~\ref{sec:expissues}.
\item
{\underline {Background Estimates}}\\*[2mm]
The primary background in all of the studies of exclusive
vector--meson production save one has been from processes where the proton
dissociates, since the fiducial coverage in the forward direction
in each experiment does not provide a good measurement of the dissociative
final state, being used primarily as a veto in the trigger and offline selection
cuts. These background estimates ranged between 10\% and 40\%, depending on
the exclusive process under study.
Indeed the {\em uncertainty} in the background estimations, which were
based on simulations of poorly--known physics processes, often
dominated the uncertainty in the measured cross sections. 
The exception was the measurement which employed the ZEUS LPS
to measure the final--state proton trajectory~\cite{dr_96_183}, allowing
the reduction of background contamination from this source to a level
of tenths of a percent.
\end{itemize}
\subsubsection{Experimental Issues Particular to the Decay Channels Investigated}
\label{sec:expissues}
\begin{itemize}
\item
\underline {$\rhoz \rightarrow \pi^+\pi^-$}\\*[2mm]
This process was the first exclusive process to be studied at 
HERA~\cite{zfp_69_39}, due to its high rate and clean signature in the
photopro\-duction data sample. Trigger 
conditions requiring an energy deposit of 0.5 GeV in the calorimeter, at
least one
track in the central tracking chamber
and a veto on energy deposits in the forward direction sufficed
to reduce the 
trigger rate to an acceptable level. The trigger efficiency was about
40\% and the geometrical acceptance 7\%. Figure~\ref{fig:zeusspprho} shows the
mass spectrum obtained. The 
sum of a relativistic Breit--Wigner
resonance term, a nonresonant dipion background term, and an interference 
term~\cite{pl_19_702,pr_5_545,nc_34_1644,prl_5_278} was used to fit this
distribution. Further models for this line shape~\cite{pr_149_1172,*pr_9_126,*pr_2_1859,*prl_25_485,*prl_25_704,*pr_3_2671}
are discussed in the review of Bauer {\em et al.}~\cite{rmp_50_261}.
This measurement covered
the range \mbox{60 $< W_{\gamma^* p} <$ 80~{\gev}} and $|t| <$ 0.5~{\gevsq}. The 
background contamination was estimated to be about 15\%, dominated by
the contribution from proton dissociative processes. The systematic uncertainty
in the total elastic 
cross section determination was 17\%, with the primary contribution
coming from the uncertainty in the acceptance calculation.
\begin{figure}[htbp]
\begin{center}
\includegraphics[width=0.7\linewidth,bb=127 268 433 760,clip=]{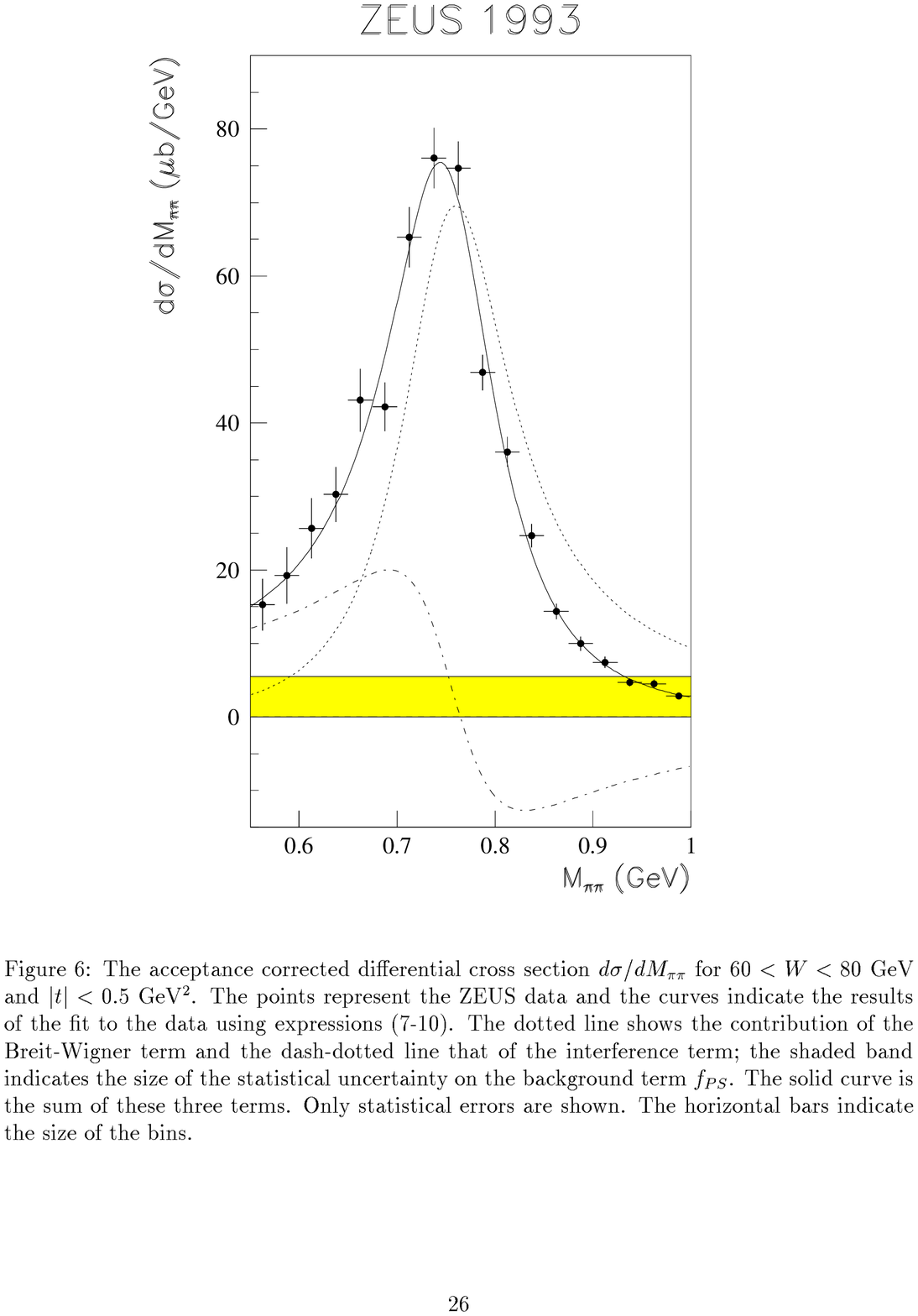}
\end{center}
\caption
{
\label{fig:zeusspprho}
\it 
The differential elastic 
$\pi^+\pi^-$ production cross section $d\sigma / dM_{\pi \pi}$
for the photopro\-duction study by the ZEUS collaboration using data recorded 
in 1993. The curves indicate the results of a fit to the data with contributions
from a relativistic Breit--Wigner resonant term interfering with a nonresonant
background term. The dotted line shows the Breit--Wigner term, the dash--dotted
line the interference term, and the shaded band indicates the statistical
uncertainty in the determination of the background term. Only statistical
errors are shown. The horizontal error bars indicate the size of the bins. 
}
\end{figure}

The H1 collaboration 
performed a similar study with the 1993 data and supplemented
it with data from a dedicated run during the 1994 data--taking period, when
the interaction vertex was shifted 67~cm in the forward direction and
the trigger was based on tagging the scattered electron in the luminosity
monitor~\cite{np_463_3}. This method allows measurement of the exclusive
cross section at values of $W_{\gamma^* p}$ similar to those of the total
cross section measurements: \mbox{164 $< W_{\gamma^* p} <$ 212~{\gev}}. 
The background
of ($31 \pm 8$)\% was dominated by dissociative processes and the systematic
error in the cross section measurement of 18\% arose primarily from the 
uncertainty in the background subtraction.

In 1994 the ZEUS LPS began operation and a measurement of elastic
{\rhoz} photoproduction with a direct and accurate measurement of the momentum
transfer at the proton vertex was thus made possible~\cite{dr_96_183}. 
The acceptance of the LPS restricted the
data to the kinematic region \mbox{$0.073 < |t| < 0.40\; {\gevsq}$}.
A fit to the 
invariant mass spectrum as in the earlier ZEUS study is shown 
in Fig.~\ref{fig:zeuslpsrho}. The requirement of a final--state proton 
in the LPS reduced
the contribution from proton dissociative processes to less than 0.5\%.
The dominant background came from events with a {\rhoz} in the main detector
in accidental coincidence with an LPS track from a 
charged particle produced in upstream beam--gas interactions,
and was
estimated to be ($5.0 \pm 0.6$)\%. The dominant contribution to the systematic
uncertainty of 12\% in the cross section measurement was the uncertainty
in the trigger efficiency. 
\begin{figure}[htbp]
\begin{center}
\includegraphics[width=0.8\linewidth,bb=74 333 469 610,clip=]{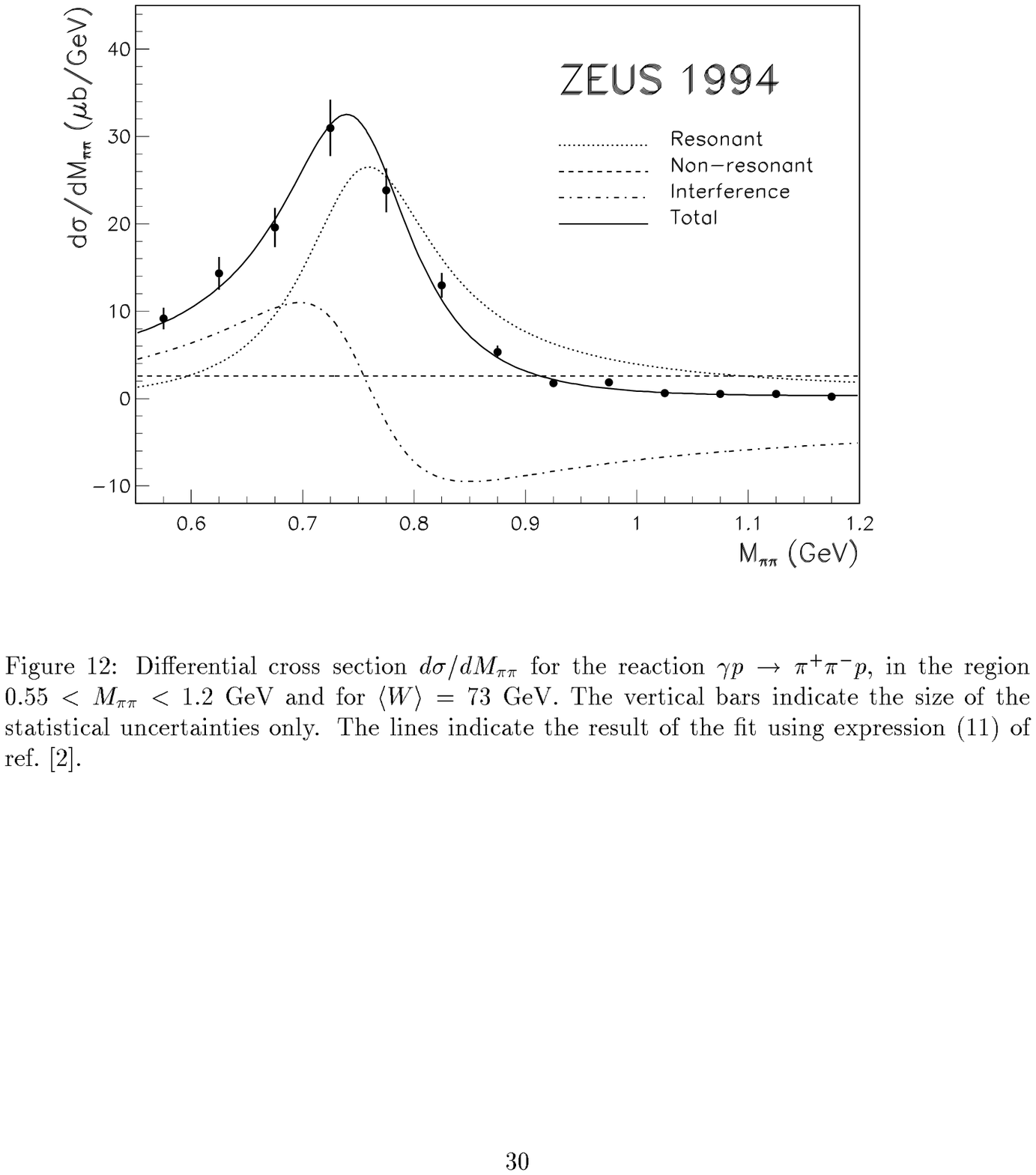}
\end{center}
\caption
{
\label{fig:zeuslpsrho}
\it 
The differential elastic 
$\pi^+\pi^-$ photoproduction cross section $d\sigma / dM_{\pi \pi}$
restricted to the kinematic region \mbox{$0.073 < |t| < 0.40\; {\gevsq}$}.
This region was defined by the acceptance of the ZEUS 
Leading Proton Spectrometer~\protect\cite{dr_96_183},
which provided a direct measurement of the final--state proton momentum. 
The solid curve indicates 
the results of a fit to the data with contributions
from a relativistic Breit--Wigner resonant term, a nonresonant
background term, and an interference term, which are shown by the broken lines.
Only statistical
errors are shown.
} 
\end{figure}

Preliminary results on {\rhoz} photopro\-duction from the 1994 data have
been presented by the ZEUS collaboration~\cite{pa02_50}, 
featuring an order--of--magnitude
increase in statistics in comparison to the earlier publication.
Figure~\ref{fig:spprhotspectra} shows the invariant mass spectra
obtained in this study for various ranges of the {\rhoz} transverse 
momentum. The statistical power of this new study shows clearly that
the relative contribution from the nonresonant dipion background is reduced
as the {\rhoz} transverse momentum increases. A subsample of this data set
for which LPS information was available served to extract an estimate
of the proton dissociative background of ($22 \pm 3$)\%. The total systematic
uncertainty in the cross section determination was about 10\%.
\begin{figure}[htbp]
\begin{center}
\includegraphics[width=0.8\linewidth,bb=73 246 507 708,clip=]{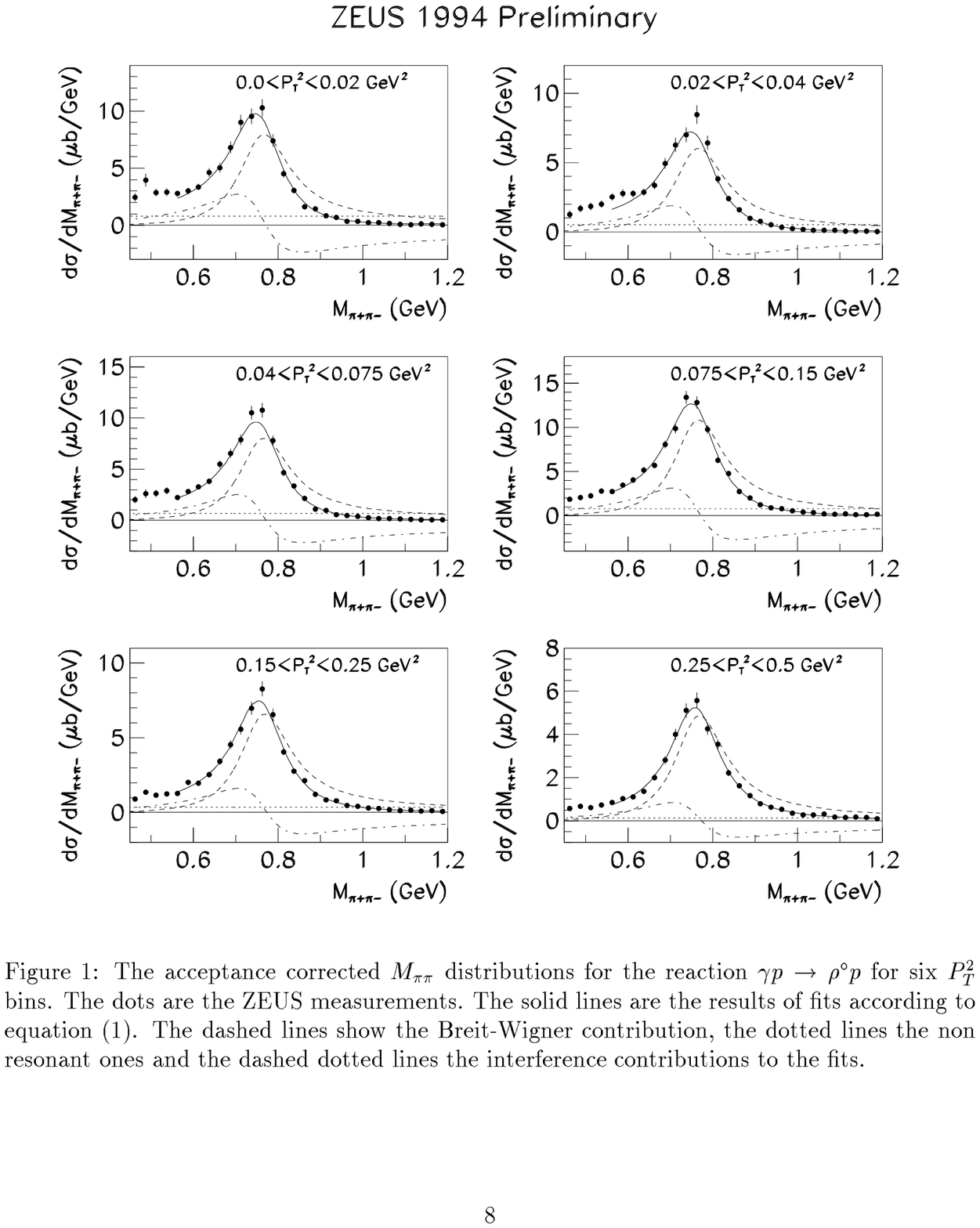}
\end{center}
\caption
{
\label{fig:spprhotspectra}
\it 
The differential elastic $\pi^+\pi^-$ 
production cross section $d\sigma / dM_{\pi \pi}$
for the photopro\-duction study by the ZEUS collaboration using data recorded 
in 1994. The solid curve indicates 
the results of a fit to the data with contributions
from a relativistic Breit--Wigner resonant term, a nonresonant
background term, and an interference term, which are shown by the broken lines.
Only statistical
errors are shown.
} 
\end{figure}

The ZEUS collaboration has presented preliminary results from the 1995
data--taking period in the low {\qsq} region 
\mbox{0.25 $< \qsq <$ 0.85~{\gevsq}~\cite{pa02_53}}. The {\qsq} acceptance
was determined by the size of the beam--pipe calorimeter used to provide
a trigger and to measure the positron scattering angle between 17 and 35~mrad
with a re\-solution of 0.3~mrad. 
Since the constrained method for determining the electron
energy yields a re\-solution of 1\% and {\qsq} depends on the square of the
scattering angle, this angular re\-solution largely determines the re\-solution
in {\qsq} of about 4\% (see eq.~\ref{eq:q2}). The BPC measurement extends the energy region
investigated at HERA to lower energies: 
\mbox{20 $< W_{{\gamma^*p}} <$ 90~{\gev}}.
Figure~\ref{fig:bpcrhospectrum} shows the acceptance and the $\pi^+\pi^-$
invariant mass spectrum, which was fit with the sum of resonant, nonresonant,
and interference terms.
The primary source of background was estimated to arise from
proton dissociative processes: 
\mbox{($27 \pm 6 \;(stat) \pm 6 \;(sys)$)\%.} The
uncertainty in this background was the dominant contribution to the
total systematic uncertainty estimate of 10\%.
\begin{figure}[htbp]
\begin{center}
\includegraphics[width=0.8\linewidth,bb=93 188 501 397,clip=]{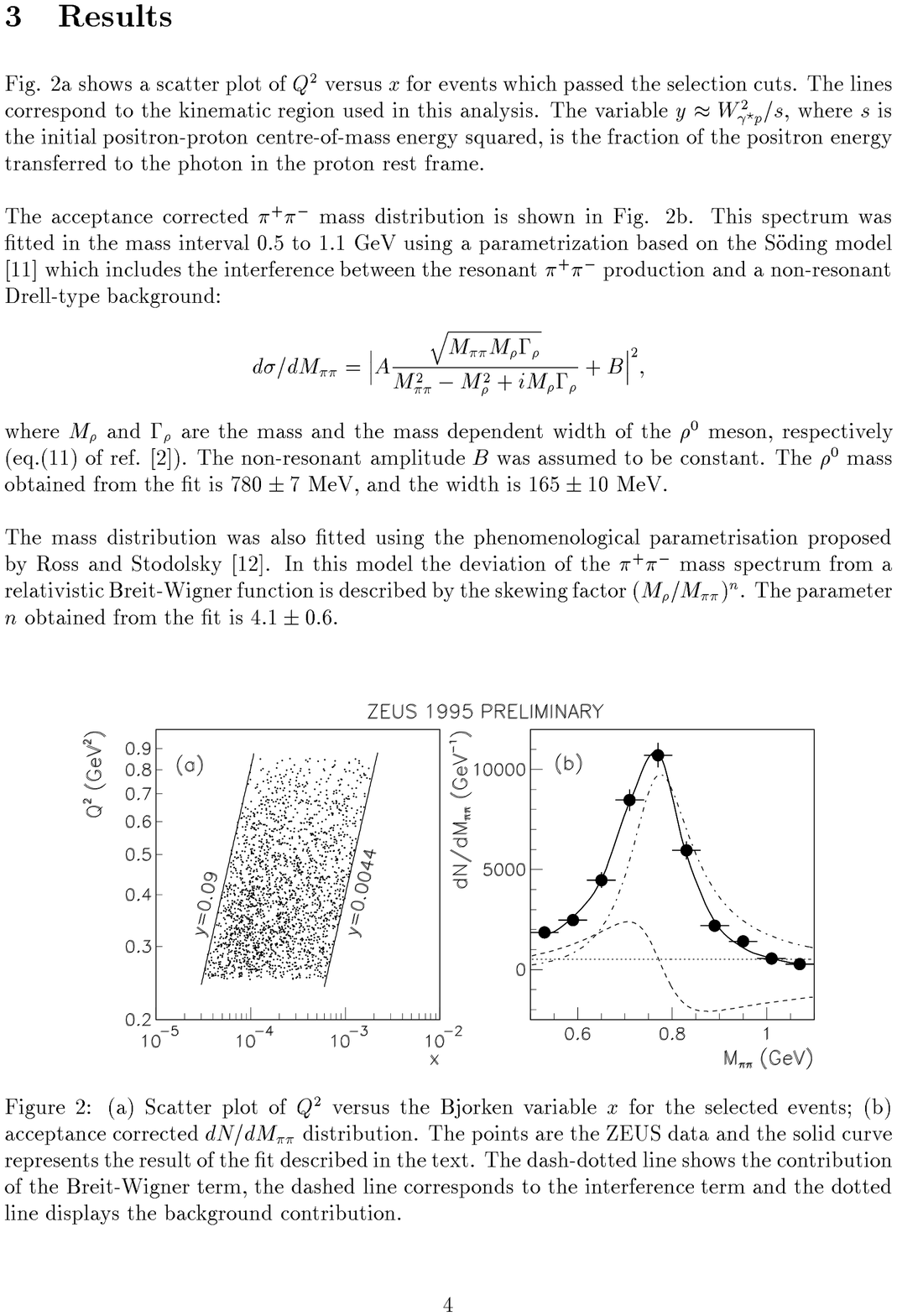}
\end{center}
\caption
{
\label{fig:bpcrhospectrum}
\it 
The kinematic region accepted and the $\pi^+\pi^-$ invariant mass distribution
for elastic {\rhoz} production at intermediate {\qsq} measured
by the ZEUS collaboration based on data recorded 
in 1995 using the newly installed beam--pipe calorimeter. 
The solid curve on the mass spectrum indicates 
the results of a fit to the data with contributions
from a relativistic Breit--Wigner resonant term, a nonresonant
background term, and an interference term, which are shown by the broken lines.
Only statistical
errors are shown.
} 
\end{figure}

The H1 and ZEUS experiments have each published observations of
elastic {\rhoz} production at values of {\qsq} near
10~{\gevsq}~\cite{pa02_28,pl_356_601,np_468_3,dr_96_162,pa02_65,pa03_48}. Trigger efficiencies
and acceptances for these processes were high ($> 50\%$) due to the
high energy and angle of the scattered electron. The data samples 
typically consisted of a few
hundred events, which sufficed to show that the mass spectra were more 
symmetric than those observed in 
photopro\-duction, and Breit--Wigner distributions 
gave reasonable fit results, the widths consistent with the Particle Data
Group value of 150~MeV. The backgrounds were around 20\%, dominated by proton
dissociation. The uncertainties in the determination of the 
exclusive cross section were around 10\%, arising from the theoretical
uncertainty in the dependence of the dissociative background rate on
the invariant mass of the dissociated final state.
\item
\underline {$\rho' \rightarrow \pi^+\pi^-\pi^+\pi^-$}\\*[2mm]
The H1 collaboration has reported preliminary results on the observation
of a resonance with mass 
\mbox{($1.57 \pm 0.02 \; (stat) \pm 0.01 \; (sys)\; {\gev}$)}
and width
($ 0.18 \pm 0.06 \;(stat) \pm 0.07 \;(sys)\; {\gev}$) 
in the \mbox{$\pi^+\pi^-\pi^+\pi^-$}
invariant mass spectrum
for \mbox{$40 < W_{{\gamma^*p}} < 140 \;{\gev}$}
and \mbox{$\qsq > 4 \; {\gevsq}$}.
A sample of 108 events was observed in the peak
shown in Fig.~\ref{fig:rhopspectrum},
of which about a third was attributed to nonresonant background. A peak
at the {\rhoz} mass was 
observed in the spectrum of $\pi^+\pi^-$ masses, of which
there are two such combinations in each of the events, indicating a decay
to $\rhoz\pi^+\pi^-$. The production rate relative to that for the {\rhoz} meson
was measured to be \mbox{($0.36 \pm 0.07 (stat) \pm 0.11 (sys)$).}
The authors offered speculation that this resonance
arises via constructive interference of the {\rhoz}[1450] and {\rhoz}[1700]
resonances.  The uncertainty in the elastic cross section was
about 30\%, with comparable 
contributions from statistics and from the uncertainty
in the nonresonant background subtraction.
\begin{figure}[htbp]
\begin{center}
\includegraphics[width=0.5\linewidth,bb=184 198 392 407,clip=]{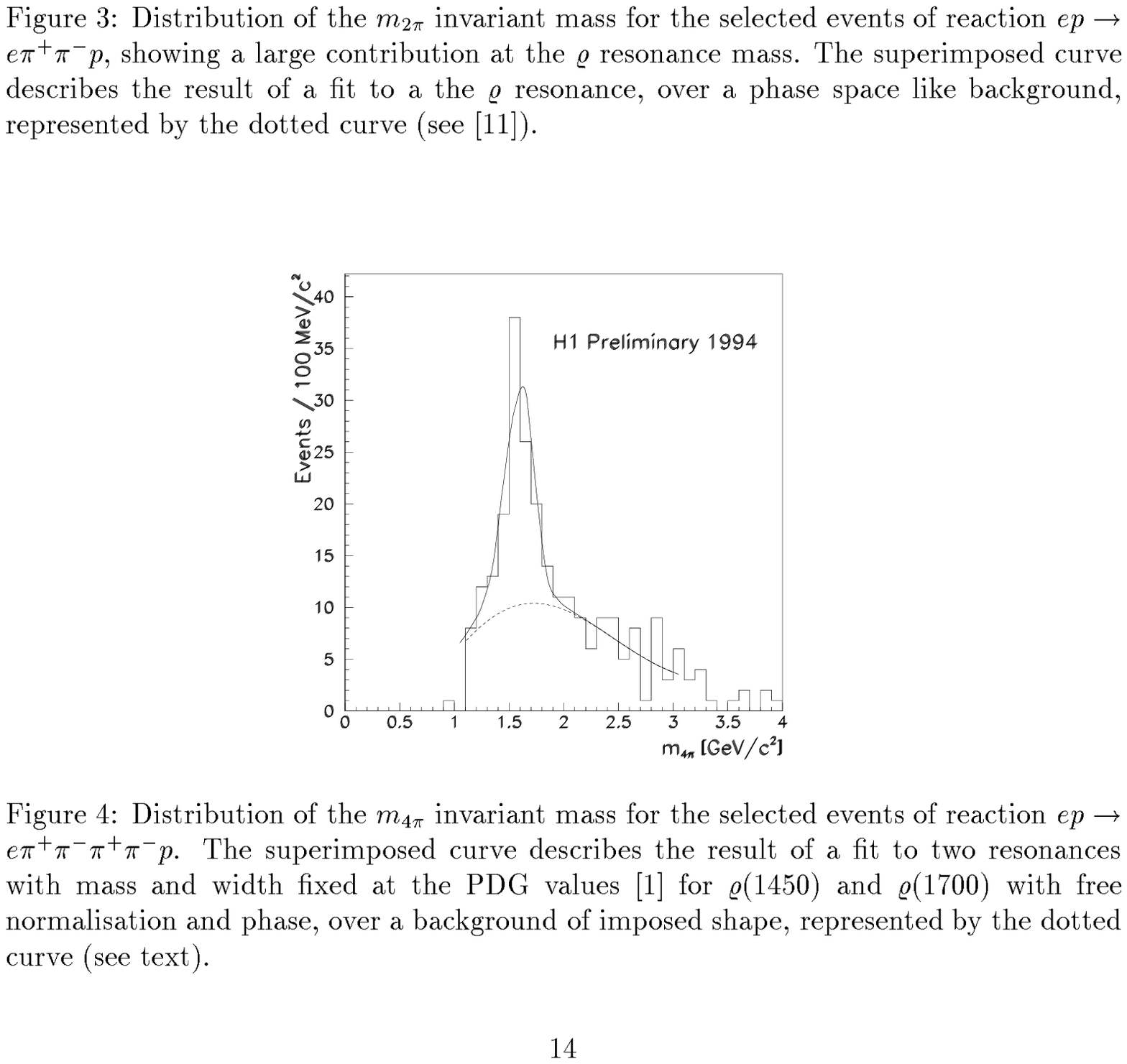}
\end{center}
\caption
{
\label{fig:rhopspectrum}
\it 
The $\pi^+\pi^-\pi^+\pi^-$ invariant mass spectrum for the $\qsq > 4 \, {\gevsq}$
data sample in 1994 from the H1 collaboration. The solid curve shows
the result of a fit to the sum of 
a single resonance and a polynomial background which is shown by the dashed line.
} 
\end{figure}
\item
\underline {$\omega \rightarrow \pi^+\pi^-\pi^0$}\\*[2mm]
The ZEUS collaboration
has reported a measurement of elastic $\omega$ production
in photopro\-duction~\cite{zfp_73_73}, where the $\pi^0$ decay photons were
detected in an array of \mbox{$3 \times 3$ cm$^2$} silicon
diodes located at a depth of 3.3 $X_0$ inside the rear calorimeter.
The average photon energy was 500~MeV with a broad spectrum of r.m.s. width
about 200~MeV. The reconstructed $\gamma \gamma$-mass spectrum is shown
in Fig.~\ref{fig:omegaspectrum}, as is the $\pi^+\pi^-\pi^0$ invariant mass
spectrum showing the $\omega$ peak. The second peak in the three--pion
invariant mass distribution was attributed to the decay 
\mbox{$\phi \rightarrow \pi^+\pi^-\pi^0$}.
The photon energy re\-solution in the rear calorimeter determined
the re\-solution in the $\gamma^* p$ center--of--mass energy, 
which was reconstructed as in eq.~\ref{eq:wrec}. 
The re\-solution in $W_{{\gamma^*p}}$ was 6\% and
that in $p_T^2$ was 0.04~{\gevsq}. The mass re\-solution for the $\omega$ was about
30~MeV. The primary contribution to 
reconstruction inefficiency was that associated with the 
reconstruction of the $\pi^0$; the product of trigger 
efficiency and acceptance was about 1\%.
The background from proton dissociation was estimated to be ($16 \pm 9$)\%.
The uncertainty in the total cross section of 20\% 
was dominated by the systematic uncertainties arising from the acceptance
calculation and the background estimation.
\begin{figure}[htbp]
\begin{center}
\includegraphics[width=0.7\linewidth,bb=115 264 493 709,clip=]{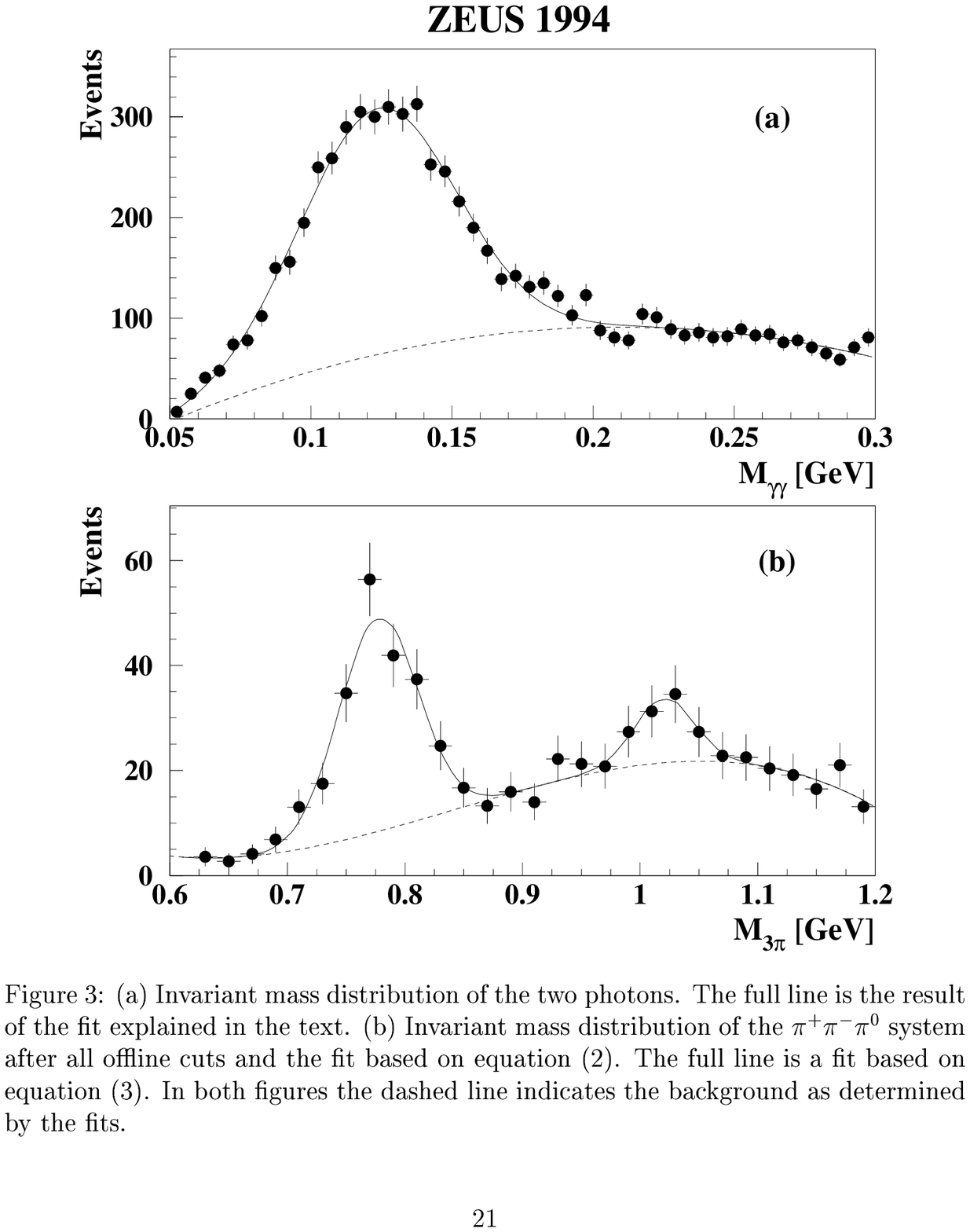}
\end{center}
\caption
{
\label{fig:omegaspectrum}
\it 
Invariant mass distributions associated with the process 
\mbox{$e^+p \rightarrow e^+p \, \pi^+\pi^- \, \gamma \gamma$}. 
(a)~The two--photon invariant mass
calculated using the photon positions as measured in the silicon array in the
rear calorimeter of the ZEUS detector. (b)~The three--pion invariant mass
distribution exhibiting peaks corresponding to the decays of $\omega$ and $\phi$
mesons. In each plot the solid line indicates the results of the full fit
used in the analysis and the dashed line shows the contribution by nonresonant
backgrounds as determined by the fit procedure.
} 
\end{figure}
\pagebreak
\item
\underline {$\phi \rightarrow K^+K^-$}\\*[2mm]
An observation of elastic $\phi$ photopro\-duction has been reported by
the ZEUS collaboration~\cite{pl_377_259}.
The triggering and event selection procedures for elastic $\phi$ production
were very similar to those for the {\rhoz} meson, since no $\pi$K--separation
was done. However, the 4.4~MeV natural width of the $\phi$
was comparable to the re\-solution in the invariant mass reconstruction using
the two charged track trajectories. Therefore the ZEUS analysis 
fit the mass spectrum with the sum
of a Gaussian--smeared Breit--Wigner function and a background polynomial,
as shown in Fig.~\ref{fig:zeussppphispectrum}. The background consisted primarily
of {\rhoz} decays where the kaon mass was mistakenly 
assigned to the decay products in the invariant mass calculation. The width
of the Breit--Wigner was fixed to the Particle Data Group value and the fit
result for the Gaussian width was found to be 4~MeV, compatible with the
expected experimental re\-solution. The detector acceptance for this process
depended strongly on the total squared transverse momentum, so the
kinematic range was restricted to \mbox{0.1 $< |t| <$ 0.5~{\gevsq}} and
\mbox{60 $< W_{{\gamma^*p}} <$ 80~{\gev}}, where the acceptance calculation
was precise. This restriction in $p_T^2$ and the acceptance resulted
in a proton dissociative background estimation of 
\mbox{($24 \pm 7\;(stat) \pm 6\;(sys)$\%)}, 
double that for the {\rhoz}. The uncertainty
in this background was included in the statistical error estimation for
the total cross section along
with that from the determination of the {\rhoz} background from the fit
to the mass peak to yield a total statistical error of 20\%. The systematic
error includes roughly equal contributions from the acceptance determination,
the trigger efficiency estimation, 
and from the extrapolation $t \rightarrow 0 \, {\gevsq}$, 
totalling approximately 20\%.
\begin{figure}[htbp]
\begin{center}
\includegraphics[width=0.5\linewidth,bb=68 245 447 626,clip=]{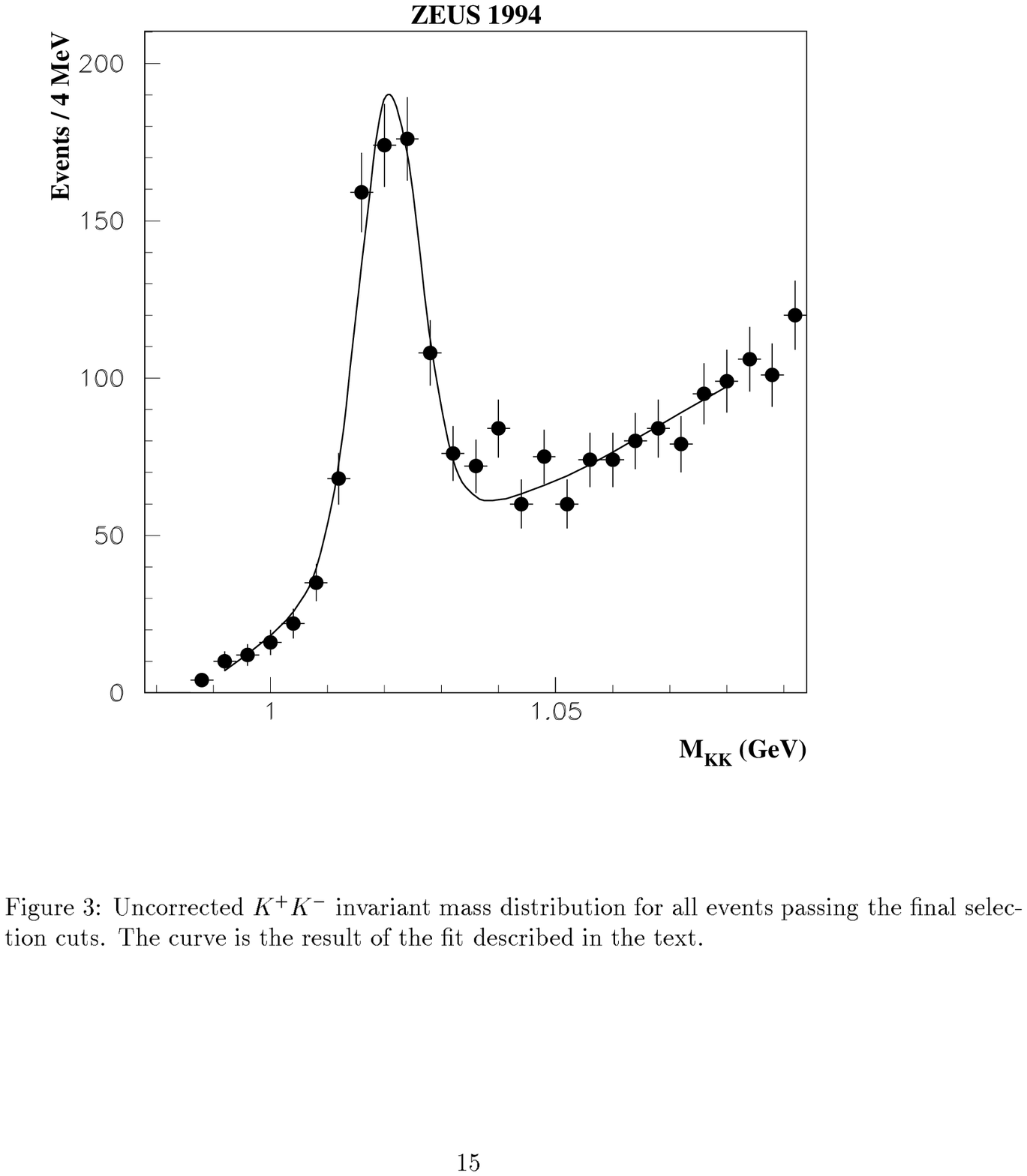}
\end{center}
\caption
{
\label{fig:zeussppphispectrum}
\it 
The $K^+K^-$ invariant mass spectrum for the photopro\-duction
data sample in 1994 from the ZEUS collaboration. The solid curve shows
the result of a fit to the sum of 
a single resonance and a background shape due primarily to {\rhoz} decays.
} 
\end{figure}

The high--{\qsq} studies of exclusive $\phi$--production by the
ZEUS~\cite{pl_380_220} and H1~\cite{pa02_64} collaborations
do not have the $t$--acceptance difficulty of
the photopro\-duction study, enjoying ${\approx}60$\% acceptance due to the
ease of triggering on the
scattered electron. The re\-solution in the mass peak was similarly ${\approx}4$ 
MeV (see Fig.~\ref{fig:h1phispectrum}),
but the background from {\rhoz} decays was smaller
than for the photoproduction results, totalling about 10\%.
The background from proton dissociation was estimated to be 
\mbox{($22 \pm 8\;(stat) \pm 15\;(sys)$\%)}
in the ZEUS study and \mbox{($9 \pm 8$)}\% by the H1 collaboration. 
The statistical uncertainties in the cross section determinations due to the
small event samples were similar 
for the two experiments at ${\approx}20$\%, while
the systematic uncertainty for H1 was estimated be roughly half of the
value of 30\% quoted by ZEUS, due to more accurate estimates of the
{\rhoz} and proton dissociation backgrounds.   

\begin{figure}[htbp]
\begin{center}
\includegraphics[width=0.5\linewidth,bb=152 330 421 595,clip=]{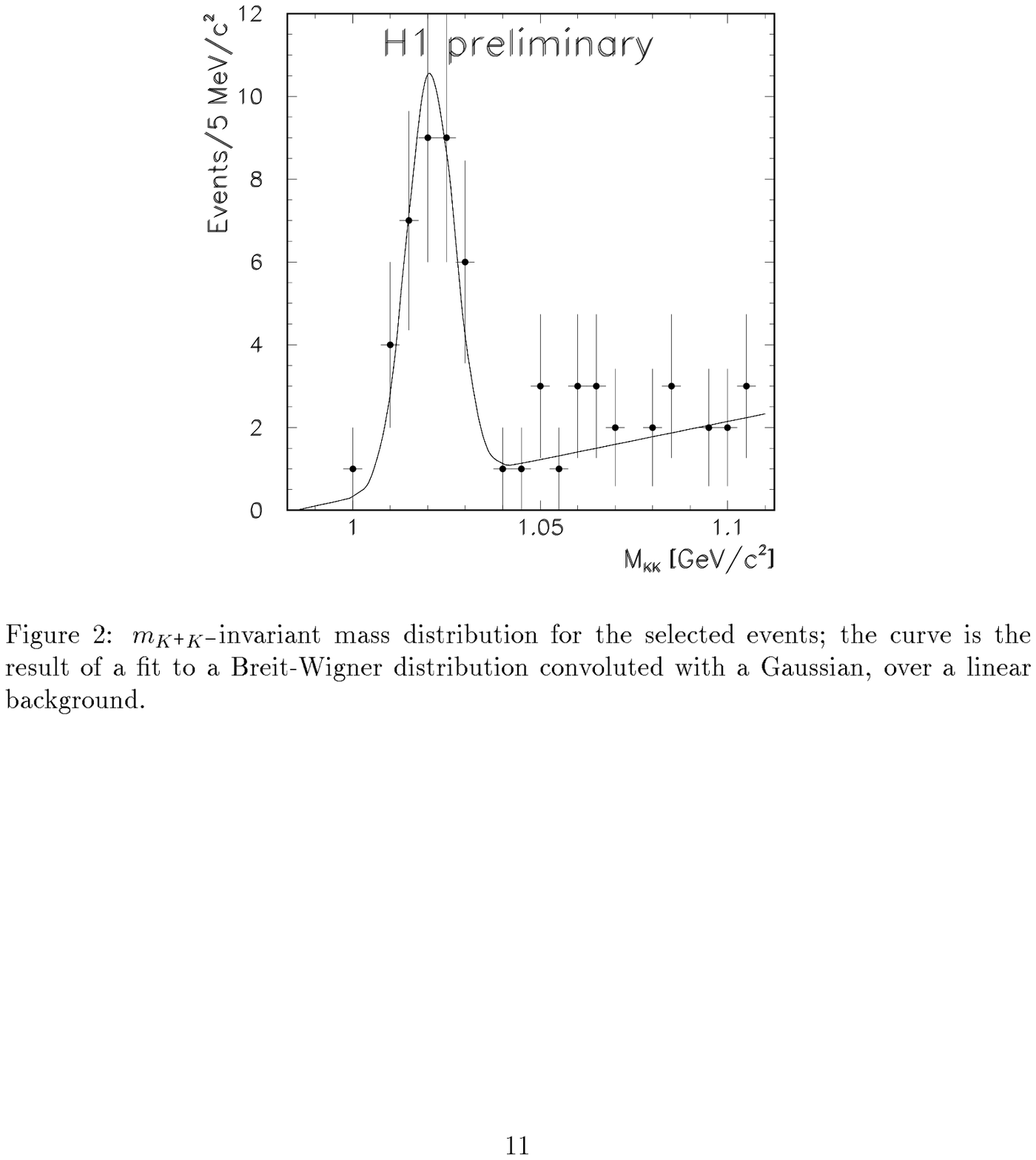}
\end{center}
\caption
{
\label{fig:h1phispectrum}
\it 
The $K^+K^-$ invariant mass spectrum for the 1994
data from the H1 collaboration for the range 
\mbox{$6 < {\qsq} < 20 \; {\gevsq}$}. The solid curve shows
the result of a fit to
a Gaussian--smeared Breit--Wigner distribution
over a linear background due primarily to {\rhoz} decays.
} 
\end{figure}
\item
\underline {$\jpsi \rightarrow e^+e^-$ and $\mu^+\mu^-$}\\*[2mm]
Measurements of elastic and inelastic photopro\-duction of {\jpsi} mesons
have been reported by both the ZEUS~\cite{pl_350_120,pa02_47} and 
the H1~\cite{pa02_85,np_472_3,pl_338_507} 
collaborations. These studies were based 
on the reconstruction of two tracks within the acceptance of the
central detectors, restricting
them to the kinematic range \mbox{$50 < W_{{\gamma^*p}} < 150\;{\gev}$},
with the exception of a sample based on the reconstruction of two
electromagnetic clusters in the rear calorimeter of the H1 detector which 
permitted a measurement in the range \mbox{$170 < W_{{\gamma^*p}} < 260\;{\gev}$}.
Figure~\ref{fig:zeusspppsispectra} shows the dilepton invariant mass spectra
obtained for the elastic sample in the ZEUS analysis. 
The $e^+e^-$ spectrum shows the asymmetry arising
from final--state radiative effects. A polynomial approximation to this shape
was folded with the Gaussian peak and with a flat background arising from
Bethe--Heitler processes to produce the total fit result shown. No such radiative correction was required for the muon spectrum. 
The widths of the spectra were consistent with the detector re\-solution
of 30~MeV. The estimate for the background from proton dissociative 
processes was 30\% with an
uncertainty of about 10\% arising from theoretical uncertainties in the
simulation of the dissociative processes. The statistical contribution to the
uncertainty in the cross sections reported in this study was about 15\%.
The systematic uncertainties were about 10\% and arose
primarily from the uncertainties in trigger efficiency and background 
estimation.
\begin{figure}[htbp]
\begin{center}
\includegraphics[width=0.6\linewidth,bb=69 314 538 800,clip=]{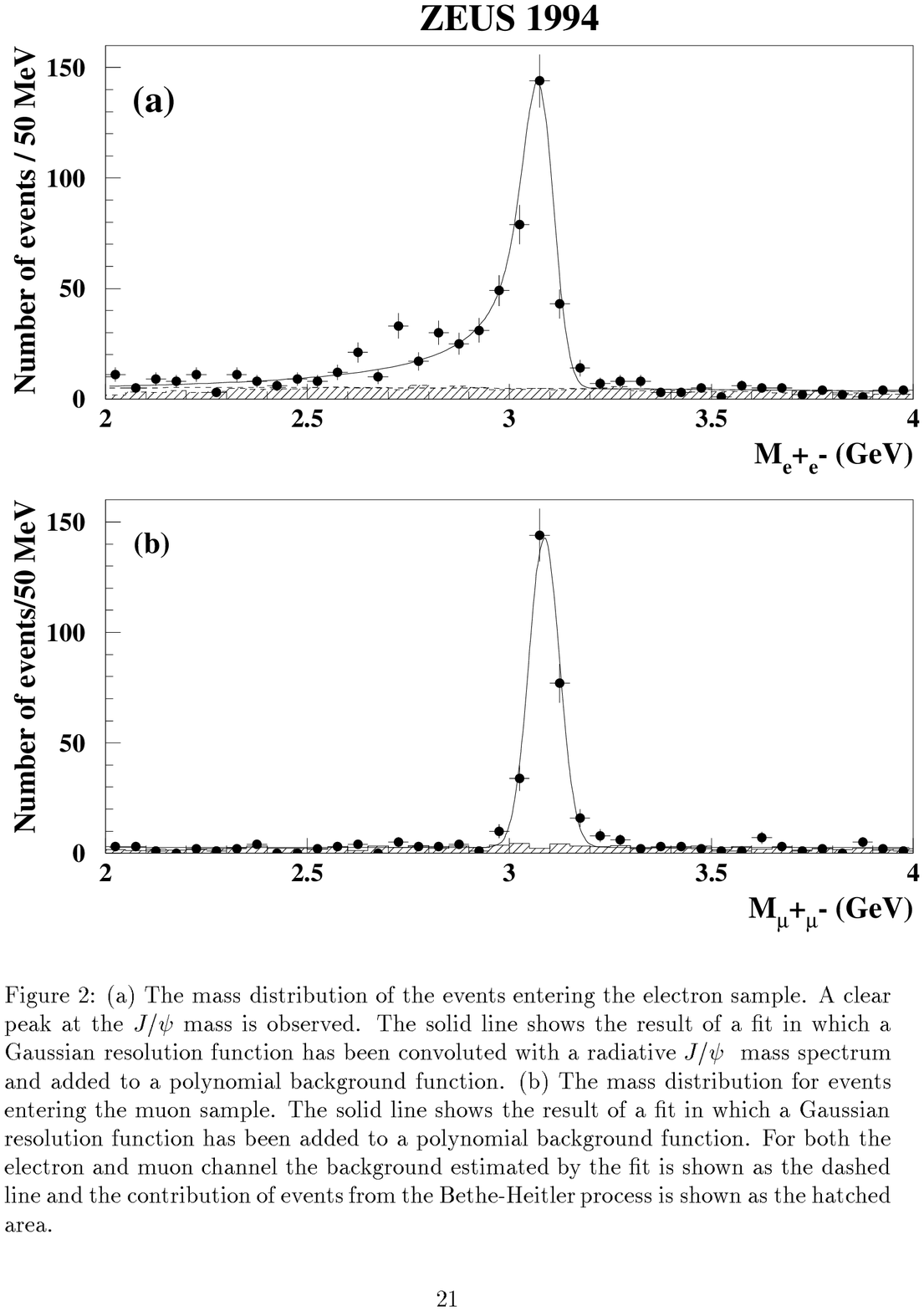}
\end{center}
\caption
{
\label{fig:zeusspppsispectra}
\it 
The dilepton invariant mass spectra in the {\jpsi} mass region
in the ZEUS elastic photopro\-duction study~\protect\cite{pa02_47} for the (a) $e^+e^-$
and (b) ${\mu}^+{\mu}^-$ decay modes. 
The $e^+e^-$ spectrum includes elastic events only
and shows the asymmetry arising
from final--state radiative effects. The radiative mass spectrum was convoluted with
a Gaussian re\-solution function and added to a flat background 
to produce the total fit result shown. 
A satisfactory fit to the distribution
was obtained with a function similar to that employed for the electron
final state but omitting the radiative contribution.
} 
\end{figure}

Figure~\ref{fig:h1spppsispectra}~shows the invariant mass spectra for electrons
and muons for elastic and inelastic processes together as presented in
the H1 study~\cite{np_472_3}, which distinguished elastic \mbox{($z$ = 1)},
elastic with proton dissociation \mbox{($z \lsim 1$)},
and inelastic channels \mbox{($z < 0.8$)} 
via energy depositions in the forward detectors. A mass re\-solution
of ${\approx}70\;\mbox{MeV}$ was indicated by the fit results,
compatible with expectations from detector simulations. The processes
could be distinguished such that background estimates were about 10\%
with an uncertainty of similar magnitude. This background uncertainty, 
together with an uncertainty of similar magnitude
in the trigger efficiency, dominated the total systematic
uncertainty. Cross sections were reported in four $W_{{\gamma^*p}}$--bins
such that the statistical uncertainties and systematic uncertainties
were of comparable magnitude.
\begin{figure}[htbp]
\begin{center}
\includegraphics[width=0.7\linewidth,bb=45 524 547 773,clip=]{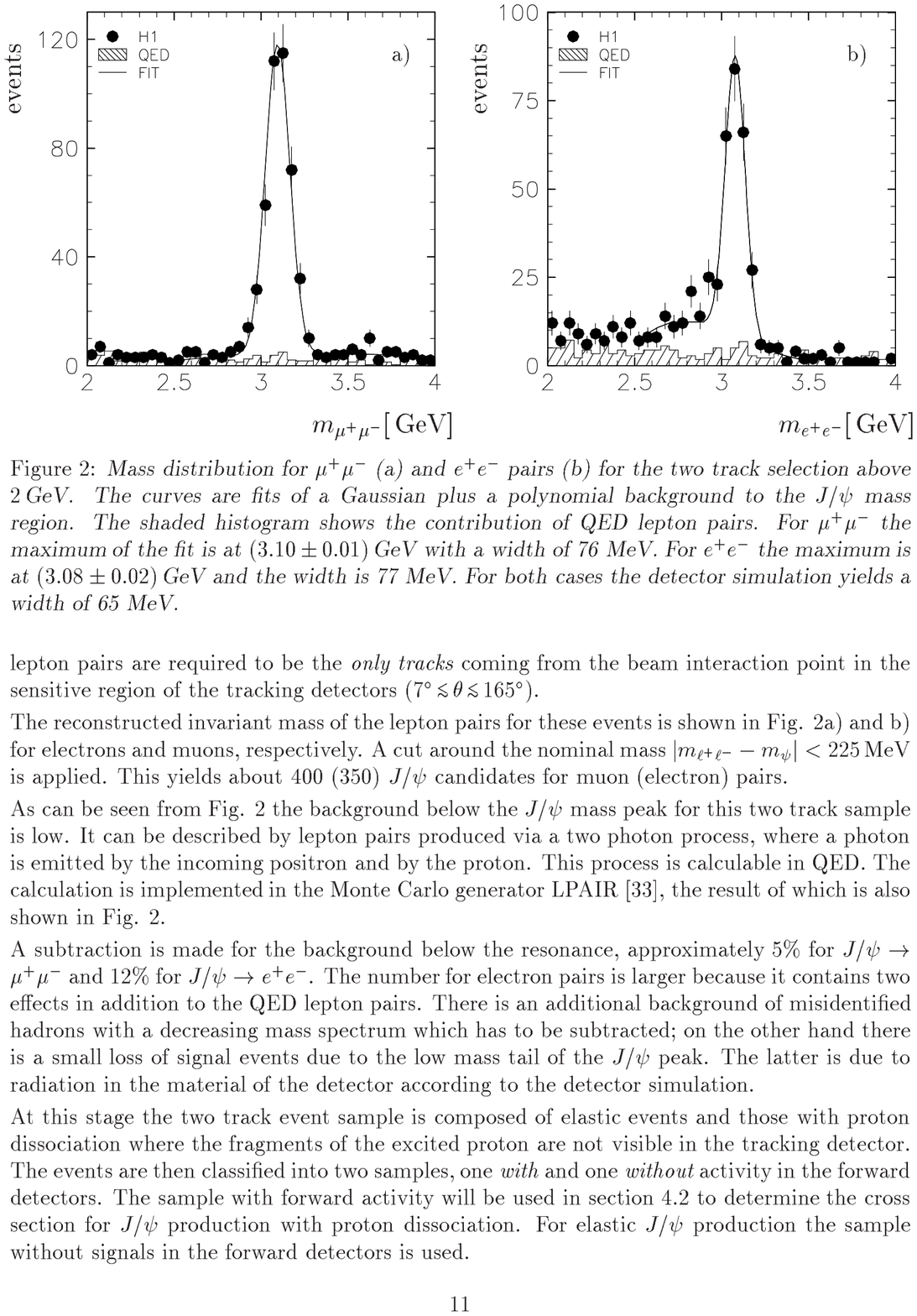}
\end{center}
\caption
{
\label{fig:h1spppsispectra}
\it 
Dilepton invariant mass distributions in the {\jpsi} photopro\-duction 
study by the
H1 collaboration~\protect\cite{np_472_3} for the (a) ${\mu}^+{\mu}^-$
and (b) $e^+e^-$ decay modes. The curves are the result of
a fit to a Gaussian peak summed with a polynomial background.
A simulation of QED Bethe--Heitler processes
is shown as a shaded histogram.
} 
\end{figure}

Clear {\jpsi} signals at $\qsq > 7 \, {\gevsq}$ have been observed 
in both the H1~\cite{np_468_3} and ZEUS~\cite{pa02_28},
experiments, 
but the conclusions of these studies were severely
limited by the available statistics. Results on the {\qsq}--dependence
of the cross section and on the rate relative to that for the {\rhoz} meson
at high {\qsq} will be discussed below.
\item
\underline {$\psi(2S) \rightarrow \jpsi \, \pi^+\pi^-$}\\*[2mm]
An observation of photopro\-duction of the $\psi(2S)$ in its decay to
$\jpsi \, \pi^+\pi^-$ has been reported by the H1 collaboration~\cite{pa02_86}.
The requirement of four charged tracks restricted the acceptance to the
kinematic region \mbox{$z > 0.95$}. The acceptance was about 7\% in the
$\mu^+\mu^-$ decay channel and about 5\% in the $e^+e^-$ channel.
Based on a sample of about two dozen events, they report a result for
the ratio of elastic $\psi(2S)$ photopro\-duction to that for the {\jpsi} 
of \mbox{0.16 $\pm$ 0.06} at a center--of--mass energy of 80~{\gev}.
\end{itemize}

\subsubsection{Elastic Production Cross Sections}
This section summarizes the HERA measurements for the exclusive vector--meson
production cross sections. 
We discuss the dependence of these cross sections 
on the pho\-ton--pro\-ton cen\-ter--of--mass energy,
on the photon virtuality, and on the momentum transferred
at the proton vertex.

\begin{figure}[htbp]
\begin{center}
\includegraphics[width=0.7\linewidth,bb=16 106 581 722,clip=]{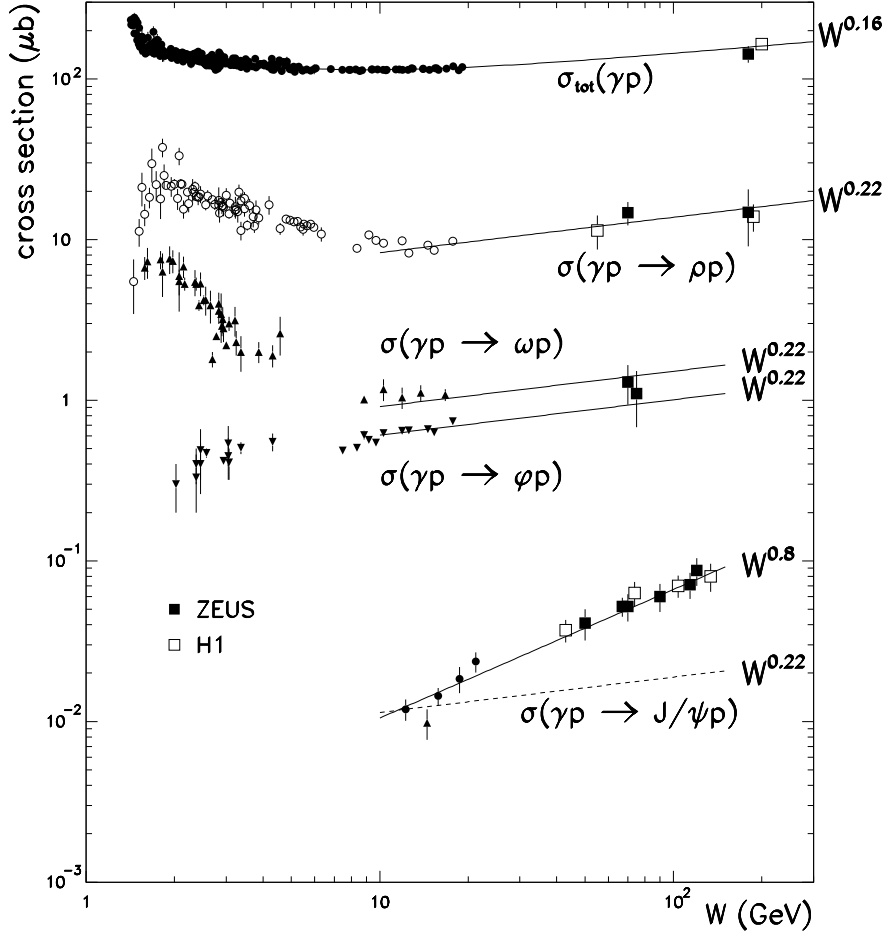}
\end{center}
\caption
{
\label{fig:sppvmwdep}
\it 
The energy dependence of the {\rhoz}, $\omega$, $\phi$
and {\jpsi} exclusive photopro\-duction cross sections measured at HERA, compared to measurements at lower energy
and to the total photoproduction cross section. 
The lines illustrate 
a comparison of various power--law energy dependences at high energy.
} 
\end{figure}

\vspace*{5mm}
\noindent
{\underline{The Energy Dependence and {\qsq}--Dependence of the Cross Section}}\\*[2mm]
\label{sec:wdep}
The various topical models for the dynamics of exclusive vector--meson
production have been outlined in section~\ref{sec:theory}. We turn now to the
cross section measurements and discuss them in the context
of the theoretical models. Figure~\ref{fig:sppvmwdep} shows the photopro\-duction
cross sections for elastic {\rhoz}, $\omega$, $\phi$, and {\jpsi} production 
measured in the two HERA experiments and compares them to measurements at
lower energy and to the total inclusive ${\gamma^*}p$ cross section.
The stark differences
between the production of the
$\omega$ and the ${\phi}$ states observed in the
low--energy resonance region (attributable to the fact that only Pomeron
exchange contributes to {$\phi$} production~\cite{nc_48_541})
are absent in the energy region measured at HERA.
The weak energy dependence associated with soft Pomeron exchange applies
to the {\rhoz}, $\omega$, $\phi$ cross sections.
The {\jpsi} cross section, however,
shows a distinctly stronger energy dependence, motivating
a modified description of the
underlying physical process. Modifications to the Pomeron--exchange
process have been suggested to account 
for the stronger energy dependence~\cite{hep95_07_394,hep96_08_384,pl_379_1}. 
Perturbative QCD calculations have also
been proposed to describe these measurements~\cite{hep95_11_228}. They
are of 
particular interest because they entail a strong sensitivity to the
gluon density in the proton (see section~\ref{sec:theory}). 

\begin{figure}[htbp]
\begin{center}
\includegraphics[width=0.7\linewidth]{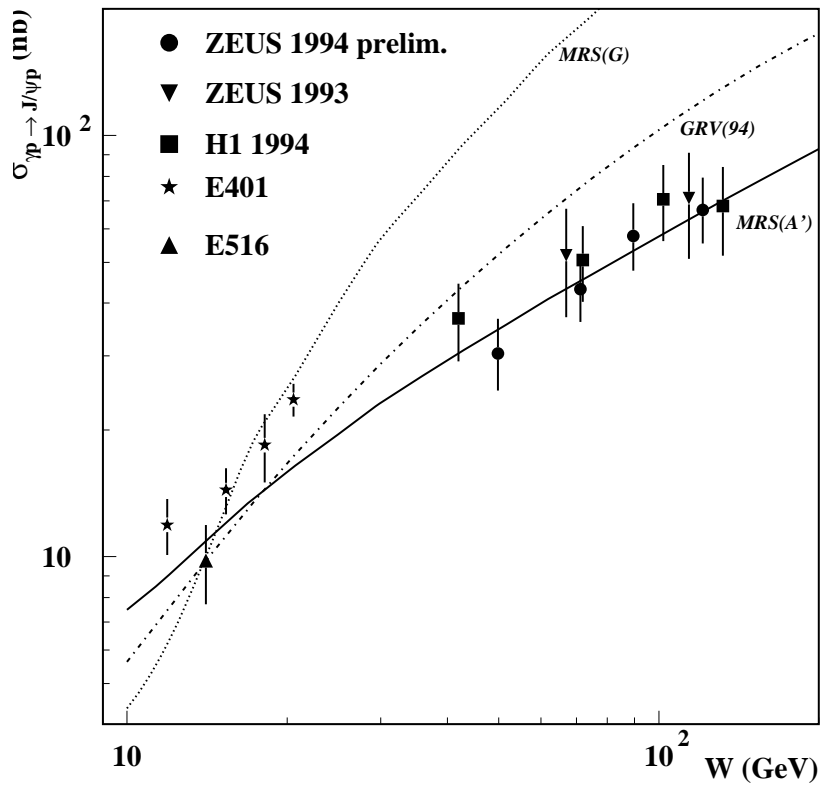}
\end{center}
\caption
{
\label{fig:psiwdep}
\it 
HERA measurements of the elastic {\jpsi} photopro\-duction cross section 
compared to those at lower energy~\protect\cite{prl_48_73,*prl_52_795}  
and to pQCD calculations employing three gluon density 
parametrizations~\protect\cite{hep95_11_228}.
} 
\end{figure}

Figure~\ref{fig:psiwdep} compares the new {\jpsi} 
measurements to those at lower 
energy~\cite{prl_48_73,*prl_52_795} and with
the results of pQCD calculations for three gluon density parametrizations,
which exhibit significantly different slopes  
in the HERA energy region and are only weakly constrained by
other measurements.

\begin{figure}[htbp]
\begin{center}
\includegraphics[width=0.445\linewidth,bb= 115 384 454 739,clip=]{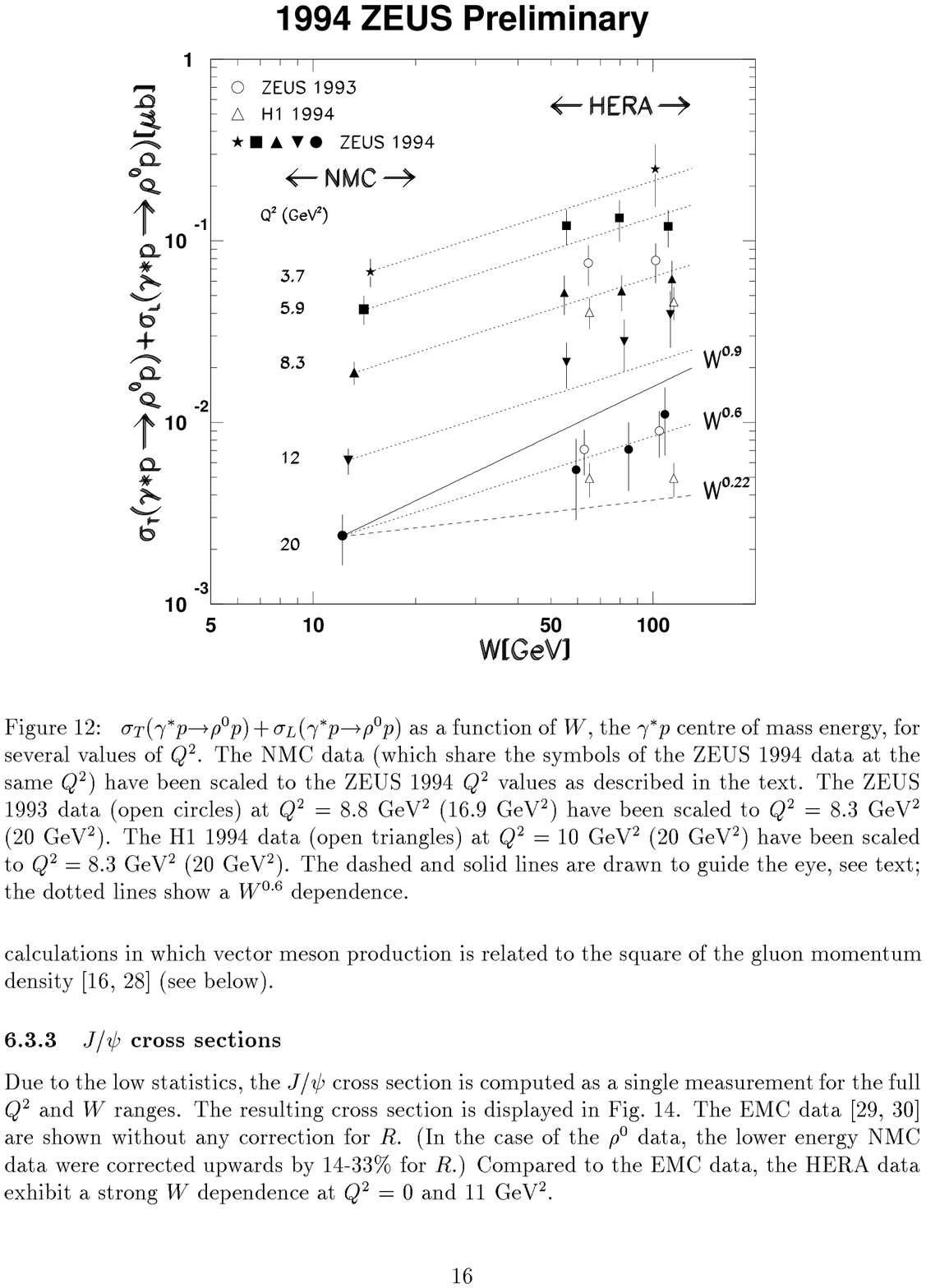}
\end{center}
\caption
{
\label{fig:rhowdep}
\it 
The elastic {\rhoz} cross sections from 
ref.~\protect\cite{pa02_28}.  The dashed lines indicate
the $W^{0.22}$ dependence expected from the exchange of the 
phenomenological soft Pomeron, 
while the
dotted line indicates the stronger dependence more consistent
with the data and the solid line an approximate upper bound.
} 
\begin{center}
\includegraphics[width=0.445\linewidth,bb=177 519 403 752,clip=]{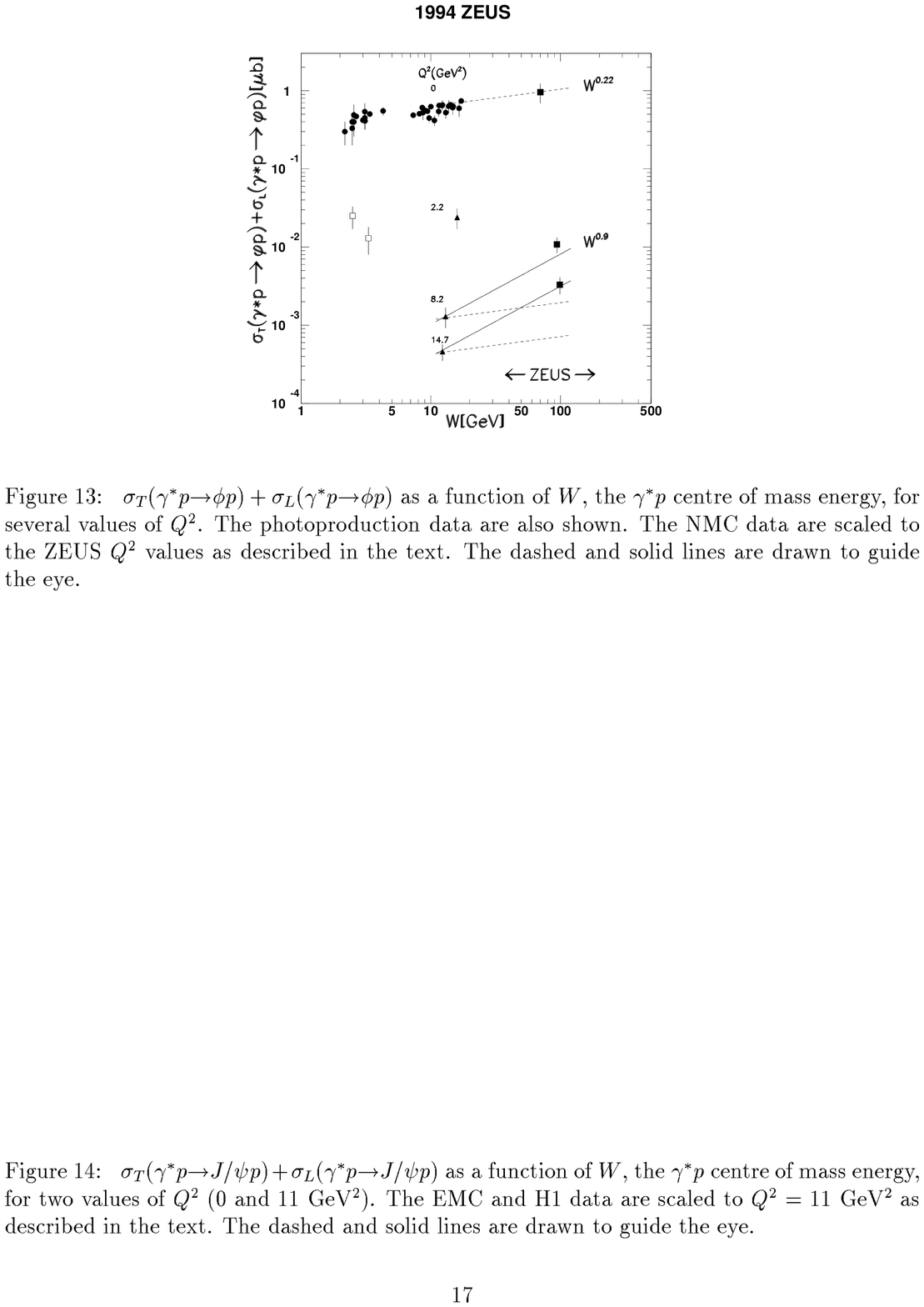}
\end{center}
\caption
{
\label{fig:phiwdep}
\it 
The elastic $\phi$ cross sections from 
ref.~\protect\cite{pa02_28}. The solid circles, open squares and solid
triangles represent low--energy
results from fixed--target experiments. 
The results from the ZEUS
collaboration are shown as solid squares. The measurements at high {\qsq}
have additional correlated systematic uncertainties of 32\% which are not 
shown. The NMC measurements (solid triangles)
have been scaled to the {\qsq}--values of the ZEUS measurements and have
20\% normalization uncertainties which are not shown. The dashed lines
indicate an energy dependence consistent with expectations of Regge 
phenomenology,
while the solid lines indicate the power--law dependence $W^{0.9}$ more
consistent with the measurements at high {\qsq}.
} 
\end{figure}

Figures~\ref{fig:rhowdep} and~\ref{fig:phiwdep} show the H1 and ZEUS results
for the energy dependences of the {\rhoz} and $\phi$ production cross
sections for {\qsq} values between 3.7 and 20~{\gevsq}, compared to
various power--law assumptions. The value of the power used to represent
expectations based on the phenomenological Pomeron exchange was obtained 
using the parameters in eqs.~\ref{eq:alpha} and averaging
over the $t$--distribution in the data.
Measurements by the NMC collaboration
in fixed--target muon scattering~\cite{np_429_503} 
are also presented, allowing comparison
over an order of magnitude in energy. The {\qsq}--dependence observed 
in the NMC results was used to scale the cross sections to the same
average {\qsq}--values as in the ZEUS measurements.
In order to account for the
different average value of the photon polarization parameter {$\epsilon$}
in the NMC data ({$\langle \epsilon \rangle \approx 0.75$), the total exclusive cross
section is calculated as follows:
\begin{eqnarray}
\sigma^T_{{{\gamma^*p}} \rightarrow Vp} +  \sigma^L_{{{\gamma^*p}} \rightarrow Vp} \, = \, \left( \frac{1+R}{1+{\epsilon}R} \right) \; \sigma^{tot}_{{{\gamma^*p}} \rightarrow Vp},
\end{eqnarray}
with
\begin{eqnarray}
R &=& \frac{\sigma^L_{{{\gamma^*p}} \rightarrow Vp}}{\sigma^T_{{{\gamma^*p}} \rightarrow Vp}},
\end{eqnarray}
where the value of $R$ is determined via the helicity analyses described in 
section~\ref{sec:helana} under the assumption that $s$--channel helicity is conserved
in the photon/vector meson transition. The photopro\-duction results included
for comparison have
been used to obtain the total cross section assuming $R$=0. 
While the uncertainties in the cross sections
prevent a firm conclusion, there does appear to be evidence for a steeper
energy dependence at high {\qsq}
than in photopro\-duction, even for these lighter mass
vector mesons. In the context of the perturbative QCD calculations such
an observation leads to the conclusion that the hard scale may indeed be set
either by the mass of the vector meson, as for the {\jpsi} in photopro\-duction,
or by the photon virtuality.

The most accurate measurements of the {\qsq}--dependence of the vector--meson
production cross sections have been presented for the {\rhoz}, due to
the greater statistical precision. This dependence is interesting as a test
of the perturbative QCD predictions~\cite{zfp_57_89,pr_50_3134}.
Calculations of the cross section for longitudinally polarized photons
to form vector mesons in interactions where the photon virtuality sets
the hard scale result in {\qsq}--dependences weaker than $Q^{-6}$, since the 
$Q^{-6}$ factor (see section~\ref{sec:theory}) 
is mitigated by the scale--breaking 
{\qsq}--dependence in the
gluon distribution functions and in the running strong coupling constant.
The models based on Regge phenomenology predict the cross section
to scale as $Q^{-6}$~\cite{pl_348_213}.

Figure~\ref{fig:rhoq2dep} shows results for the {\qsq}--dependence 
in the elastic {\rhoz} production cross section 
\mbox{$\sigma^{tot}_{{{\gamma^*p}} \rightarrow Vp}$} as measured by
the ZEUS collaboration~\cite{pa02_28} for various values of the 
center--of--mass energy and parametrized as $Q^{-2a}$:
\begin{eqnarray*}
a &=& 2.52 \pm 0.28 \; (stat) \pm 0.24 \; (sys) \hspace{15mm} \langle W \rangle = 56 \; {\gev},
\end{eqnarray*}
\begin{eqnarray*}
a &=& 2.37 \pm 0.28 \; (stat) \pm 0.25 \; (sys) \hspace{15mm} \langle W \rangle = 81 \; {\gev},
\end{eqnarray*}
\begin{eqnarray*}
a &=& 1.84 \pm 0.19 \; (stat) \pm 0.24 \; (sys) \hspace{15mm} \langle W \rangle = 110 \; {\gev}.
\end{eqnarray*}
\begin{figure}[htbp]
\begin{center}
\includegraphics[width=0.6\linewidth,bb=146 427 428 766,clip=]{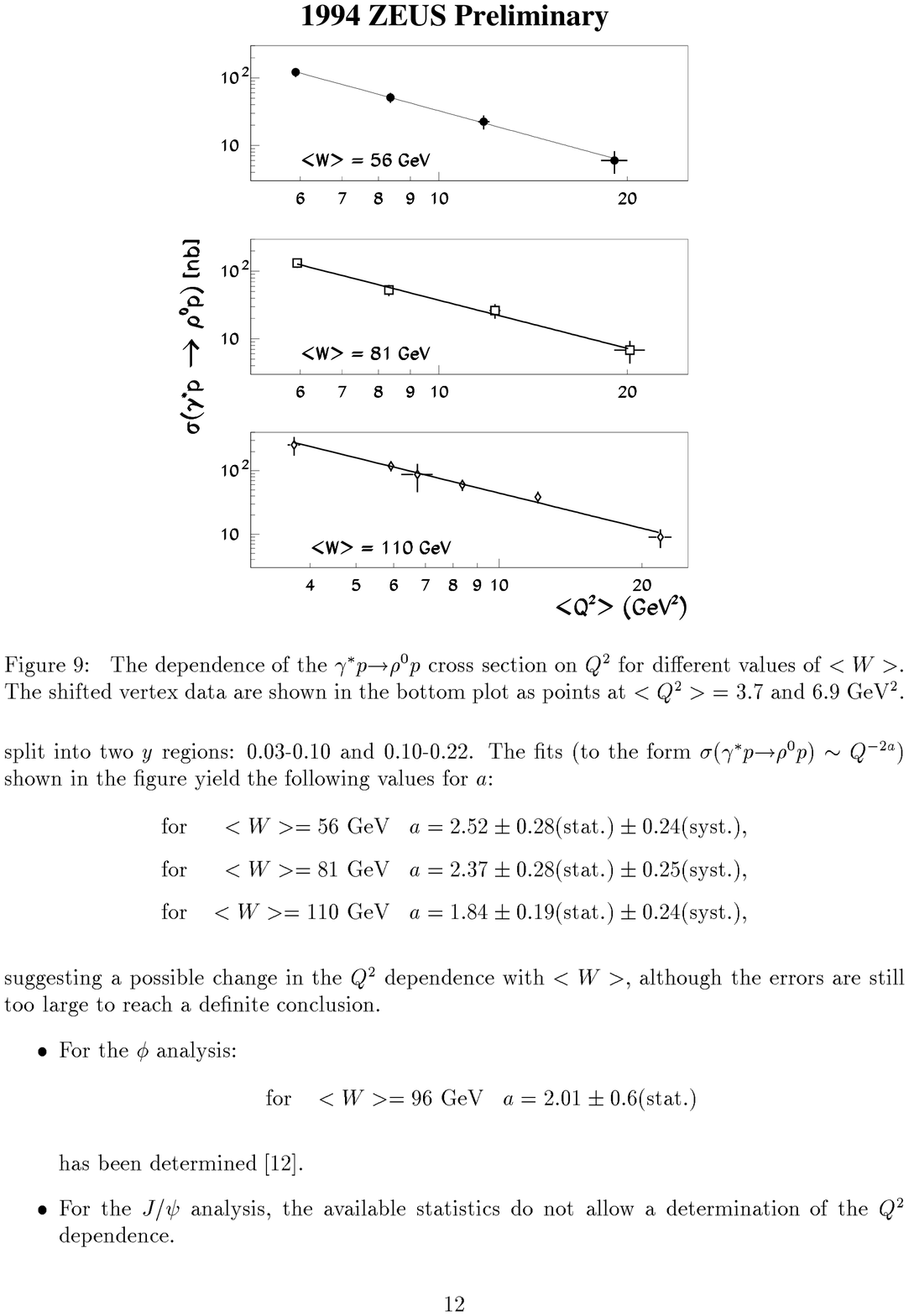}
\end{center}
\caption
{
\label{fig:rhoq2dep}
\it 
The {\qsq}--dependence of the  
elastic {\rhoz} cross section in three ranges of $W_{{\gamma^*p}}$ 
from the preliminary
ZEUS results using data from 1994~\protect\cite{pa02_28}. 
The solid circles, open squares, and open diamonds
represent data restricted to the kinematic regions 
\mbox{0.02 $< y <$ 0.05}, \mbox{0.05 $< y <$ 0.10, 0.10 $< y <$ 0.20}
respectively. The points at {\qsq} values of 3.7 and 6.9~{\gevsq} in the
lower plot were obtained from a data sample where the interaction vertex
was shifted in the forward direction to gain acceptance at lower
values of {\qsq}.
} 
\end{figure}
A result for the {\qsq}--dependence in elastic $\phi$ production was
also presented:
\begin{eqnarray*}
a &=& 2.0 \pm 0.6 \; (stat) \hspace{15mm} \langle W \rangle = 96 \; {\gev}.
\end{eqnarray*}

The H1 collaboration has published results for the {\qsq}--dependence for
{\rhoz} mesons~\cite{np_468_3}:
\begin{eqnarray*}
a &=& 2.5 \pm 0.5 \; (stat) \pm 0.2 \; (sys) \hspace{15mm} \langle W \rangle = 81 \; {\gev},
\end{eqnarray*}
and $\phi$ mesons~\cite{pa02_64} as well:
\begin{eqnarray*}
a &=& 2.0 \pm 0.6 \; (stat) \pm 0.2 \; (sys) \hspace{15mm} \langle W \rangle = 88 \; {\gev}.
\end{eqnarray*}

The H1 collaboration further published a result for the {\jpsi} which combines
their measurements at high {\qsq} with a preliminary ZEUS photopro\-duction
result and uses the parametrization \mbox{$(\qsq + M^2_{J/\psi})^{-a}$}
to obtain~\cite{np_468_3}:
\begin{eqnarray*}
a &=& 1.9 \pm 0.3 \; (stat) \hspace{15mm} \langle W \rangle = 92 \; {\gev}.
\end{eqnarray*}

These results do not allow an unambiguous
distinction between the {\qsq}--dependences
for the various vector mesons. However, there is evidence that the power--law
dependence is weaker than $Q^{-6}$, consistent with pQCD calculations
for longitudinal photons,
as discussed by Frankfurt {\em et al.}~\cite{pr_54_3194}, for example. There
are also indications that the {\qsq}--dependence weakens as the energy
increases. A straightforward comparison with the calculations is not yet
possible, since the measurements consist of a mixture of contributions
from transversely and longitudinally polarized photons 
(see section~\ref{sec:helana}). 

\vspace*{5mm}
\noindent
{\underline {The Dependence of the Cross Section on the Momentum Transfer at the Proton Vertex}}\\*[2mm]
The exponential dependence of production cross sections on the momentum transfer
at the proton vertex is a defining characteristic of diffractive processes.
In the context of Regge theory the exchange of a single trajectory results
in a $t$--dependence with a single well defined slope parameter $b$:
\begin{eqnarray}
\frac{d{\sigma}}{d|t|} = A \cdot e^{-b|t|}.
\end{eqnarray}
This slope
parameter is expected to increase logarithmically with the interaction energy,
a phenomenon sometimes referred to 
as ``shrinkage'' (see section~\ref{sec:theory}), since the $t$--dependence
steepens with increasing energy, narrowing the foward diffractive peak. 
In diffractive processes this slope can be 
directly related to the size of the interaction, which is related
to the size of the vector--meson wave packet. The
pQCD--inspired calculations of ref.~\cite{pr_50_3134}, however, while predicting
slight shrinkage effects as well, point to a universality of the 
$t$--dependence, since it depends only on the two--gluon form factor 
of the proton at high
{\qsq} and is therefore expected to be independent of the 
type of vector meson
produced. These
calculations furthermore predict the rapid disappearance of shrinkage effects
as {\qsq} increases. For elastic scattering the $t$--slope is limited
to values greater than the proton size, i.e. greater than 
about $4 \; {\gev}^{-2}$.

The ZEUS Leading Proton Spectrometer provided a direct measurement of 
the proton transverse momentum for the data recorded in 1994, 
achieving a re\-solution of 0.01 {\gevsq} in the squared 
momentum transfer~\cite{dr_96_183} and reducing the background from
processes with proton dissociation to less than 0.5\%.
The result for the slope parameter in elastic {\rhoz} photoproduction was
\begin{eqnarray*}
b = 9.8 \pm 0.8 \; (stat) \pm 1.1 \; (sys) \; {\gev}^{-2},
\end{eqnarray*}
where the systematic error was primarily determined by the acceptance of
the LPS detector and the influence of the intrinsic transverse momentum of
the proton beam.

All other H1 and ZEUS analyses use
the total transverse momentum observed in the final state as an estimator
for the momentum transferred to the proton. 
For analyses in which the scattered electron was not detected, the
{\qsq}--dependence in the relationship between $p_T^2$ and $t$ was accounted for
in the final results via Monte Carlo simulation and unfolding techniques.
The most accurate measurements of the slope parameter from HERA have been
presented by the ZEUS collaboration using a high statistics $\pi^+\pi^-$ sample
from the 1994 data--taking period~\cite{pa02_50} where neither the electron
nor the proton was detected in the final state. A deviation from a purely
exponential dependence was observed, motivating to the parametrization
\begin{eqnarray}
\label{eq:tdep}
\frac{d{\sigma}}{d|t|} = A \cdot e^{-b_{\pi \pi}|t|+c_{\pi \pi}t^2}.
\end{eqnarray}
Figure~\ref{fig:spprhotslope} shows the $t$--dependence of the measured cross section
together with the results of the fit:
\begin{eqnarray*}
b_{\pi \pi} = 11.6 \pm 0.2 \; (stat) \; {\gev}^{-2}, \hspace{1cm}
c_{\pi \pi} = 4.2 \pm 0.5 \; (stat) \; {\gev}^{-4}.
\end{eqnarray*}
\begin{figure}[htbp]
\begin{center}
\includegraphics[width=0.5\linewidth,bb=159 197 419 762,clip=]{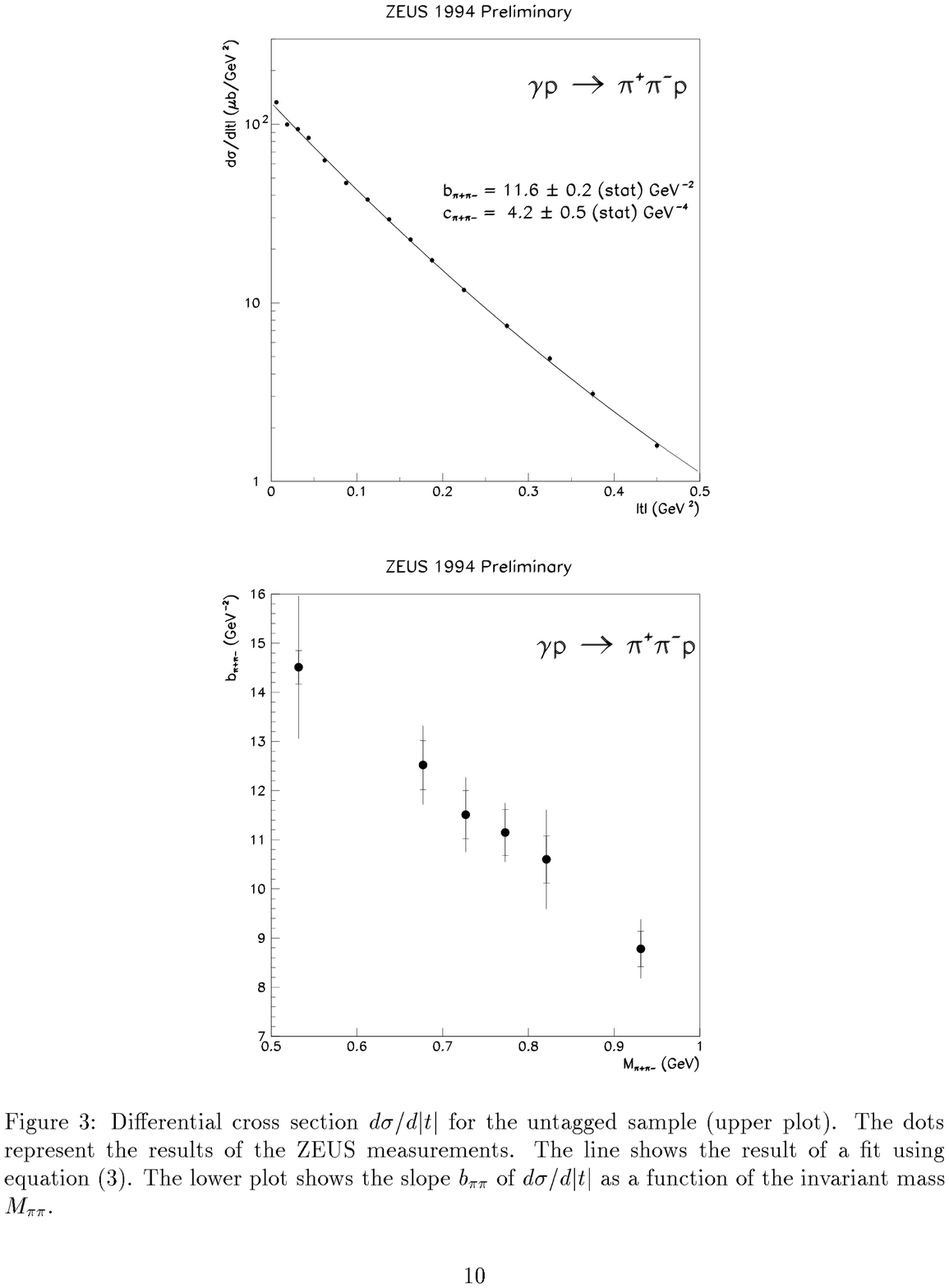}
\end{center}
\caption
{
\label{fig:spprhotslope}
\it 
The $t$--dependence of the  
elastic $\pi^+\pi^-$ photopro\-duction cross section from 
ref.~\protect\cite{pa02_50}. Results for the fit parameters described in
the text are shown in the upper plot. The lower plot shows the linear
slope parameter value as a function of the dipion invariant mass.
The inner error bars represent
the statistical uncertainties and the outer error bars represent the 
quadratic sum of statistical and systematic uncertainties.
} 
\end{figure}

The slope parameter clearly depends
on the range of dipion invariant mass (M$_{\pi \pi}$) 
included in the fit, an effect observed in earlier photopro\-duction 
studies~\cite{pr_5_545,pr_175_1669}.
As described
in section~\ref{sec:expissues} the mass spectrum is influenced by the contribution
of an interference term from a nonresonant dipion background,
leading to the
dependence of the slope parameter on M$_{\pi \pi}$ shown in 
Fig.~\ref{fig:spprhotslope}, and to the necessity for the nonlinear term
in the exponent
in eq.~\ref{eq:tdep}. Further preliminary results
concerning 
the asymmetry in the mass spectrum 
which extend the $t$--range to
\mbox{0.5 $< t < 4.0$ {\gevsq}} (albeit for a sample
dominated by proton-dissociative events)   
show the nonresonant contribution
to decrease sharply with increasing momentum transfer~\cite{pa02_51}.

Photopro\-duction results for the $\omega$~\cite{zfp_73_73} and $\phi$~\cite{pl_377_259} mesons
have been published by the ZEUS collaboration including measurements
of the slope parameter $b$: 
\begin{eqnarray*}
b &=& 10.0 \, \pm \, 1.2 \; (stat) \pm 1.3 \; (sys) \; {\gev}^{-2}
\end{eqnarray*}
for the $\omega$ meson, and 
\begin{eqnarray*}
b &=& 7.3\, \pm\, 1.0 \; (stat) \pm 0.8 \; (sys) \; {\gev}^{-2}
\end{eqnarray*}
for the $\phi$ meson. 
These results indicate that the interaction
radius for the $\phi$ meson 
is smaller than those for the {\rhoz} and $\omega$ mesons.

The ZEUS~\cite{pa02_47} and H1~\cite{np_472_3} 
collaborations have each presented results on the slope
parameter for {\jpsi} photopro\-duction:
\begin{eqnarray*}
\mbox{ZEUS}: b &=& 4.0 \pm 0.4 \; (stat)\;_{-0.7}^{+0.6} \; (sys) \; {\gev}^{-2},\\
\hspace*{5mm} \mbox{H1}: b &=& 4.0 \pm 0.2 \; (stat) \pm 0.2 \; (sys) \; {\gev}^{-2}.
\end{eqnarray*}
The forward detectors were used by the H1 collaboration 
to exclude contamination
from proton dissociative processes, which have a weaker $t$--dependence.
These small values for the slope parameter are consistent with that expected
from the proton alone, indicating that the contribution
from the {\jpsi} wave function is smaller than that from
the proton.

All HERA studies of the photopro\-duction slope parameters covered in this
article lack
the precision necessary to identify deviations from 
the logarithmic energy dependence expected
in soft diffractive processes governed by exchange of the phenomenological
Pomeron. 
The comparisons to measurements at lower energy
also lack such precision, despite the wide range of energy covered, due
to contributions from the uncertainties in the earlier measurements. Probably
the best hope for future information on the topic of shrinkage is
measurement at HERA of the slope parameter
over an extended range of energy (50 -- 200~{\gev})
with an accuracy small compared
to the rise of approximately 0.7~{\gev}$^{-2}$ expected from the
phenomenological Pomeron trajectory over this range.

The status of the slope parameter measurements at high {\qsq} for {\rhoz}
production~\cite{pa02_28} is shown
in Fig.~\ref{fig:bvalues}. 
The precision is limited by statistics,
and is hence significantly better for the {\rhoz} meson than for the $\phi$ and
{\jpsi} mesons, for which slope determinations have also been presented. The
preliminary ZEUS results from the 1994 data--taking period cover the
kinematic region
\mbox{$5 < \qsq < 30 \; {\gevsq}$} and 
\mbox{$42 < W_{{\gamma^*p}} < 134 \; {\gev}$}.
The slopes were determined by fits to the distributions in 
\mbox{$(\vec{p}_T(e)+\vec{p}_T({\rhoz}))^2$}. The inner error bars show the
contribution from statistics, and the outer error bars represent the 
quadratic sum of statistical and systematic uncertainties, excluding a
correlated contribution of 1.5 {\gev}$^{-2}$. 
\begin{figure}[htbp]
\begin{center}
\includegraphics[width=0.6\linewidth,bb=138 158 403 458,clip=]{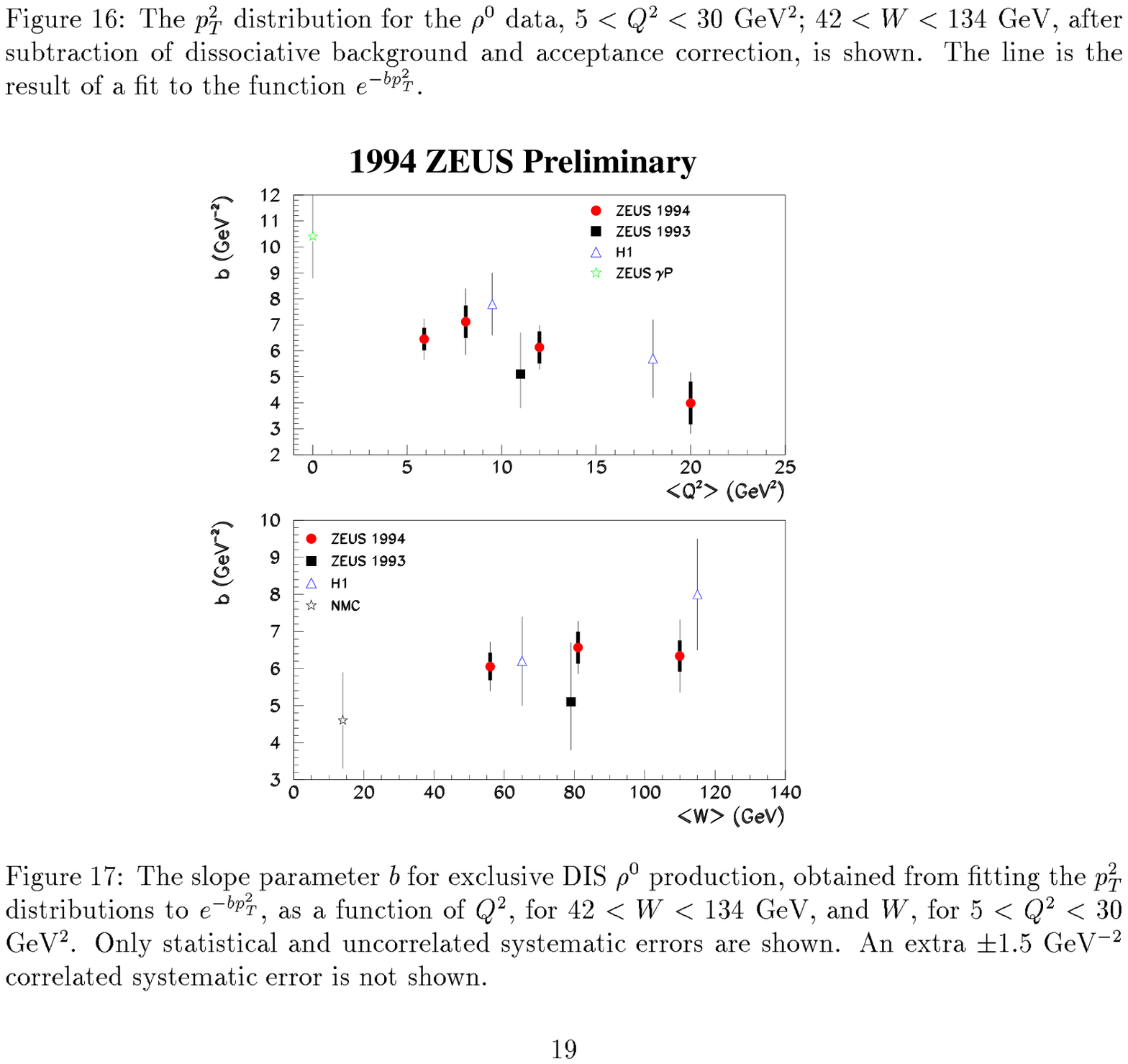}
\end{center}
\caption
{
\label{fig:bvalues}
\it 
A comparison of slope parameter determinations at high 
{\qsq}~\protect\cite{pa02_28}. Preliminary results from the ZEUS collaboration
from the 1994 data--taking period are shown, covering the energy range
\mbox{$42 < W_{{\gamma^*p}} < 134 \; {\gev}$} in the upper plot
and the range
\mbox{$5 < \qsq < 30 \; {\gevsq}$} in the lower plot.
The measurements are compared to published results from
the H1~\protect\cite{np_468_3} and NMC~\protect\cite{np_429_503} 
experiments, and to the ZEUS photopro\-duction result
from ref.~\protect\cite{zfp_69_39}.
The inner error bars on the ZEUS points show the
contribution from statistics, and the outer error bars represent the 
quadratic sum of statistical and systematic uncertainties, excluding a
correlated contribution of 1.5~${\gev}^{-2}$.
} 
\end{figure}
The measurements of the H1~\cite{np_468_3} and NMC~\cite{np_429_503} 
experiments are also shown for comparison, as is the ZEUS photopro\-duction
result. The preliminary conclusions from the observed {\qsq} and W$_{{\gamma^*p}}$
dependences are that there is evidence for a decrease in the slope
parameter with increasing 
{\qsq}, but that the present level of uncertainty excludes
confirmation of any shrinkage effect in the energy dependence. 
The decrease
of the slope with {\qsq} is indicative of a reduction in size of the $q\bar{q}$
wave packet with {\qsq} and the value of about 4~{\gev}$^{-2}$ at high {\qsq}
is comparable to the size of the proton alone as measured in its hadronic
interactions.
For discussions of theoretical ideas 
concerning the {\qsq}--dependence of photon size, see
refs.~\cite{pr_3_1382} and~\cite{rmp_50_261}.
Improvement in these results is a matter of increasing
their statistical precision, for which a gain of a factor of approximately 
five can be expected from analyses of 1995 and 1996 data presently in progress.

\subsubsection{Production Ratios}
The low--energy measurements of $\phi$ photopro\-duction cross sections  
yielded results more than a factor of two lower 
than expected from simple quark counting and a flavor--independent
production mechanism
when compared to {\rhoz} photopro\-duction~\cite{rmp_50_261}. 
The photopro\-duction of {\jpsi} mesons
was shown to be suppressed by nearly three orders of magnitude compared to
that of the {\rhoz} at 
\mbox{$\langle W_{\gamma^* p} \rangle \, = \, 15 \; {\gev}$} 
(see Fig.~\ref{fig:sppvmwdep}). 
These measurements have now been complemented by
the HERA results at high energy and high {\qsq}. 
References~\cite{dr_95_47,pr_50_3134} include discussions of flavor dependence,
or lack thereof, expected in models based on pQCD.

Figure~\ref{fig:phiratio} shows the HERA measurements for the
ratio of the exclusive $\phi$ and {\rhoz} production cross sections
as a function of {\qsq}.  
\begin{figure}[htbp]
\begin{center}
\includegraphics[width=0.6\linewidth,bb=153 168 420 430,clip=]{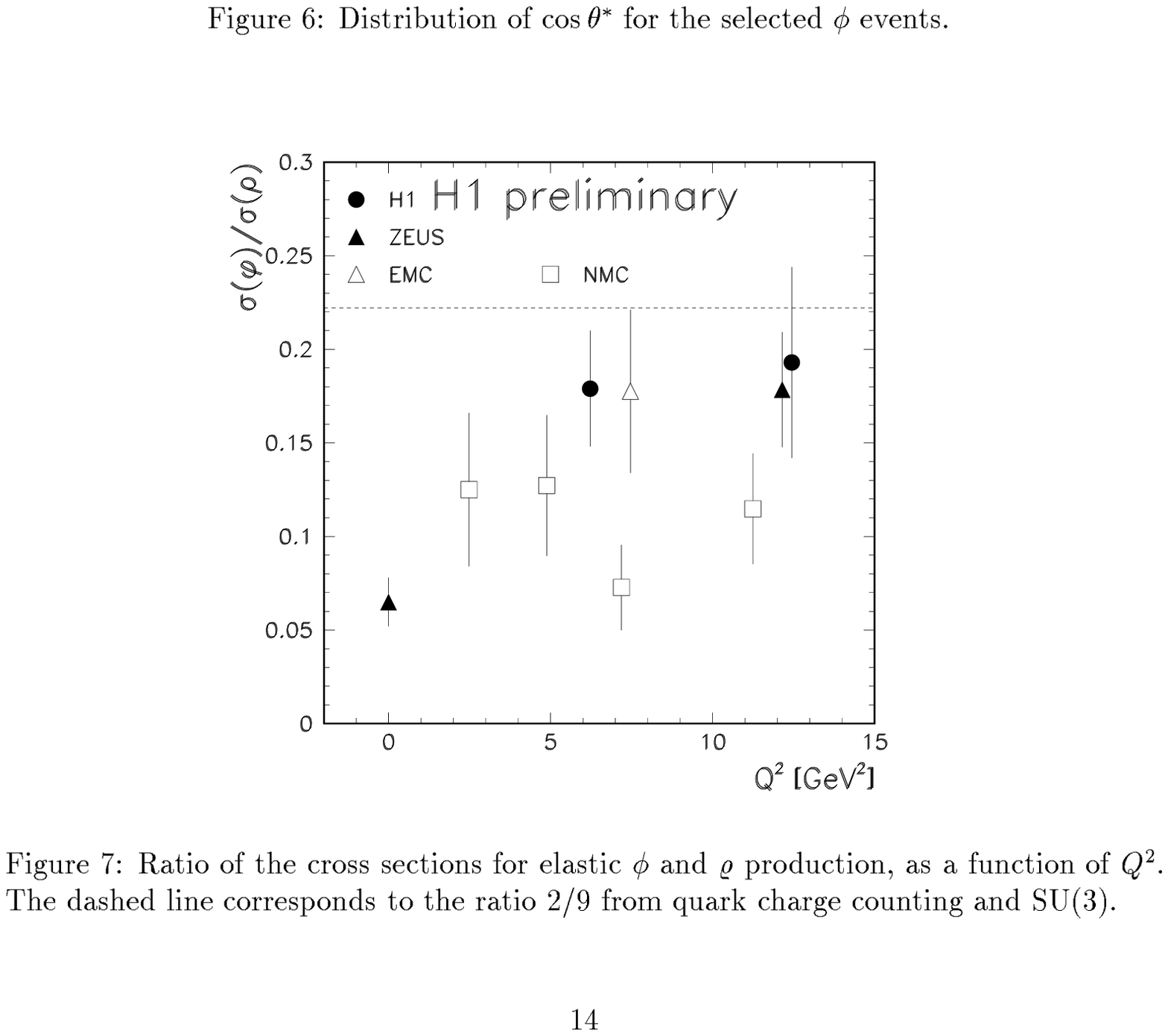}
\end{center}
\caption
{
\label{fig:phiratio}
\it 
The {\qsq}--dependence of the $\phi$--to--{\rhoz} production 
ratio. The results from the ZEUS~\protect\cite{pl_380_220} 
and H1~\protect\cite{pa02_64} collaborations
(shown as the solid circles and triangles)
were obtained at an average photon--proton center--of--mass energy of 
about 90~{\gev}, while those from the
fixed--target muon experiments EMC and NMC (shown as the open squares and triangle) 
were obtained at 15~{\gev}.
The dashed line corresponds to the value for the ratio of 2/9 derived
from a simple quark--counting rule under the assumption of a flavor--independent
production process.
} 
\end{figure}
The results from the ZEUS~\cite{pl_380_220} 
and H1 collaborations~\cite{pa02_64}
were obtained at an average photon--proton center--of--mass energy 
of about 90~{\gev}, while those from the
fixed--target muon experiments EMC~\cite{zfp_39_169} 
and NMC~\cite{np_429_503} were obtained at 15~{\gev}. The data support
a conclusion that at high energy and high {\qsq} the value of
2/9 expected from a quark--counting rule and flavor symmetry is reached.
The ZEUS collaboration has included a discussion of $\phi$ and $\omega$
exclusive photopro\-duction relative to that for the {\rhoz} in their
article on $\omega$ photopro\-duction~\cite{zfp_73_73}, concluding that
the observed energy dependence is consistent with expectations arising
from a diffractive process mediated by Pomeron exchange.

As discussed in section~\ref{sec:wdep}, 
the energy dependence
observed in the elastic {\jpsi} photopro\-duction 
cross section is much stronger
than  that observed in {\rhoz} photopro\-duction. 
The {\jpsi} cross section rises by
nearly an order of magnitude as W$_{\gamma^* p}$ increases from 10 to 
100~{\gev}, while the energy dependence of the {\rhoz} cross section
is consistent with the weak energy dependence typical of soft Pomeron 
exchange.
The {\jpsi}/{\rhoz} 
photoproduction ratio is thus steeply rising with energy, though
production of the {\jpsi} is 
still relatively 
suppressed by two orders of magnitude at HERA energies. This suppression
stands in sharp contrast to the results presented at high {\qsq}. The ZEUS
collaboration has presented a result for the ratio of 
\mbox{($0.35 \pm 0.11 \;(stat)$)}
for the range  
\mbox{7 $< \qsq <$ 25~{\gevsq}~\cite{pa02_28}}, confirming
the earlier 
measurement
of comparable cross sections for exclusive {\rhoz} and {\jpsi}
production at high {\qsq} 
published by the H1 collaboration~\cite{np_468_3}. These results 
represent further evidence for an underlying flavor--independent
production process at high {\qsq}. In the context of QCD, the
dominance of {\rhoz} production at low {\qsq} can be interpreted as due
to nonperturbative effects associated with the large value of the strong
coupling constant in the absence of a hard scale.

\subsubsection{Helicity Analyses}
\label{sec:helana}
The ratio of the vector--meson production cross section for longitudinally
polarized photons to that for transversely polarized photons is a quantity
for which there are specific predictions which characterize several of
the models employed to describe the high--{\qsq} measurements. A firm prediction
of the pQCD calculations is that the process be dominated by the production
of longitudinally polarized vector mesons from longitudinally polarized
photons at high {\qsq}~\cite{pr_50_3134,hep96_11_433}. The applicability of the
calculations depends crucially on this prediction, since the production by
transversely polarized photons is not included. Models based on Pomeron exchange
similarly predict the predominance of longitudinally polarized vector--meson
production at high {\qsq}~\cite{pl_348_213,np_336_1}. The color dipole 
model of ref.~\cite{hep96_05_231} also offers
a specific prediction for the {\qsq}--dependence of the 
cross section ratio: the ratio of the
longitudinal contribution to the transverse contribution
is expected to rise more slowly than \mbox{${\qsq}/M_V^2$}.

The polarization of the vector meson is experimentally accessible via the
decay angular distributions, accurately measured in the central tracking
chambers of the HERA experiments. A general geometrical description of the
process~\cite{collins} 
involves the definition of three planes: \mbox{1) the} electron
scattering plane, \mbox{2) the} plane containing the photon and vector--meson
momentum vectors (called the the vector--meson production plane), 
and \mbox{3)} the vector--meson decay plane. The vector--meson
production mechanism then defines which choice of reference frame is best
adapted to a simple description of the process. Bubble chamber studies
of
{\rhoz} photopro\-duction~\cite{pr_5_545,pr_175_1669} 
have shown that the spin--density matrix
elements defined in the Gottfried--Jackson and Adair reference systems
were strongly dependent on $t$, thus excluding $t$--channel helicity conserving
and spin--independent processes as candidates for the dominant underlying
production mechanism. (Ref.~\cite{pr_5_545} includes a nice discussion
of these reference frames.) The $s$--channel 
helicity frame, i.e. the {\rhoz} meson rest frame with the quantization axis chosen
as the 
direction of the {\rhoz} meson in the $\gamma{p}$ center--of--mass system,
proved to yield spin--density matrix elements independent of the momentum 
transfer. This result pointed to 
an underlying process in which helicity is conserved
in the photon---vector--meson transition for
amplitudes defined in the $s$--channel (SCHC). 
(Theoretical discussions of $s$--channel helicity conservation can
be found in 
refs.~\cite{pl_31_387,pl_65_463,*pr_15_2503}.) 
A streamer chamber experiment studying
{\rhoz} electroproduction on a hydrogen target at DESY
also concluded the validity of SCHC~\cite{np_113_53}.
Such results have led to the helicity frame being chosen as reference
in the HERA studies at both low and high {\qsq}, though present experimental
precision has not allowed a full spin--density analysis of the type performed
for the low--energy measurements. The spin--density matrix can be
decomposed into even/odd (referred to as ``natural'' and ``unnatural'')
parity--exchange components when the amplitudes are expressed in the
$t$--channel~\cite{collins,np_61_381}. Interference between these two
components decreases with energy. Linear photon polarization
allows the experimental distinction of natural and unnatural
parity--exchange components. Together with the assumption of SCHC, the
transformation properties of the underlying production process
under $t$--channel parity exchange completely determine the spin--density
matrix elements.

A complete characterization of the process (see, for example, 
refs.~\cite{np_61_381} and~\cite{manybody}) requires the definition of
three angles in the vector-meson rest frame. Two of these are canonically
chosen as 
the polar (\thetah) and azimuthal (\phih) angles of the 
momentum of the 
positively charged vector--meson decay product in the coordinate system where
the vector--meson momentum 
in the $\gamma{p}$ center--of--mass frame
is the $z$ axis and the $y$ axis is perpendicular to
the vector--meson production
plane. The third is the
angle between the vector--meson production plane and the electron
scattering plane (\Phih). For small values of the momentum transfer
at the proton vertex, $|t| \gsim |t|_{min}$, the
photon and vector meson are nearly collinear and the angle {\phih} 
becomes ill--defined and poorly measured. 
The assumption of $s$--channel helicity conservation 
constrains the azimuthal distribution to depend solely on the difference
of the two azimuthal angles: $\psih = {\Phih} - {\phih}$. 
The angle {\Phih} can be determined only if
the direction of the 
final--state electron momentum 
is measured, so the untagged photopro\-duction experiments
necessarily average over it. Furthermore, the
angle {\phih} can be determined only if the photon direction of flight 
is known,
but in practice 
it turns out that for the photopro\-duction processes studied at HERA 
the flight direction of the photon
is a good approximation to that of the initial--state 
electron,
so the azimuthal decay angular distribution can be determined even when
the final--state electron is not detected.

The three--dimensional angular distribution for the case of an 
unpolarized lepton beam is given by~\cite{np_61_381}:

\vspace*{5mm}
\noindent
$W(\cos{\thetah},\phih,\Phih) \; =$
\begin{equation}
\label{eq:threedang}
\begin{array}{ll}
\frac{3}{4\pi}\biggl[\biggr.&\hspace*{-2mm}
\frac{1}{2} \left(1-r_{00}^{04}\right)+\frac{1}{2}\left(3r_{00}^{04}-1\right)\cos^2{\thetah}-\sqrt{2}\;\Rej\,r_{10}^{04}\;\sin{2\thetah}\cos{\phih}-r_{1-1}^{04}\sin^2{\thetah}\cos{2\phih}\\[3mm]
&\hspace*{-7mm}- \epsilon \cos{2\Phih}\left\{r_{11}^1\sin^2{\thetah}+r_{00}^1\cos^2{\thetah}-\sqrt{2}\;\Rej\,r_{10}^1\;\sin{2\thetah}\cos{\phih}-r_{1-1}^1\sin^2{\thetah}\cos{2\phih}\right\}\\[3mm]
&\hspace*{-7mm}-\epsilon\sin{2\Phih}\left\{\sqrt{2}\;\Imj\,r_{10}^2\;\sin{2\thetah}\sin{\phih}+\Imj\,r_{1-1}^2\sin^2{\thetah}\sin{2\phih}\right\}\\[3mm]
&\hspace*{-7mm}+\sqrt{2\epsilon(1+\epsilon)}\cos{\Phih}\left\{r_{11}^5\sin^2{\thetah}+r_{00}^5\cos^2{\thetah}-\sqrt{2}\;\Rej\,r_{10}^5\;\sin{2\thetah}\cos{\phih}-r_{1-1}^5\sin^2{\thetah}\cos{2\phih}\right\}\\[3mm]
&\hspace*{-7mm}\biggl.+\sqrt{2\epsilon(1+\epsilon)}\sin{\Phih}\left\{\sqrt{2}\;\Imj\,r_{10}^6\;\sin{2\thetah}\sin{\phih}+\Imj\,r_{1-1}^6\sin^2{\thetah}\sin{2\phih}\right\}
\hspace*{2mm}\biggr],
\end{array}
\end{equation}

\vspace*{3mm}
\noindent
where $\epsilon$ is the virtual photon polarization parameter defined in
eq.~\ref{eq:epsilon}. The subscripts $i$ and $k$ run over the three
possible helicity states $-1$, $0$, and $+1$.
The quantities $r_{ik}^{\alpha}, r_{ik}^{04}$, are
linear combinations of the spin--density matrix 
elements $\rho_{ik}^{\alpha}$, ($\alpha$=0,1,2,4,5,6),
which themselves depend on the helicity amplitudes
and are sensitive to the dynamics of the production process:
\begin{eqnarray}
\label{eq:rtorho1}
r^{04}_{ik} &=& \frac{\rho^0_{ik} \, + \, \epsilon R \rho^{4}_{ik}}{1 \, + \, \epsilon R} \\[5mm]
\label{eq:rtorho2}
r^{\alpha}_{ik} &=& \, \left\{ \begin{array}{ll} \frac{\rho^{\alpha}_{ik}}{1 \, + \, \epsilon R}, & {\alpha}=1,2\\[5mm] \frac{\sqrt{R} \; \rho^{\alpha}_{ik}}{1 \, + \, \epsilon R}, & {\alpha}=5,6 \end{array} \right.
\end{eqnarray}
The superscript $0$ corresponds to the case of unpolarized transverse
photons; the superscripts $1$ and $2$ indicate terms arising in the case
of linearly polarized transverse photons; the superscript $4$ represents the
contribution from longitudinally polarized photons; and the terms with
superscripts $5$ and $6$ indicate the interference between the longitudinal and
transverse amplitudes.
The variable $R$ is the ratio of the elastic {\rhoz} production cross section 
for longitudinal photons to that for transverse photons.

If the helicity amplitudes obey
SCHC and natural parity exchange, then the only nonzero density matrix elements
are 
$\rho^4_{00}=1, \Rej\,\rho_{1-1}^1=-\Imj\,\rho_{1-1}^2=\frac{1}{2}, \Rej\,\rho_{10}^5=-\Imj\,\rho_{10}^6=-\cos{\delta}/\sqrt{8}$
and the angular distribution reduces to
\begin{eqnarray}
\label{eq:schcang}
\begin{array}{lll}
W(\cos{\thetah},\psih) = & \frac{3}{8\pi}\;\frac{1}{1+\epsilon R}\;
\biggl[ \biggr. &
\sin^2{\thetah}(1+\epsilon\cos{2\psih})+2\epsilon R \cos^2{\thetah}\\[3mm]
& & \biggl.-\sqrt{2R\epsilon(1+\epsilon)}\,\cos{\delta}\sin{2\thetah}\cos{\psih}
\hspace*{5mm} \biggr],
\end{array}
\end{eqnarray}
where the parameter $\delta$ is the relative phase angle between the
longitudinal and transverse amplitudes.
The value of $R$ can be deduced directly from the value of
$r_{00}^{04}$ determined by the polar angle dependence when
averaging over the azimuthal dependence:
\begin{eqnarray}
\label{eq:r04tor}
R = \frac{1}{\epsilon} \frac{r_{00}^{04}}{1-r_{00}^{04}}.
\end{eqnarray}

The first measurements at HERA of the spin--density matrix elements derived from
vector--meson decay angular distributions~\cite{zfp_69_39} were based
on the observation of {\rhoz} mesons produced in pho\-to\-pro\-duction by the
ZEUS collaboration. The
polar and azimuthal angular distributions are 
shown in Fig.~\ref{fig:spprhohelicity}. 
The observed 
$\sin^2{\thetah}$--dependence indicates that the {\rhoz} mesons were produced
in a transversely polarized state, consistent with the expectations arising
from SCHC. The matrix element was determined to be 
$r_{00}^{04}=(0.055 \pm 0.028)$. The final--state electron is not
measured and the resulting average over {\Phih} 
eliminates the possibility of a contribution from
$r_{1-1}^{1}$, so the hypothesis of a contribution from $r_{1-1}^{04}$ to
the azimuthal distribution was tested. It was found to be small: 
$r_{1-1}^{04}=(0.008 \pm 0.014)$, consistent with the value derived from SCHC, 
since SCHC requires the matrix elements 
${\rho}_{1-1}^{0}$ and ${\rho}_{1-1}^{4}$ to vanish.
From the value of $r_{00}^{04}$ a value for $R$ of ($0.06 \pm 0.03$) was
calculated based on eq.~\ref{eq:r04tor}, commensurate with the value
of 0.1 expected in the Vector--Dominance Model.

\begin{figure}[htbp]
\begin{center}
\includegraphics[width=0.8\linewidth,bb=64 263 504 646,clip=]{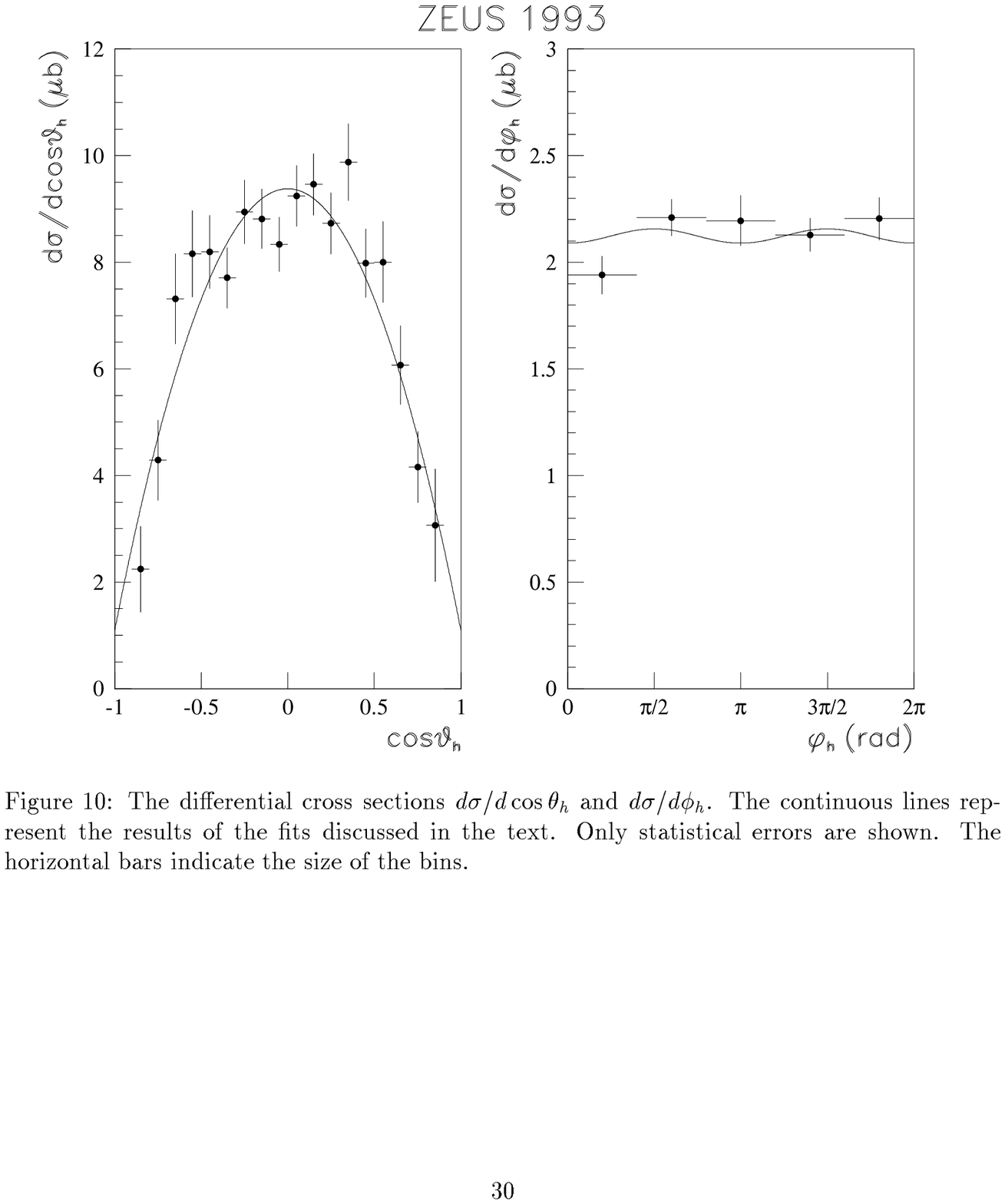}
\end{center}
\caption
{
\label{fig:spprhohelicity}
\it 
The helicity angle distributions observed in 
photopro\-duction of the {\rhoz} meson by the ZEUS 
collaboration. The dominance of the $\sin^2{\thetah}$--dependence
and the lack of $\phih$--dependence are consistent with the expectations
based on $s$--channel helicity conservation.
} 
\end{figure}

We turn now to the helicity analyses in which the final--state electron
was detected, allowing the measurement of the angle between the
electron scattering plane and the {\rhoz} production plane.
Under the assumption of SCHC the azimuthal distribution is dependent
only upon the difference of {\phih} and {\Phih}. Calculating the projections
from eq.~\ref{eq:schcang}
we write the angular distributions
\begin{eqnarray}
\frac{1}{N} \frac{dN}{d(\cos{\thetah})}
&=&\frac{3}{4}[1-r_{00}^{04}+(3r_{00}^{04}-1)\cos^2{\thetah}],
\end{eqnarray}
\begin{eqnarray}
\frac{1}{N} \frac{dN}{d\psih} &=&
\frac{1}{2\pi}(1+2{\epsilon} \, r_{1-1}^{1}\cos2\psih).
\end{eqnarray}
The spin--density matrix element $r_{00}^{04}$ represents the probability that
the vector meson be produced with helicity zero (i.e. longitudinally 
polarized) by either transversely or longitudinally polarized virtual photons. 
According to eqs.~\ref{eq:rtorho1} and~\ref{eq:rtorho2}
\begin{eqnarray}
r_{00}^{04} = \frac{1}{1+{\epsilon}R} \; ({\rho}_{00}^{0}+{\epsilon}R{\rho}_{00}^{4}), \hspace{5mm}
r_{1-1}^{1}= \frac{1}{1+{\epsilon}R} \; {\rho}_{1-1}^{1},
\end{eqnarray}
where ${\rho}_{00}^{0}$ is the probability to produce a longitudinally polarized
vector meson from a transverse photon, ${\rho}_{00}^{4}$ is the probability
to produce such a longitudinally polarized vector meson from a longitudinal
photon, and ${\rho}_{1-1}^{1}$ is sensitive to the interference between
nonflip and double--flip amplitudes from transverse photons. Thus
one can see that a nonzero value for $r_{00}^{04}$ indicates the production
of longitudinally polarized vector mesons, but does not distinguish between
the production by transverse and longitudinal photons, whereas a 
nonzero
value for $r_{1-1}^{1}$, signaled by variation with {\psih},
provides an unambiguous sign of a contribution from transverse photons.

The assumption of SCHC and natural parity exchange determines the 
vector--meson spin--density matrix elements to be: 
\mbox{${\rho}_{00}^{0}=0$}, \mbox{${\rho}_{00}^{4}=1$}, 
\mbox{${\rho}_{1-1}^{1}=\frac{1}{2}$} 
and the matrix element $r_{1-1}^{1}$ is
directly related to $r_{00}^{04}$:
\begin{eqnarray}
\label{eq:r1-1tor04}
r_{1-1}^{1} = \frac{1}{2} (1-r_{00}^{04}).
\end{eqnarray}

An helicity analysis for {\rhoz} production for $\qsq > 8~{\gevsq}$ 
has
been presented by the H1 collaboration~\cite{pa03_48}, which includes
a clear indication for a nonzero value for the matrix elements governing
both the polar and azimuthal angular dependences. Figure~\ref{fig:rhohelicity}
shows the observed angular distributions. The polar angle dependence
clearly exhibits the predominantly longitudinal {\rhoz} polarization, 
and a value for $R$ was determined from 
the value for $r_{00}^{04}$: 
\begin{eqnarray*}
r_{00}^{04} &=& 0.73 \pm 0.05 \;(stat) \pm 0.02 \;(sys), \hspace{5mm}
R = 2.7^{+0.7}_{-0.5} \;(stat) ^{+0.3}_{-0.2} \;(sys).
\end{eqnarray*}
Since the assumption of SCHC requires 
$r_{1-1}^{04}$ to vanish, eliminating its contribution
to the azimuthal angular distribution (see eq.~\ref{eq:threedang}), a value for 
$r_{1-1}^{1}$ could be determined by a fit to the distribution,
with the result:
\begin{eqnarray*}
r_{1-1}^{1} &=& 0.14 \pm 0.05 \;(stat) \pm 0.01 \;(sys).
\end{eqnarray*}
The con\-sistency of this result with the in\-de\-pen\-dent
ex\-pec\-tation from eq.~\ref{eq:r1-1tor04} 
\begin{eqnarray*}
r_{1-1}^{1} &=& 0.14 \pm 0.03 \;(stat) \pm 0.01 \;(sys)
\end{eqnarray*}
sup\-ports the validity of the
assumption of $s$--channel helicity
conservation with natural parity exchange in the $t$--channel.
\begin{figure}[htbp]
\begin{center}
\includegraphics[width=0.8\linewidth,bb=90 464 489 655,clip=]{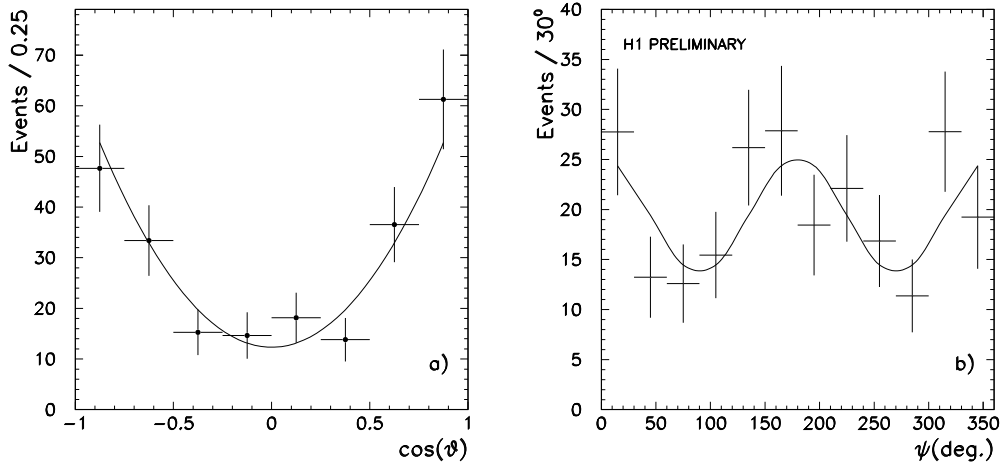}
\end{center}
\caption
{
\label{fig:rhohelicity}
\it 
The helicity angle distributions for elastic {\rhoz} production obtained
by the H1 collaboration for $\qsq > 8$~{\gevsq}.
} 
\end{figure}

The ZEUS collaboration has used data from its beam--pipe calorimeter
to obtain results on {\rhoz} production in the range 
\mbox{$0.25 < {\qsq} < 0.85$ {\gevsq}}
 which exhibit a large 
longitudinal component, as well as a nonzero value for $r_{1-1}^{1}$.
This analysis furthermore considers two separate ranges of {\qsq},
since there is other evidence that this region is sensitive to the
transition between soft and hard physics (cf. section~\ref{sec:gammapxsect}), 
and
found an appreciable increase of the longitudinal contribution with {\qsq}.
Figure~\ref{fig:bpcrhohelicity} shows the angular distributions from which
the following values for the matrix elements were derived:
\begin{eqnarray*}
\langle \qsq \rangle &=& 0.34 \, \gevsq: \hspace{5mm} r_{00}^{04} = 0.21 \pm 0.03, \hspace{5mm} 
r_{1-1}^{1} = 0.38 \pm 0.03, \hspace{5mm} R = 0.26 \pm 0.04\\ 
\langle \qsq \rangle &=& 0.61 \, \gevsq: \hspace{5mm} r_{00}^{04} = 0.36 \pm 0.04, \hspace{5mm} 
r_{1-1}^{1} = 0.31 \pm 0.03, \hspace{5mm} R = 0.56 \pm 0.09.
\end{eqnarray*}
\begin{figure}[htbp]
\begin{center}
\includegraphics[width=0.72\linewidth,bb=120 510 470 778,clip=]{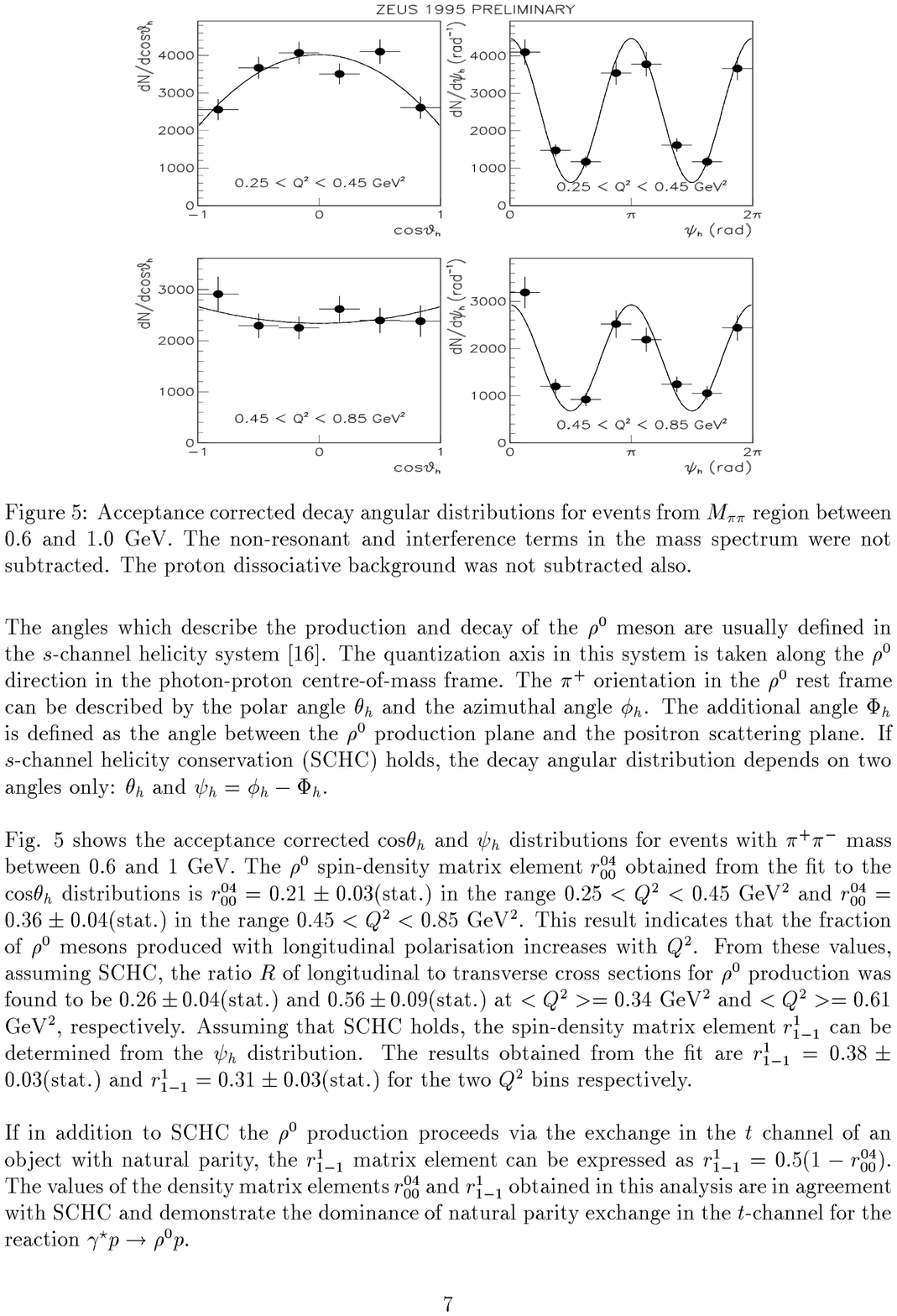}
\end{center}
\caption
{
\label{fig:bpcrhohelicity}
\it 
The {\rhoz} helicity angle distributions obtained from data recorded
in 1995 with the ZEUS beam--pipe calorimeter, separated into
two {\qsq} regions with average {\qsq} values of 0.34 and 0.61~{\gevsq}.
} 
\end{figure}

The present status of the helicity angle analyses for the higher mass
vector mesons suffers from more stringent statistical constraints. 
The ZEUS collaboration has measured matrix elements
for $\phi$ production both in photopro\-duction~\cite{pl_377_259}:
\begin{eqnarray*}
r_{00}^{04} &=& -0.01 \pm 0.04 \\
r_{1-1}^{04} &=& +0.03 \pm 0.05 
\end{eqnarray*}
and for $\langle {\qsq} \rangle = 12.3\;{\gevsq}$~\cite{pl_380_220}:
\begin{eqnarray*}
r_{00}^{04} &=& 0.76^{+0.11}_{-0.16} \;(stat) \pm 0.12 \;(sys).
\end{eqnarray*}
The photopro\-duction 
results are consistent with an underlying diffractive process
and the results at both the low and high {\qsq} are consistent with
the expectations based on SCHC. Similar conclusions
were drawn from the $\omega$ photopro\-duction study~\cite{zfp_73_73}, 
where the following values for the matrix elements were determined:
\begin{eqnarray*}
r_{00}^{04} &=& +0.11 \pm 0.08 \;(stat) \pm 0.26 \;(sys)\\
r_{1-1}^{04} &=& -0.04 \pm 0.08 \;(stat) \pm 0.12 \;(sys).
\end{eqnarray*}

The H1 collaboration has published an 
helicity analysis for the polar angle in {\jpsi} 
photopro\-duction~\cite{np_472_3} and find the  dependence
\begin{eqnarray}
W(\cos{\thetah}) &\propto& 1+\cos^2{\thetah}
\end{eqnarray}
consistent with the expectations of SCHC for the decay to 
fermions~\cite{pl_65_463,*pr_15_2503}.

We conclude this section with a compilation of results for 
$R = \sigma_L/\sigma_T$ in the studies at high {\qsq} where the predominance
of the longitudinal part is an essential prediction of the perturbative
models. Figure~\ref{fig:rhorvsq2} shows the results obtained by the HERA
experiments for {\rhoz} production and compares them to the result from
the NMC experiment in fixed--target muon beam studies. While these results
clearly show values for $R$ inconsistent with the small value measured in
photopro\-duction both at low energies and at HERA, conclusive results on
the {\qsq}--dependence of $R$ await studies availed of higher statistical
precision.
\begin{figure}[htbp]
\begin{center}
\includegraphics[width=0.7\linewidth]{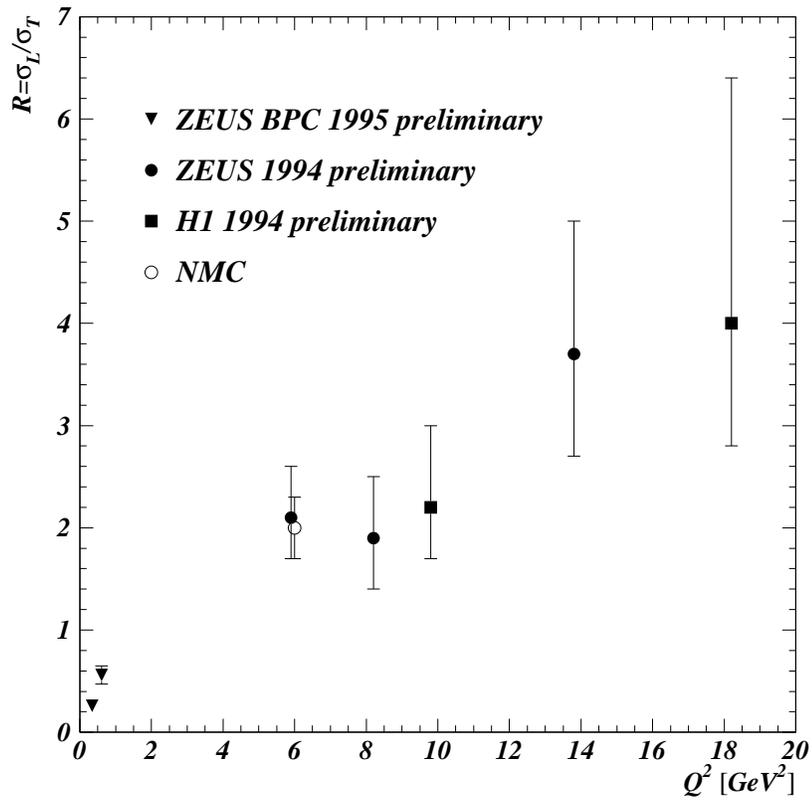}
\end{center}
\caption
{
\label{fig:rhorvsq2}
\it 
The {\qsq}--dependence of the ratio of longitudinal to transverse
elastic {\rhoz} production cross sections from the HERA experiments
compared to the fixed--target muon beam result from
the NMC experiment~\protect\cite{np_429_503}. 
The error bars represent the statistical errors alone.
} 
\end{figure}

\vspace*{5mm}
The near future is very bright for the continuation of these helicity
analyses in elastic vector--meson production at HERA. Soon more results
from the high statistics {\rhoz} photopro\-duction sample from the 1994
and 1995 data--taking periods will become available, which should provide 
information on the $t$--dependence of the matrix elements, a definitive
indicator for the applicability of SCHC. For the high--{\qsq} {\rhoz}
investigations, samples with an order of magnitude greater statistics
are now under study. By the end of 1997 the studies in the transition region of
\mbox{0.2 $< {\qsq} <$ 0.9~{\gevsq}} will also have nearly an order--of--magnitude
increase in the number of {\rhoz} decays available. By that time
precise measurements for the {\jpsi} should also be possible and the value
of $R$ so crucial to the theoretical understanding of these processes will
be accessible with better precision over a greater kinematic region and
for a wider variety of vector mesons.
\subsection{Vector--Meson Production with Proton Dissociation}

The H1 collaboration has isolated a {\jpsi} data set with proton 
dissociation~\cite{np_472_3} in photopro\-duction
as described in section~\ref{sec:expvm}. 
The authors report a magnitude for the dissociative cross section
comparable to that for the elastic cross section, and a
significantly steeper energy dependence in
the dissociative sample. The dissociative sample exhibited
a much flatter $t$--dependence, as shown
in Fig.~\ref{fig:h1psipdiss}. The result of the exponential fit to the
$p_T^2$--slope for the elastic sample was 
\begin{eqnarray*}
b &=& 4.0 \pm 0.2 \; (stat) \pm 0.2 \; (sys) \; {\gev}^{-2}
\end{eqnarray*}
while for the dissociative sample a slope of 
\begin{eqnarray*}
b &=& 1.6 \pm 0.3 \; (stat) \pm 0.1 \; (sys) \; {\gev}^{-2}
\end{eqnarray*}
was found. An additional uncertainty due to the approximation of $t$ with
$p_T^2$ was estimated to be 7\%.
\begin{figure}[htbp]
\begin{center}
\includegraphics[width=0.8\linewidth,bb=46 525 535 767,clip=]{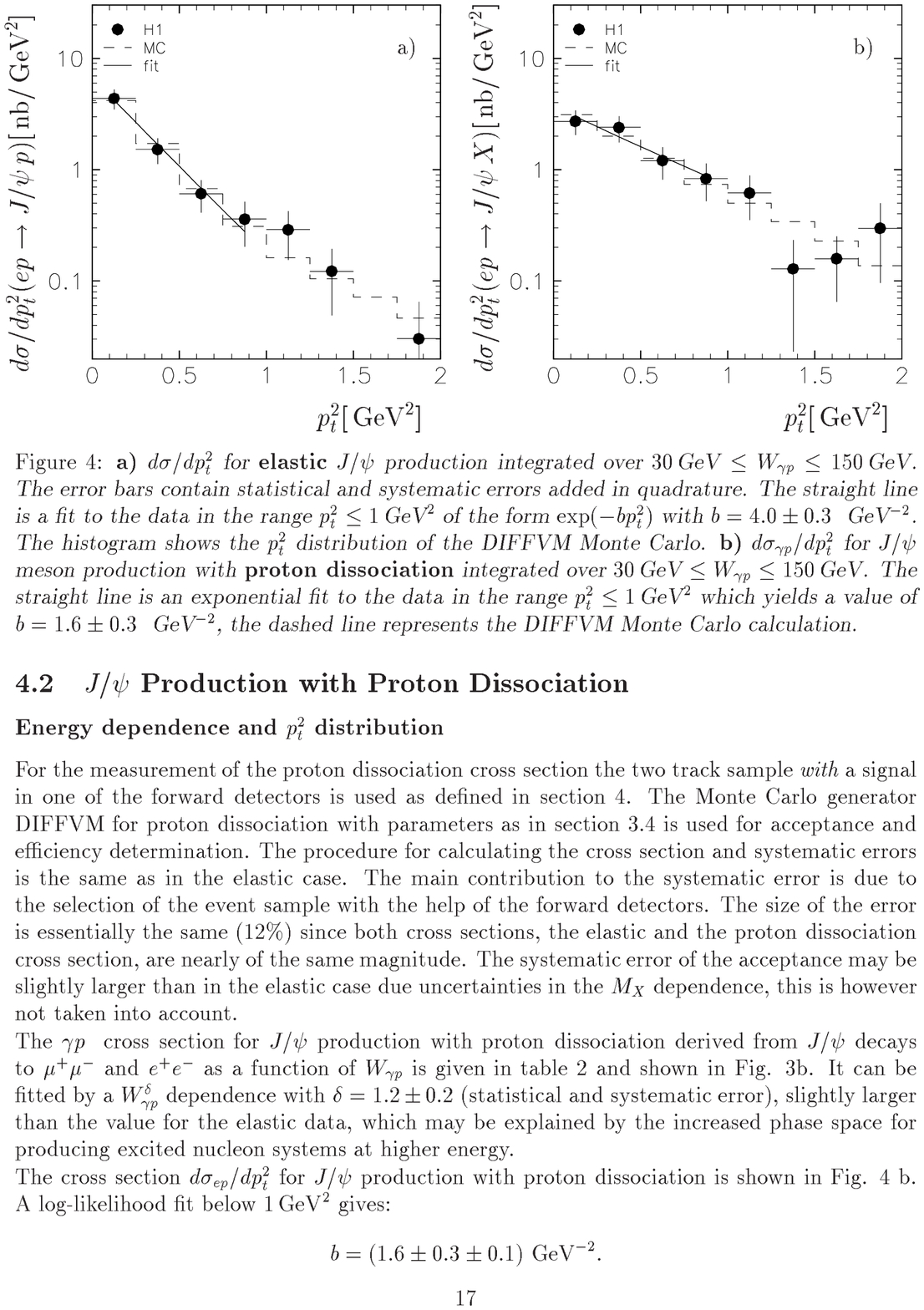}
\end{center}
\caption
{
\label{fig:h1psipdiss}
\it 
The $t$--dependence of the cross section for
elastic {\jpsi} photoproduction (a)~compared to that for 
proton dissociative {\jpsi} production (b)~from 
ref.~\protect\cite{np_472_3}. The error bars show the systematic and
statistical uncertainties added in quadrature. The solid lines are
the fits assuming an exponential dependence as described in the text.
The dashed lines indicate the results of  simulations.} 
\end{figure}

The H1 collaboration has also
presented results from a study of {\rhoz} production with proton dissociation
in the {\qsq} range \mbox{7 $ < {\qsq} < $ 36~{\gevsq}}
for pho\-ton--pro\-ton cen\-ter--of--mass en\-er\-gies 
\mbox{60 $ < W_{\gamma^* p} < $ 180~{\gev}}~\cite{pa02_65}. 
The data sample was restricted to transverse momenta less than 0.8~{\gevsq}.
They found the ra\-tio
of the dis\-soc\-ia\-tive
cross sec\-tion to that for e\-las\-tic pro\-duc\-tion to be 
\begin{eqnarray*}
\mbox{}&\mbox{}&0.63\; \pm \;0.11\;(stat)\; \pm\; 0.11\;(sys). 
\end{eqnarray*}
The $t$--slope was measured to be 
\begin{eqnarray*}
b &=& 2.0 \pm 0.5 \; (stat) \pm 0.5 \; (sys) \; {\gev}^{-2}.
\end{eqnarray*}
This value is comparable to the value for the $t$--slope found in {\jpsi}
photopro\-duction with proton dissociation and significantly smaller than
the value determined in elastic {\rhoz} production at high {\qsq}. Such small
values for the $t$--slope ($< 4 \; {\gev}^{-2}$) indicate that the
interaction investigated has a substantial contribution from processes which
resolve the proton.

A fit to the {\qsq}--dependence $Q^{-2a}$ measured in diffractive
{\rhoz} production accompanied by proton dissociation yielded the result
\begin{eqnarray*}
a &=& 2.8 \pm 0.6 \; (stat) \pm 0.3 \; (sys)
\end{eqnarray*}
similar to that for elastic production. An helicity analysis determined
a value for the matrix element $r_{00}^{04}$:
\begin{eqnarray*}
r_{00}^{04} &=& 0.78 \pm 0.10 \;(stat) \pm 0.05 \;(sys),
\end{eqnarray*}
consistent with the $\cos{\thetah}$ dependence found in the case of
elastic production. The latter results encourage the conclusion that
the {\rhoz} production process and the proton dissociative process
with little momentum transferred to the proton
are weakly correlated, if at all.

The ZEUS collaboration has employed two methods to identify  samples of
{\rhoz} photoproduction events in which the proton dissociates using the
data recorded in 1994~\cite{pa02_50}. 
The first relied on the observation of energy
depositions in forward scintillation counters (PRT)
while the second was based on detection of a proton in the LPS 
carrying less than 98\% of the beam momentum, inconsistent
with an elastic scatter. Figure~\ref{fig:pdisstslope} shows the $p_T^2$--distributions corresponding to these two independent data samples, as well as
the result of an exponential fit to the scintillation--counter--based sample
which determined the $t$--slope to be
\begin{eqnarray*}
b &=& 5.3 \pm 0.3 \; (stat) \pm 0.7 \; (sys) \; {\gev}^{-2}.
\end{eqnarray*}
The sample identified by the direct detection of a forward proton 
exhibited a slope of
\begin{eqnarray*}
b &=& 5.3 \pm 0.8 \; (stat) \pm 1.1 \; (sys) \; {\gev}^{-2}.
\end{eqnarray*}
These slopes are each consistent with a value approximately half that
observed for {\rhoz} photoproduction without proton dissociation,
similar to the slopes measured in {\jpsi} photoproduction and in
{\rhoz} production at high {\qsq}. 
\begin{figure}[htbp]
\begin{center}
\includegraphics[width=0.6\linewidth,bb=98 268 480 674,clip=]{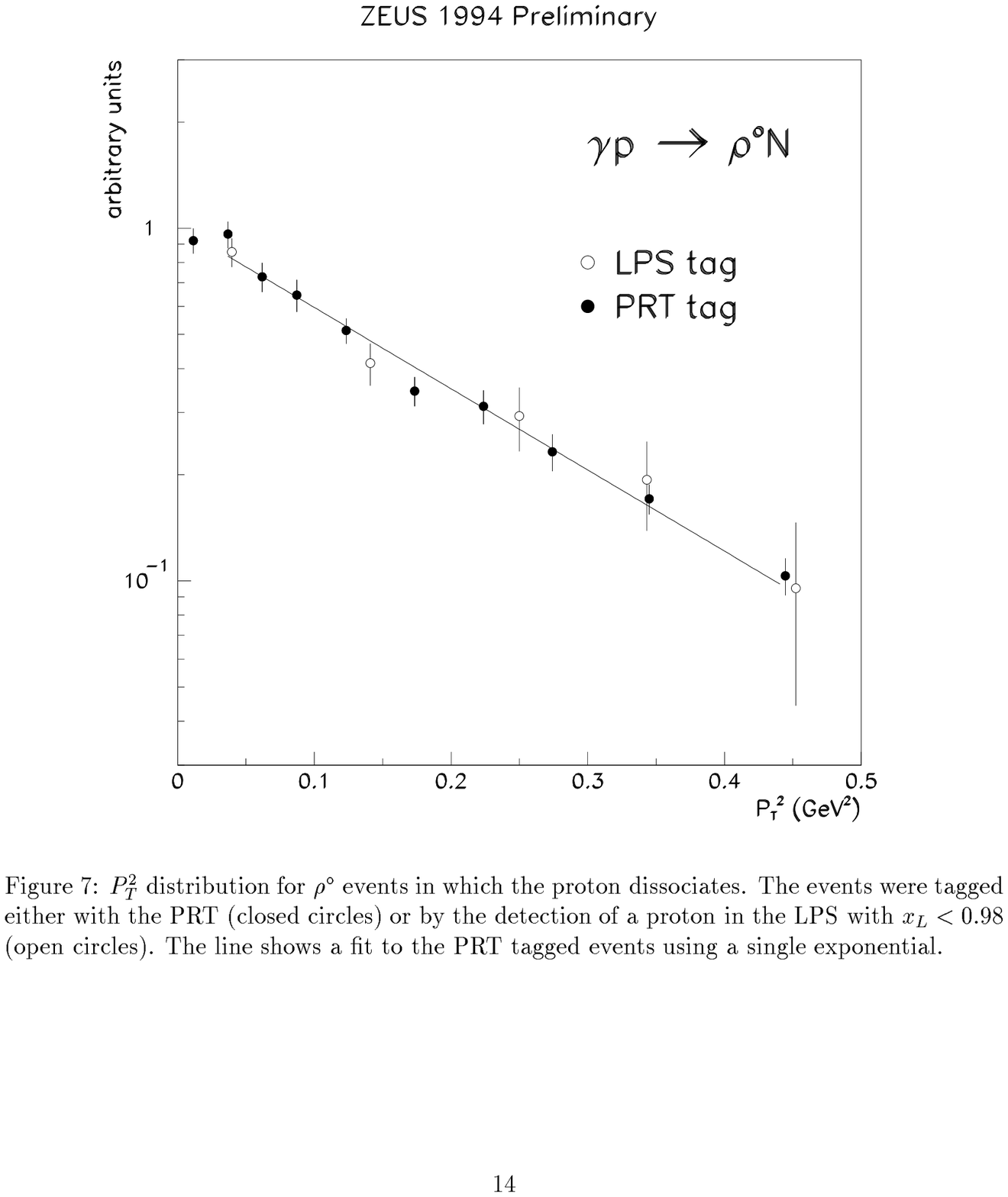}
\end{center}
\caption
{
\label{fig:pdisstslope}
\it 
The $t$--dependence of the cross section for
{\rhoz} photoproduction accompanied by dissociation of the proton
as measured by the ZEUS collaboration with data recorded in 
1994~\protect\cite{pa02_50}. The closed circles indicate a data sample
based on the observation of depositions in forward scintillation
counters (PRT). The open circles represent a sample in which a proton
was detected in the Leading Proton Spectrometer with momentum inconsistent
with that of an elastic scatter. 
The line shows the results of an exponential fit
to the PRT--selected data.
}
\end{figure}

A preliminary helicity analysis in {\rhoz} photopro\-duction
at high $|t|$ (\mbox{0.5 $< |t| < 4.0$ {\gevsq}}) has been presented by the 
ZEUS collaboration~\cite{pa02_51}. Given
such high momentum transfer at the proton vertex, the dominant contribution
is expected from simulation studies to
be from proton--dissociative events. 
The identification of the high--$|t|$ event sample was made possible 
by a small--angle electron tagger installed 44~m from the interaction point
in the positron flight direction prior to the 1995 running period. 
This photon--tagging facility limited the photon virtuality to
\mbox{${\qsq} < 10^{-2}\;{\gevsq}$}, 
rendering its contribution to $t$ negligible
compared to that of the transverse momentum of the {\rhoz} measured in
the central detector. The polar decay angle distributions obtained
indicated that the {\rhoz} mesons are predominantly transversely polarized.
The analysis averaged over the angle {\Phih}, eliminating any contribution
to the azimuthal angular dependence from the linear photon polarization
(eq.~\ref{eq:threedang}).
A significant azimuthal angular dependence was observed, however,
forcing the preliminary conclusion of a nonzero value for $r_{1-1}^{04}$ and
pointing to a possible breakdown of $s$--channel helicity conservation 
in the case of 
high momentum transfer at the proton vertex.

\cleardoublepage
\section{Future Developments}
\setcounter{equation}{0}
\label{sec:future}
An appreciation for the richness of the future HERA research program is
given by a perusal of the proceedings of the HERA workshop of 
1995/96~\cite{futurephysics}. The contents make it clear that the
investigations published until now represent the tip of an iceberg.
The purpose of this section is to give the reader a quantitative idea of which
topics in the subfields of total photon--proton cross sections and exclusive
vector--meson production will be extended and of those which will become newly
accessible.
\subsection{Luminosity Schedule}
Aside from a few specialized investigations which have been brought
to conclusion, the data recorded during the
1995 and 1996 running periods have been under active analysis
since they were obtained. Thus a fairly mature
knowledge of the systematic issues associated with these
data has been obtained within the ZEUS and H1 collaborations. Conclusive
results with the integrated luminosity of ${\approx}$35~pb$^{-1}$, compared to
those covered in this article based on a luminosity of 7~pb$^{-1}$, are now
being published. 

Prior to the 1996/97 winter shutdown in HERA operation an agreement between
the HERA collaborations and the DESY 
directorate to continue operation in two-year cycles
was reached. This decision was motivated in part by the experience in past
years of impressive improvement in accelerator operation during the
running periods due to an increased familiarity with tuning algorithms and
hardware reliability, which became obsolete during the shutdowns. The decision
was also influenced by the scheduling of two extensive improvement projects
scheduled to take place in the coming years: an upgrade of the electron
ring vacuum system, which will allow the reintroduction of electrons with
little loss in beam lifetime, and an ambitious luminosity upgrade system
involving the installation of quadrupole magnets near the interaction
region within the H1 and ZEUS detectors. The upgrade of the electron ring
will occur during the 1997/98 shutdown and can be expected 
to yield an increase within a 
single year of more than an order of magnitude over the 2~pb$^{-1}$ 
of electron--proton data presently available. 

A design study for a luminosity 
upgrade~\cite{lumiupgrade} concluded that the present HERA design is
limited to an integrated luminosity of 40--50~pb$^{-1}$ per year, and
proposed new optics, including low--$\beta$ quadrupole magnets in the
detectors, which result in an increase in instantaneous luminosity
by more than a factor of four. Integrated luminosities of 100~pb$^{-1}$ in
the first year of operation after the upgrade and 1~fb$^{-1}$
by the end of the year 2005 appear feasible according to the results of
this study. The installation of quadrupole magnets near the interaction regions
of the H1 and ZEUS experiments, which will take place during the 1999/2000
shutdown, entails severe consequences for the operation of small--angle
detectors. Remarkable among these is the necessary increase in the transverse
momentum spread in the the proton beam, which will increase by nearly a factor
of two. The emittance of the
beam is furthermore a function of the time since injection, and will therefore
introduce severe systematic uncertainties in the measurement of 
$t$--distributions via direct observation of the scattered proton
in investigations of diffractive processes. 
This luminosity upgrade is, however, of decisive importance for the
statistically--limited studies at high {\qsq} and for those of
the photopro\-duction of heavy vector mesons.
\subsection{Beam Energies}
The opportunity to explore various kinematic regions by varying the HERA
beam energies has not yet been exploited, 
aside from the minor upgrade of the electron beam 
energy from 26.7 to 27.5~{\gev} in 1994.
The most frequently discussed motivation for varying the beam momenta has
been the systematically difficult measurement of the longitudinal proton
structure function {\fl}(\xbj,\qsq). 
Tentative plans to take data with a yet--unspecified
lower proton beam energy have been made for the 1997 running period. Further
impetus to this decision has been provided by the possibility of extending
the energy range covered for the total photopro\-duction cross section
measurement using the new electron--tagging facilities installed in the
ZEUS experiment prior to the 1995 running period. With the data recorded
in 1995 and 1996 at the nominal
proton energy of 820 {\gev} it is expected that the energy range 
\mbox{80 $< W_{{\gamma^* p}} <$ 300~{\gev}} can be covered and with appropriately
lower proton energies in 1997 the range 
\mbox{55 $< W_{{\gamma^* p}} <$ 210~{\gev}} 
would become accessible, increasing
sensitivity to the value of ${\alpha}'_{\Pma}$ in the phenomenological Pomeron
trajectory.

The primary considerations providing upper bounds on the HERA beam
energies are those of consequence for the reliability of HERA operation,
an essential contributor to limits on the integrated luminosity over the
first five years. To increase the lepton beam energy from 27.5~{\gev} to
the design value of 30~{\gev}, improvements in the reliability of the
superconducting RF cavities would be 
necessary. All proton beam superconducting
dipoles have been tested at currents corresponding to a proton beam
energy exceeding 1~TeV. However, no information concerning their reliability
under long--term operation at such currents is available.
In the absence of physics arguments which are convincing even under 
conditions of reduced machine reliability, no consensus on upgrading
HERA to higher energies has been reached.
\subsection{Detector Upgrades}
The ZEUS electron tagger~\cite{pa02_51} is an example of several ZEUS and
H1 detector
upgrades for which data is already available and from which results can be
expected during the coming year. Of note is the scintillating--fiber forward
proton detector of H1, which will provide clean identification of 
elastic proton interactions, as has the ZEUS leading proton spectrometer.
Prior to the 1998 running period the H1 collaboration will install
a small calorimeter in the rear direction to perform measurements in a 
{\qsq}--range comparable to that covered by the ZEUS BPC. 
\subsection{Beam Polarization}
At present it is planned to install electron spin rotators in the electron
beam line during the 1997/98 shutdown, permitting the longitudinal polarization
of the lepton beams. Extensive studies of polarization algorithms during
HERA operation since 1993 have resulted in the routine achievement of transverse
polarizations in excess of 50\%. While transverse polarization has negligible
influence on the decay angular distributions used for the helicity analyses
in vector--meson production, longitudinal polarization plays an important
r\^ole in that it represents the only way to distinguish the components
of the spin density matrix corresponding to 
circular photon polarization~\cite{np_61_381}.

A discussion of the possibility of running polarized proton beams at HERA
can be found in ref.~\cite{polprot}, but no definitive 
decision concerning such a project
has been made.
\subsection{Nuclear Beams}
Simple model--independent considerations easily show why there has 
been such great theoretical interest in the introduction of 
nuclear beams at HERA. (For a compilation, see ref.~\cite{heranuc}.)
Our calculations of virtual photon lifetime in section~\ref{sec:gammapxsect}
showed that the bulk of the HERA data cover a kinematic region where the
photon lifetime exceeds the interaction time associated with the size of
the proton in its rest frame. 
The interaction time 
increases with the extent of the hadronic target and so for nuclear
beams the threshold region moves toward the kinematic region where HERA
provides greater statistical power. The result is that the
nuclear target size
represents a further parameter which can be tuned to scan the space--time
structure of deep--inelastic scattering, along with the photon virtuality
and the mass of diffractively--produced vector mesons. Striking
effects are predicted in the {\qsq}--dependence of diffractive production
of radially--excited vector mesons off 
nuclei~\cite{hep96_05_208}. Further
arguments are based on the precepts of QCD, which result in the
expectation of strong nuclear
enhancements of non-linear effects at 
low {\xbj}~\cite{pr_50_3134,pr_54_3194,pl_309_179,*pl_382_6,*pl_324_469,*pl_383_362}. Extensive 
treatment of the topic of the electroproduction of vector mesons off nuclei
may be found in the review of Bauer {\em et al.}~\cite{rmp_50_261}, who
consider such investigations important as tests of vector--dominance models.

The investigations by the Light and Heavy Nuclei Working Group in
the Future Physics at HERA workshop concluded that substantial investment
in the development of an ion source would be necessary to introduce nuclear
beams. Despite strong interest in a deuteron beam from other working
groups and the apparent feasibility of nuclear beams of nuclear weights
as high as that of sulfur, this project remains only at the discussion stage. 
\cleardoublepage
\section{Summary}
\setcounter{equation}{0}
\label{sec:summary}
Results obtained by the ZEUS and H1 col\-la\-bo\-ra\-tions from 
in\-ves\-tiga\-tions
of high--energy elec\-tron--pro\-ton interactions during the first five years
of operation of the HERA  collider have demonstrated the 
versatility of this experimental technique. These studies
have yielded quantititative
insight into questions of proton structure, electroweak cross sections
at the unification scale, the hypothesized 
existence of particles and interactions
unforeseen in the Standard Model, 
diffractive processes at high energy and the hadronic interactions
of photons both virtual and real. This article began with an historical
perspective on the decision to build an electron--proton collider, and
described the operation of the HERA accelerator complex during its first years
of operation. Following a description of the 
experimental apparatus employed by the ZEUS and H1 
collaborations, a general overview of the physics topics
addressed in the investigations of photon--proton interactions at HERA
was given. 
Our introduction concluded by placing the recent studies of 
diffractive vector--meson production at HERA
in the rich historical context of past and present 
experimental investigations of vector--meson lepto-- and photoproduction.

\vspace*{2mm}
The recently obtained wealth of new information has motivated 
a remarkable 
variety of new theoretical calculations. These span a wide range
of physical concepts, requiring tools and ideas developed separately
for the description of soft diffractive hadronic processes and for that of 
hard processes subject to the application of perturbative
calculational techniques. In section~\ref{sec:theory}
we presented a discussion of general theoretical considerations 
followed by synopses of topical models in this field. 
 
\vspace*{2mm}
Section~\ref{sec:experiment} addressed
the studies at HERA of the interactions of photons with
protons in a range of photon virtuality {\qsq} between 10$^{-10}$ and
10$^3\;{\gevsq}$, and of the Bjorken scaling variable 
{\xbj} between 10$^{-5}$ and 10$^{-1}$.
After showing that in the low--{\xbj} kinematic region covered by
the HERA data the virtual
photon lifetime far exceeds the interaction time, we 
defined the total virtual--photon--proton
cross section and
presented measurements of it 
for energies of the photon in the proton rest frame
between 400 and 40000~{\gev}.  We concluded that a region of
transition from energy dependence characteristic of 
phenomenological Regge--theory--based models
to that described by QCD evolution lies in the region of photon
virtuality between 0.1 and 3~{\gevsq}. These preliminary studies
will be extended in order to further investigate the underlying dynamics
governing the transition region.

\vspace*{2mm}
This brief summary of issues 
concerning the total inclusive virtual--photon--proton cross section
served to introduce our main topic, the exclusive production of
single neutral vector mesons in photon--proton interactions at high energy.
A comprehensive discussion of results published by the H1 and ZEUS
collaborations prior to January 1, 1997 was presented. The experimental
issues of technique, re\-solution, efficiencies, and backgrounds were
covered for the measurements of {\rhoz}, {$\omega$}, {$\phi$}, and {\jpsi}
mesons, and for the initial observations of {$\rho'$} and {${\psi}(2S)$} mesons.
The energy dependence of the elastic cross sections at high photon
virtuality was shown to be steep, similar to that observed for the inclusive
cross section. In photopro\-duction the energy dependence was found to
be much weaker,
comparable to that for the total photopro\-duction cross section, excepting 
that for {\jpsi} production. These observations are consistent with
calculations performed in the context of perturbative QCD,
in which the necessary hard scale is given either by the photon virtuality
or by the vector--meson mass. We continued our description of results
by considering the {\qsq}--dependence of the cross section, which was seen
to be consistent with a dependence of the form
$({\qsq}+M_V^2)^{-a}$, with \mbox{$2< a < 3$}, 
but for which the experimental
uncertainties are at a level which does not clearly distinguish the exponent
for the various vector mesons. There are, however, significant
indications that the
{\qsq}--dependence is weaker than $Q^{-6}$, in accordance with
expectations based on
perturbative QCD calculations. This comparison is mitigated by the ambiguity
introduced due to the unseparated
contributions in the measured cross sections from
transversely and longitudinally polarized photons. 

\vspace*{2mm}
The
exponential slopes of the
differential cross sections as functions of the momentum transfer
at the proton vertex indicated that the interaction sizes for
the {$\phi$} and {\jpsi} mesons are distinctly smaller than those for
the {\rhoz} and $\omega$ mesons 
in photopro\-duction. There are also preliminary indications
that the interaction size for elastic {\rhoz} production decreases
with {\qsq}. The measurements cover a {\qsq}--range in which the
exponential slope approaches a value consistent with that expected
from the proton alone. 

\vspace*{2mm}
Comparisons of the production rates for $\phi$ and {\rhoz} mesons
for photon virtuality exceeding 5 {\gevsq} have shown that
the ratio is consistent
with the value of 2/9 expected from a simple quark--counting rule
and a flavor--independent production mechanism, contrasting the 
relative suppression of approximately a factor of three in $\phi$ 
photoproduction. Even more impressive are the results for
{\jpsi} production at high {\qsq}, which show that the suppression
by two orders of magnitude relative to the {\rhoz} as 
measured in photoproduction is reduced to a mere factor of 
about three.

\vspace*{2mm}
The helicity 
analyses of the vector--meson decay angle distributions were considered
in detail,
culminating in a preliminary determination of the ratio of longitudinal
to transverse photon cross sections as a function of {\qsq} in exclusive
{\rhoz} production under the assumption of $s$--channel helicity conservation. 
This ratio rises from a value consistent with
zero at the level of a few percent in photopro\-duction to a value
exceeding unity for {\qsq} greater than 5~{\gevsq}. This {\qsq}--dependence
is particularly sensitive to the mixture of point--like and nonperturbative
processes in the vector--meson production mechanism.

\vspace*{2mm}
Our consideration of experimental results concluded
with a  brief
description of measurements of vector--meson production in processes where
dissociation of the proton is observed. Such data sets have been identified
in {\rhoz} and {\jpsi} photoproduction and in {\rhoz} production
for ${\qsq} > 7~\gevsq$. A much weaker dependence on
the momentum transfer at the proton vertex $|t|$ characterizes these processes,
the $t$--slope measured to be only about half that of the purely elastic
interaction.  Helicity analyses applied to the high {\qsq} {\rhoz} production
at low $|t|$ showed characteristics similar to those observed for
event samples without proton dissociation, indicating that the {\rhoz}
production mechanism and the dissociation of the proton proceed independently
in this low $|t|$ region. However, a preliminary helicity analysis applied to
an event sample which was restricted to high $|t|$ ($|t| > 0.5~\gevsq$) 
showed evidence for the violation of $s$--channel helicity conservation.

\vspace*{2mm}
The H1 and ZEUS collaborations are presently completing analyses of
data sets obtained from a combined integrated luminosity approximately
a factor of five greater than that covered by the analyses described in
this article. Furthermore, a
luminosity upgrade, as well as upgraded apparatus and polarization of
electron and positron beams, is planned for the coming years. The 
introduction of nuclear beams is under consideration.

\section{Concluding Remarks}
\setcounter{equation}{0}
\label{sec:conclusions}
The introduction of  nonabelian field theory in the 1970s 
as a description of the strong interaction
has proven immensely successful 
at transverse momentum scales exceeding the confinement scale.
Soft hadronic processes are well described by phenomenological
fits inspired by Regge theory. But there has been little progress
in developing a theoretical picture which offers a unified understanding
of the relationship between these two distinct dynamical descriptions.
The early investigations of high--energy electron--proton interactions
at HERA described in this article yield new information on the transition
region between the hard and soft hadronic interactions of the photon. 
The high virtual photon energies at HERA provide a clean
environment for the study of diffractive processes. An example of
such interactions is the elastic production
of vector mesons, which has been shown to provide a further handle on 
the transition between hard and soft processes, as the photopro\-duction
of {\jpsi} mesons exhibits an energy dependence characteristic of 
hard processes, while that of the lighter vector mesons 
has been shown to obey scaling laws typical of soft interactions. 
Variation
of photon virtuality and vector--meson mass has  been used to
scan the space--time structure of the photon--proton interaction,
yielding information on the wave function of the virtual photon and on that
of the vector meson. 
Measurements of vector--meson production ratios 
have shown that flavor asymmetries 
decrease with increasing photon virtuality.
Such revelations, together with the future schedule of HERA,
justify confidence in the potential of
these investigations to further contribute to our understanding of the
strong interaction.
\cleardoublepage
\section*{Acknowledgements}
The preparation of this article was immeasurably assisted by many
pleasant discussions with my colleagues at DESY.
The final version bears the influence of constructive criticism from
M.~Arneodo, S.~Bhadra, E.~Gallo, U.~Katz, R.~Klanner, P.~Marage,
A.S.~Proskuryakov, 
and A.~Quadt.
The difficult task of coordinating the efforts in the ZEUS diffractive
physics working group was accomplished by M.~Arneodo, S.~Bhadra, E.~Gallo,
and G.~Iacobucci. 
I extend thanks to T.~Doeker, T.~Monteiro, and Q.~Zhu for
explanations concerning the data from the ZEUS beam--pipe calorimeter.
I found my discussions of the helicity analyses with H.~Beier,
K.~Piotrzkowski, A.S.~Proskuryakov, D.~Westphal,
and G.~Wolf particularly enlightening.
\mbox{A.~Levy} was of great assistance on the subject of photon--proton
total cross sections. A.~Quadt provided Fig.~\ref{fig:xq2plane}.
Discussions with B.~Burow on the vagaries
of virtual photon flux definitions served as a calming influence.

I am further grateful to B.~Kopeliovich and L.~Frankfurt
for answering questions pertaining
to the description of vector--meson production in the framework
of perturbative QCD calculations.

The research reported in this article was made possible by
the heroic efforts of the HERA
accelerator physicists and the constant support provided to 
the HERA experiments by the DESY directorate. 

This work was performed in the context of the research group
led by Prof.~Erwin Hilger in the 
Physics Institute of the University of Bonn.
I would like to acknowledge his unflagging
encouragement and helpful commentary on several drafts of the article.

The research activities of the ZEUS group in Bonn are supported by
the German Federal Ministry for Education and Science, 
Research and Technology (BMBF).

\cleardoublepage
\bibliographystyle{zeusstylem}
\bibliography{zeuspubs,h1pubs,schriftrefs}

\begin{mcbibliography}{100}

\bibitem{epseminar73}
{\em Proceedings of the Seminar on e$^-$p and e$^-$e$^+$ Storage Rings}, edited
  by J.K.~Bienlein, I.~Dammann, H.~Wiedemann  (DESY, Hamburg, Germany,
  1973)\relax
\relax
\bibitem{dr_72_22}
H. Gerke {\em et al.},
\newblock  DESY Report {\bf H--72--22}  (1972)\relax
\relax
\bibitem{cheer}
S. Conetti {\em et al.}, in {\em Proceedings of the 11th Int. Conf. on High
  Energy Accelerators}, edited by W.S. Newman  (Birkh\"auser, Geneva,
  Switzerland, 1980), p. 189\relax
\relax
\bibitem{ecfahera}
{\em Proceedings of the Study of an ep--Facility for Europe}, edited by U.
  Amaldi  (DESY, Hamburg, Germany, 1979)\relax
\relax
\bibitem{heraprop}
DESY Report {\bf H--81/10}  (1981)\relax
\relax
\bibitem{hera1}
B.H. Wiik, in {\em Proceedings of the 11th Int. Conf. on High Energy
  Accelerators}, edited by W.S. Newman  (Birkh\"auser, Geneva, Switzerland,
  1980), p. 120\relax
\relax
\bibitem{hera2}
G.--A. Voss, in {\em Proceedings of the 12th Int. Conf. on High Energy
  Accelerators}, edited by F.T. Cole and R. Donaldson  (FNAL, Batavia,
  Illinois, 1983), p.~29\relax
\relax
\bibitem{arnps_44_413}
G.--A. Voss and B.H. Wiik,
\newblock  Ann. Rev. Nucl. Part. Sci. {\bf 44}  (1994) 413\relax
\relax
\bibitem{ijmp_2a_28}
F. Willeke,
\newblock  Int. J. Mod. Phys. {\bf Proc.Suppl. 2A}  (1993) 28\relax
\relax
\bibitem{dr_95_08a}
R. Brinkmann,
\newblock  DESY Report {\bf M--95--08A}  (1995)\relax
\relax
\bibitem{dr_96_001}
The H1 Collaboration,
\newblock  DESY Report {\bf 96--001}  (1996)\relax
\relax
\bibitem{nim_279_217}
J. Buerger {\em et al.},
\newblock  Nucl. Instrum. Methods {\bf A279}  (1989) 217\relax
\relax
\bibitem{nim_336_460}
The H1 Calorimeter Group,
\newblock  Nucl. Instrum. Methods {\bf A336}  (1993) 460\relax
\relax
\bibitem{zeusdetector}
The ZEUS Collaboration,
\newblock  The ZEUS Detector,
\newblock  Status Report, 1993\relax
\relax
\bibitem{nim_265_200}
R. Klanner,
\newblock  Nucl. Instrum. Methods {\bf A 265}  (1988) 200\relax
\relax
\bibitem{nim_289_115}
U. Behrens {\em et al.},
\newblock  Nucl. Instrum. Methods {\bf A 289}  (1990) 115\relax
\relax
\bibitem{nim_290_95}
A. Andresen {\em et al.},
\newblock  Nucl. Instrum. Methods {\bf A 290}  (1990) 95\relax
\relax
\bibitem{nim_336_23}
A. Bernstein {\em et al.},
\newblock  Nucl. Instrum. Methods {\bf A 336}  (1993) 23\relax
\relax
\bibitem{nim_238_489}
J.E. Brau and T.E. Gabriel,
\newblock  Nucl. Instrum. Methods {\bf A 238}  (1985) 489\relax
\relax
\bibitem{nim_259_389}
R. Wigmans,
\newblock  Nucl. Instrum. Methods {\bf A 259}  (1987) 389\relax
\relax
\bibitem{nim_263_136}
H. Br\"uckmann {\em et al.},
\newblock  Nucl. Instrum. Methods {\bf A 263}  (1988) 136\relax
\relax
\bibitem{nim_309_77}
M. Derrick {\em et al.},
\newblock  Nucl. Instrum. Methods {\bf A 309}  (1991) 77\relax
\relax
\bibitem{nim_309_101}
A. Andresen {\em et al.},
\newblock  Nucl. Instrum. Methods {\bf A 309}  (1991) 101\relax
\relax
\bibitem{vcalhep}
J.A. Crittenden, in {\em Proceedings of the Fifth International Conference on
  Calorimetery in High Energy Physics, Brookhaven National Laboratory}, edited
  by H.A. Gordon and D. Rueger  (World Scientific, 1994), p.~58\relax
\relax
\bibitem{nim_279_290}
N. Harnew {\em et al.},
\newblock  Nucl. Instrum. Methods {\bf A 279}  (1989) 290\relax
\relax
\bibitem{nim_283_473}
C.B. Brooks {\em et al.},
\newblock  Nucl. Instrum. Methods {\bf A 283}  (1989) 473\relax
\relax
\bibitem{nim_338_181}
B. Foster {\em et al.},
\newblock  Nucl. Instrum. Methods {\bf A 338}  (1993) 181\relax
\relax
\bibitem{nim_382_419}
A. Bamberger {\em et al.},
\newblock  Nucl. Instrum. Methods {\bf A 382}  (1996) 419\relax
\relax
\bibitem{nim_333_342}
G. Abbiendi {\em et al.},
\newblock  Nucl. Instrum. Methods {\bf A 333}  (1993) 342\relax
\relax
\bibitem{nim_321_356}
A. Caldwell {\em et al.},
\newblock  Nucl. Instrum. Methods {\bf A 321}  (1992) 356\relax
\relax
\bibitem{nim_277_217}
W. Buttler {\em et al.},
\newblock  Nucl. Instrum. Methods {\bf A 277}  (1989) 217\relax
\relax
\bibitem{dr_92_130}
A. Caldwell {\em et al.},
\newblock  DESY Report {\bf 92--130}  (1992)\relax
\relax
\bibitem{hervasthesis}
L. Hervas,
\newblock  {\em The Readout for the ZEUS Calorimeter},
\newblock  Ph.D. thesis, Universidad Aut\'onoma de Madrid, 1991,
\newblock  DESY F35D--91--01\relax
\relax
\bibitem{ieee_36_465}
W. Sippach {\em et al.},
\newblock  IEEE Trans. {\bf NS 36}  (1989) 465\relax
\relax
\bibitem{nim_355_278}
W.H. Smith {\em et al.},
\newblock  Nucl. Instrum. Methods {\bf A 355}  (1995) 278\relax
\relax
\bibitem{dr_96_183}
The ZEUS Collaboration,
\newblock  DESY Report {\bf 96--183}  (1996),
\newblock  accepted by Z.Phys. MS 489\relax
\relax
\bibitem{sacchithesis}
R. Sacchi,
\newblock  {\em Studio di eventi diffrattivi con uno spettrometro per protoni a
  ZEUS},
\newblock  Ph.D. thesis, University of Torino, 1996\relax
\relax
\bibitem{prl_23_935}
M. Breidenbach {\em et al.},
\newblock  Phys. Rev. Lett. {\bf 23}  (1969) 935\relax
\relax
\bibitem{pr_179_1547}
J.D. Bjorken,
\newblock  Phys. Rev. {\bf 179}  (1969) 1547\relax
\relax
\bibitem{photonhadron}
R.P. Feynman, {\em Photon--Hadron Interactions} (W.A. Benjamin, Inc.,
  1972)\relax
\relax
\bibitem{prl_23_1415}
R.P. Feynman,
\newblock  Phys. Rev. Lett. {\bf 23}  (1969) 1415\relax
\relax
\bibitem{pr_1_2901}
J.B. Kogut and D.E. Soper,
\newblock  Phys. Rev. {\bf D 1}  (1970) 2901\relax
\relax
\bibitem{prl_22_744}
S.D. Drell, D.J. Levy, and T.M. Yan,
\newblock  Phys. Rev. Lett. {\bf 22}  (1969) 744\relax
\relax
\bibitem{pr_3_1382}
J.D. Bjorken, J.B. Kogut, and D.E. Soper,
\newblock  Phys. Rev. {\bf D 3}  (1971) 1382\relax
\relax
\bibitem{byckling}
E. Byckling and K. Kajantie, {\em Particle Kinematics} (John Wiley and Sons,
  1973)\relax
\relax
\bibitem{pl_76_89}
G. Altarelli and G. Martinelli,
\newblock  Phys. Lett. {\bf B 76}  (1978) 89\relax
\relax
\bibitem{pl_299_385}
The H1 Collaboration,
\newblock  Phys. Lett. {\bf B 299}  (1993) 385\relax
\relax
\bibitem{np_407_515}
The H1 Collaboration,
\newblock  Nucl. Phys. {\bf B 407}  (1993) 515\relax
\relax
\bibitem{pl_321_161}
The H1 Collaboration,
\newblock  Phys. Lett. {\bf B 321}  (1994) 161\relax
\relax
\bibitem{pl_303_183}
The ZEUS Collaboration,
\newblock  Phys. Lett. {\bf B 303}  (1993) 183\relax
\relax
\bibitem{np_439_471}
The H1 Collaboration,
\newblock  Nucl. Phys. {\bf B 439}  (1995) 471\relax
\relax
\bibitem{pl_316_412}
The ZEUS Collaboration,
\newblock  Phys. Lett. {\bf B 316}  (1993) 412\relax
\relax
\bibitem{zfp_65_379}
The ZEUS Collaboration,
\newblock  Z. Phys. {\bf C 65}  (1995) 379\relax
\relax
\bibitem{zfp_72_399}
The ZEUS Collaboration,
\newblock  Z. Phys. {\bf C 72}  (1996) 399\relax
\relax
\bibitem{pl_364_107}
The NMC Collaboration,
\newblock  Phys. Lett. {\bf B 364}  (1995) 107\relax
\relax
\bibitem{pl_332_393}
K. Prytz,
\newblock  Phys. Lett. {\bf 332}  (1994) 393\relax
\relax
\bibitem{sjnp_15_438}
V.N. Gribov and L.N. Lipatov,
\newblock  Sov. J. Nucl. Phys. {\bf 15}  (1972) 438\relax
\relax
\bibitem{sjnp_15_675}
V.N. Gribov and L.N. Lipatov,
\newblock  Sov. J. Nucl. Phys. {\bf 15}  (1972) 675\relax
\relax
\bibitem{sjnp_20_95}
L.N. Lipatov,
\newblock  Sov. J. Nucl. Phys. {\bf 20}  (1975) 95\relax
\relax
\bibitem{jetp_46_641}
Yu.L Dokshitzer,
\newblock  Sov. Phys. JETP {\bf 46}  (1977) 641\relax
\relax
\bibitem{np_126_298}
G. Altarelli and G. Parisi,
\newblock  Phys. Lett. {\bf 126}  (1977) 298\relax
\relax
\bibitem{pr_10_1649}
A. De R\'ujula {\em et al.},
\newblock  Phys. Rev. {\bf D 10}  (1974) 1649\relax
\relax
\bibitem{prl_44_1118}
C. L\'opez and F.J. Yndur\'ain,
\newblock  Phys. Rev. Lett. {\bf 44}  (1980) 1118\relax
\relax
\bibitem{dr_96_87}
F. Barreiro, C. L\'opez, and F.J. Yndur\'ain,
\newblock  Z. Phys. {\bf C 72}  (1996) 561\relax
\relax
\bibitem{np_470_3}
The H1 Collaboration,
\newblock  Nucl. Phys. {\bf B 470}  (1996) 3\relax
\relax
\bibitem{pl_354_494}
The H1 Collaboration,
\newblock  Phys. Lett. {\bf B 354}  (1995) 494\relax
\relax
\bibitem{np_449_3}
The H1 Collaboration,
\newblock  Nucl. Phys. {\bf B 449}  (1995) 3\relax
\relax
\bibitem{pl_345_576}
The ZEUS Collaboration,
\newblock  Phys. Lett. {\bf B 345}  (1995) 576\relax
\relax
\bibitem{dr_96_236}
The H1 Collaboration,
\newblock  DESY Report {\bf 96--236}  (1996),
\newblock  submitted to Phys. Lett\relax
\relax
\bibitem{pl_298_469}
The H1 Collaboration,
\newblock  Phys. Lett. {\bf B 298}  (1993) 469\relax
\relax
\bibitem{zfp_61_59}
The H1 Collaboration,
\newblock  Z. Phys. {\bf C 61}  (1994) 59\relax
\relax
\bibitem{zfp_63_377}
The H1 Collaboration,
\newblock  Z. Phys. {\bf C 63}  (1994) 377\relax
\relax
\bibitem{pl_346_415}
The H1 Collaboration,
\newblock  Phys. Lett. {\bf B 346}  (1995) 415\relax
\relax
\bibitem{np_445_3}
The H1 Collaboration,
\newblock  Nucl. Phys. {\bf B 445}  (1995) 3\relax
\relax
\bibitem{pl_356_118}
The H1 Collaboration,
\newblock  Phys. Lett. {\bf B 356}  (1995) 118\relax
\relax
\bibitem{pl_358_412}
The H1 Collaboration,
\newblock  Phys. Lett. {\bf B 358}  (1995) 412\relax
\relax
\bibitem{zfp_72_573}
The H1 Collaboration,
\newblock  Z. Phys. {\bf C 72}  (1996) 573\relax
\relax
\bibitem{dr_96_215}
The H1 Collaboration,
\newblock  DESY Report {\bf 96--215}  (1996),
\newblock  submitted to Nucl. Phys\relax
\relax
\bibitem{pl_306_158}
The ZEUS Collaboration,
\newblock  Phys. Lett. {\bf B 306}  (1993) 158\relax
\relax
\bibitem{zfp_59_231}
The ZEUS Collaboration,
\newblock  Phys. Lett. {\bf C 59}  (1993) 231\relax
\relax
\bibitem{zfp_67_81}
The ZEUS Collaboration,
\newblock  Z. Phys. {\bf C 67}  (1995) 81\relax
\relax
\bibitem{zfp_67_93}
The ZEUS Collaboration,
\newblock  Z. Phys. {\bf C 67}  (1995) 93\relax
\relax
\bibitem{pl_363_201}
The ZEUS Collaboration,
\newblock  Phys. Lett. {\bf B 363}  (1995) 201\relax
\relax
\bibitem{np_480_3}
The H1 Collaboration,
\newblock  Nucl. Phys. {\bf B 480}  (1996) 3\relax
\relax
\bibitem{zfp_68_29}
The ZEUS Collaboration,
\newblock  Z. Phys. {\bf C 68}  (1995) 29\relax
\relax
\bibitem{zfp_72_593}
The H1 Collaboration,
\newblock  Z. Phys. {\bf C 72}  (1996) 593\relax
\relax
\bibitem{pl_338_483}
The ZEUS Collaboration,
\newblock  Phys. Lett. {\bf B 338}  (1994) 483\relax
\relax
\bibitem{zfp_70_1}
The ZEUS Collaboration,
\newblock  Z. Phys. {\bf C 70}  (1996) 1\relax
\relax
\bibitem{np_429_477}
The H1 Collaboration,
\newblock  Nucl. Phys. {\bf B 429}  (1994) 477\relax
\relax
\bibitem{pl_348_681}
The H1 Collaboration,
\newblock  Phys. Lett. {\bf B 348}  (1995) 681\relax
\relax
\bibitem{zfp_70_609}
The H1 Collaboration,
\newblock  Z. Phys. {\bf C 70}  (1996) 609\relax
\relax
\bibitem{zfp_68_569}
The ZEUS Collaboration,
\newblock  Z. Phys. {\bf C 68}  (1995) 569\relax
\relax
\bibitem{zfp_70_391}
The ZEUS Collaboration,
\newblock  Z. Phys. {\bf C 70}  (1996) 391\relax
\relax
\bibitem{pl_315_481}
The ZEUS Collaboration,
\newblock  Phys. Lett. {\bf B 315}  (1993) 481\relax
\relax
\bibitem{pl_384_388}
The ZEUS Collaboration,
\newblock  Phys. Lett. {\bf B 384}  (1996) 388\relax
\relax
\bibitem{dr_94_215}
B.D. Burow {\em et al.},
\newblock  DESY Report {\bf 94--215}  (1994)\relax
\relax
\bibitem{np_407_539}
G. Schuler and T. Sj\"ostrand,
\newblock  Nucl. Phys. {\bf B 407}  (1993) 539\relax
\relax
\bibitem{pl_297_205}
The H1 Collaboration,
\newblock  Phys. Lett. {\bf B 297}  (1992) 205\relax
\relax
\bibitem{pl_314_436}
The H1 Collaboration,
\newblock  Phys. Lett. {\bf B 314}  (1993) 436\relax
\relax
\bibitem{pl_328_176}
The H1 Collaboration,
\newblock  Phys. Lett. {\bf B 328}  (1994) 176\relax
\relax
\bibitem{np_445_195}
The H1 Collaboration,
\newblock  Nucl. Phys. {\bf B 445}  (1995) 195\relax
\relax
\bibitem{zfp_70_17}
The H1 Collaboration,
\newblock  Z. Phys. {\bf C 70}  (1996) 17\relax
\relax
\bibitem{pl_297_404}
The ZEUS Collaboration,
\newblock  Phys. Lett. {\bf B 297}  (1992) 404\relax
\relax
\bibitem{pl_322_287}
The ZEUS Collaboration,
\newblock  Phys. Lett. {\bf B 322}  (1994) 287\relax
\relax
\bibitem{pl_342_417}
The ZEUS Collaboration,
\newblock  Phys. Lett. {\bf B 342}  (1995) 417\relax
\relax
\bibitem{pl_354_163}
The ZEUS Collaboration,
\newblock  Phys. Lett. {\bf B 354}  (1995) 163\relax
\relax
\bibitem{pl_384_401}
The ZEUS Collaboration,
\newblock  Phys. Lett. {\bf B 384}  (1995) 401\relax
\relax
\bibitem{pl_348_665}
The ZEUS Collaboration,
\newblock  Phys. Lett. {\bf B 348}  (1995) 665\relax
\relax
\bibitem{np_435_3}
The H1 Collaboration,
\newblock  Nucl. Phys. {\bf B 435}  (1995) 3\relax
\relax
\bibitem{zfp_67_227}
The ZEUS Collaboration,
\newblock  Z. Phys. {\bf C 67}  (1995) 227\relax
\relax
\bibitem{pl_346_399}
The ZEUS Collaboration,
\newblock  Phys. Lett. {\bf B 346}  (1995) 399\relax
\relax
\bibitem{pl_356_129}
The ZEUS Collaboration,
\newblock  Phys. Lett. {\bf B 356}  (1995) 129\relax
\relax
\bibitem{pl_369_55}
The ZEUS Collaboration,
\newblock  Phys. Lett. {\bf B 369}  (1995) 55\relax
\relax
\bibitem{np_472_32}
The H1 Collaboration,
\newblock  Nucl. Phys. {\bf B 472}  (1996) 32\relax
\relax
\bibitem{pl_349_225}
The ZEUS Collaboration,
\newblock  Phys. Lett. {\bf B 349}  (1995) 225\relax
\relax
\bibitem{rmp_50_261}
T.H. Bauer {\em et al.},
\newblock  Rev. Mod. Phys. {\bf 50}  (1978) 261\relax
\relax
\bibitem{pr_5_545}
J. Ballam {\em et al.},
\newblock  Phys. Rev. {\bf D 5}  (1972) 545\relax
\relax
\bibitem{np_113_53}
P. Joos {\em et al.},
\newblock  Nucl. Phys. {\bf B 113}  (1976) 53\relax
\relax
\bibitem{sjnp_9_69}
V.L. Auslander {\em et al.},
\newblock  Sov. J. Nucl. Phys. {\bf 9}  (1969) 69\relax
\relax
\bibitem{np_29_349}
K.C. Moffeit {\em et al.},
\newblock  Nucl. Phys. {\bf B 29}  (1971) 349\relax
\relax
\bibitem{pl_39_289}
D. Benaksas {\em et al.},
\newblock  Phys. Lett. {\bf 39}  (1972) 289\relax
\relax
\bibitem{np_256_365}
L.M. Barkov {\em et al.},
\newblock  Nucl. Phys. {\bf B 256}  (1985) 365\relax
\relax
\bibitem{prl_43_657}
R.M. Egloff {\em et al.},
\newblock  Phys. Rev. Lett. {\bf 43}  (1979) 657\relax
\relax
\bibitem{prl_43_1545}
R.M. Egloff {\em et al.},
\newblock  Phys. Rev. Lett. {\bf 43}  (1979) 1545\relax
\relax
\bibitem{np_209_56}
D. Aston {\em et al.},
\newblock  Nucl. Phys. {\bf B 209}  (1982) 56\relax
\relax
\bibitem{np_36_404}
J. Park {\em et al.},
\newblock  Nucl. Phys. {\bf B 36}  (1972) 404\relax
\relax
\bibitem{pr_8_687}
J.T. Dakin {\em et al.},
\newblock  Phys. Rev. {\bf D 8}  (1973) 687\relax
\relax
\bibitem{pr_10_765}
J. Ballam {\em et al.},
\newblock  Phys. Rev. {\bf D 10}  (1974) 765\relax
\relax
\bibitem{pr_19_1303}
C. del Papa {\em et al.},
\newblock  Phys. Rev. {\bf D 19}  (1979) 1303\relax
\relax
\bibitem{pr_24_2787}
D.G. Cassel {\em et al.},
\newblock  Phys. Rev. {\bf D 24}  (1981) 2787\relax
\relax
\bibitem{pr_25_634}
I. Cohen {\em et al.},
\newblock  Phys. Rev. {\bf D 25}  (1982) 634\relax
\relax
\bibitem{prl_38_633}
W.R. Francis {\em et al.},
\newblock  Phys. Rev. Lett. {\bf 38}  (1977) 633\relax
\relax
\bibitem{pr_26_1}
W.D. Shambroom {\em et al.},
\newblock  Phys. Rev. {\bf D 26}  (1982) 1\relax
\relax
\bibitem{pl_161_203}
The EMC Collaboration,
\newblock  Phys. Lett. {\bf B 161}  (1985) 203\relax
\relax
\bibitem{zfp_39_169}
The EMC Collaboration,
\newblock  Z. Phys. {\bf C 39}  (1988) 169\relax
\relax
\bibitem{zfp_54_239}
The NMC Collaboration,
\newblock  Z. Phys. {\bf C 54}  (1992) 239\relax
\relax
\bibitem{np_429_503}
The NMC Collaboration,
\newblock  Nucl. Phys. {\bf B 429}  (1994) 503\relax
\relax
\bibitem{mpi_97_3}

\newblock  The E665 Collaboration, Preprint MPI--PhE/97--03 (1997),
\newblock  submitted to Z.Phys.\relax
\relax
\bibitem{pa03_09}
The E665 Collaboration,
\newblock  PA03--009, XXVIII International Conference on High Energy Physics,
  Warsaw, July 25--31, 1996\relax
\relax
\bibitem{zfp_58_375}
G.T. Jones {\em et al.},
\newblock  Z. Phys. {\bf C 58}  (1993) 375\relax
\relax
\bibitem{ijmp_9_513}
B.~Kopeliovich and P.~Marage,
\newblock  Int. J. Mod. Phys. {\bf A 8}  (1993) 513\relax
\relax
\bibitem{prl_35_1616}
B. Gittelman {\em et al.},
\newblock  Phys. Rev. Lett. {\bf 35}  (1975) 1616\relax
\relax
\bibitem{prl_34_1040}
B. Knapp {\em et al.},
\newblock  Phys. Rev. Lett. {\bf 34}  (1975) 1040\relax
\relax
\bibitem{prl_36_1233}
T.Nash {\em et al.},
\newblock  Phys. Rev. Lett. {\bf 36}  (1976) 1233\relax
\relax
\bibitem{prl_35_483}
U. Camerini {\em et al.},
\newblock  Phys. Rev. Lett. {\bf 35}  (1975) 483\relax
\relax
\bibitem{arnps_35_397}
S.D. Holmes, W. Lee, and J.E. Wiss,
\newblock  Ann. Rev. Nucl. Part. Sci. {\bf 35}  (1985) 397\relax
\relax
\bibitem{prl_48_73}
M.Binkley {\em et al.},
\newblock  Phys. Rev. Lett. {\bf 48}  (1982) 73\relax
\relax
\bibitem{prl_52_795}
B.H. Denby {\em et al.},
\newblock  Phys. Rev. Lett. {\bf 52}  (1984) 795\relax
\relax
\bibitem{prl_43_187}
A.R. Clark {\em et al.},
\newblock  Phys. Rev. Lett. {\bf 43}  (1979) 187\relax
\relax
\bibitem{np_213_1}
The EMC Collaboration,
\newblock  Nucl. Phys. {\bf B 213}  (1983) 1\relax
\relax
\bibitem{pl_332_195}
The NMC Collaboration,
\newblock  Phys. Lett. {\bf B 332}  (1994) 195\relax
\relax
\bibitem{zfp_33_505}
R. Barate {\em et al.},
\newblock  Z. Phys. {\bf C 33}  (1987) 505\relax
\relax
\bibitem{pl_316_197}
P.L. Frabetti {\em et al.},
\newblock  Phys. Lett. {\bf B 316}  (1993) 197\relax
\relax
\bibitem{dr_95_47}
H. Abramowicz, L.L. Frankfurt, and M. Strikman,
\newblock  DESY Report {\bf 95--047}  (1995),
\newblock  HEP--PH/95-03-437, publ. in SLAC Summer Inst., 1994\relax
\relax
\bibitem{perl}
M.L. Perl, {\em High Energy Hadron Physics} (John Wiley \& Sons, 1974)\relax
\relax
\bibitem{collins}
P.D.B. Collins, {\em An Introduction to Regge Theory and High Energy Physics}
  (Cambridge University Press, 1977)\relax
\relax
\bibitem{hep96_11_433}

\newblock  J.C. Collins {\em et al.}, Preprint HEP--PH/96-11-433 (1996)\relax
\relax
\bibitem{pl_348_213}
A. Donnachie and P.V. Landshoff,
\newblock  Phys. Lett. {\bf B 348}  (1995) 213\relax
\relax
\bibitem{pl_296_227}
A. Donnachie and P.V. Landshoff,
\newblock  Phys. Lett. {\bf B 296}  (1992) 227\relax
\relax
\bibitem{rmp_68_611}
R.M. Barnett {\em et al.},
\newblock  Rev. Mod. Phys. {\bf 68}  (1996) 611\relax
\relax
\bibitem{np_14_543}
H. Fraas and D. Schildknecht,
\newblock  Nucl. Phys. {\bf B 14}  (1969) 543\relax
\relax
\bibitem{prl_22_981}
J.J. Sakurai,
\newblock  Phys. Rev. Lett. {\bf 22}  (1969) 981\relax
\relax
\bibitem{sakurai}
J.J. Sakurai, {\em Currents and Mesons} (University of Chicago Press,
  1969)\relax
\relax
\bibitem{hep95_07_394}

\newblock  L.P.A. Haakman {\em et al.}, Preprint HEP--PH/95-07-394 (1995)\relax
\relax
\bibitem{hep96_08_384}

\newblock  L.L. Jenkovszky {\em et al.}, Preprint HEP--PH/96-08-384
  (1996)\relax
\relax
\bibitem{pl_379_1}
M.A. Pichowsky and T.S.H. Lee,
\newblock  Phys. Lett. {\bf 379}  (1996) 1\relax
\relax
\bibitem{hep96_08_203}

\newblock  H.G. Dosch {\em et al.}, Preprint HEP--PH/96-08-203 (1996)\relax
\relax
\bibitem{pl_389_157}
G. Niesler {\em et al.},
\newblock  Phys. Lett. {\bf 389}  (1996) 157\relax
\relax
\bibitem{zfp_69_39}
The ZEUS Collaboration,
\newblock  Z. Phys. {\bf C 69}  (1995) 39\relax
\relax
\bibitem{zfp_57_89}
M.G. Ryskin,
\newblock  Z. Phys. {\bf C 57}  (1993) 89\relax
\relax
\bibitem{cpc_100_195}
M. Arneodo, L. Lamberti, and M. Ryskin,
\newblock  Comp. Phys. Commun. {\bf 100}  (1996) 195\relax
\relax
\bibitem{hep95_11_228}

\newblock  M.G. Ryskin {\em et al.}, Preprint HEP--PH/95-11-228 (1995)\relax
\relax
\bibitem{pr_50_3134}
S.J. Brodsky {\em et al.},
\newblock  Phys. Rev. {\bf D 50}  (1994) 3134\relax
\relax
\bibitem{pr_22_2157}
S.J. Brodsky and G.P. Lepage,
\newblock  Phys. Rev. {\bf D 22}  (1980) 2157\relax
\relax
\bibitem{pr_54_3194}
L. Frankfurt {\em et al.},
\newblock  Phys. Rev. {\bf D 54}  (1996) 3194\relax
\relax
\bibitem{hep96_11_207}

\newblock  P. Hoodbhoy, Preprint HEP--PH/96-11-207 (1996)\relax
\relax
\bibitem{hep96_09_448}

\newblock  A. Martin {\em et al.}, Preprint HEP--PH/96-09-448 (1996)\relax
\relax
\bibitem{pl_374_199}
J. Nemchik {\em et al.},
\newblock  Phys. Lett. {\bf 374}  (1996) 199\relax
\relax
\bibitem{pl_341_228}
J. Nemchik {\em et al.},
\newblock  Phys. Lett. {\bf 341}  (1994) 228\relax
\relax
\bibitem{hep96_05_231}

\newblock  J. Nemchik {\em et al.}, Preprint HEP--PH/96-05-231 (1996)\relax
\relax
\bibitem{hep96_10_276}

\newblock  L.N. Lipatov, Preprint HEP--PH/96-10-276 (1996)\relax
\relax
\bibitem{sjnp_23_642}
L.N. Lipatov,
\newblock  Sov. J. Nucl. Phys. {\bf 23}  (1976) 642\relax
\relax
\bibitem{pl_60_50}
V.S. Fadin {\em et al.},
\newblock  Phys. Lett. {\bf B 60}  (1975) 50\relax
\relax
\bibitem{jetp_44_443}
E.A. Kuraev {\em et al.},
\newblock  Sov. Phys. JETP {\bf 44}  (1976) 443\relax
\relax
\bibitem{jetp_45_199}
E.A. Kuraev {\em et al.},
\newblock  Sov. Phys. JETP {\bf 45}  (1977) 199\relax
\relax
\bibitem{sjnp_28_822}
Ya.Ya. Balitsky and L.N. Lipatov,
\newblock  Sov. J. Nucl. Phys. {\bf 28}  (1978) 822\relax
\relax
\bibitem{hep96_01_298}

\newblock  J. Amundson {\em et al.}, Preprint HEP--PH/96-01-298 (1996)\relax
\relax
\bibitem{pr_51_1125}
G.T. Bodwin {\em et al.},
\newblock  Phys. Rev. {\bf D 51}  (1995) 1125\relax
\relax
\bibitem{pr_54_5523}
I.F. Ginzburg and D.Yu. Ivanov,
\newblock  Phys. Rev. {\bf D 54}  (1996) 5523\relax
\relax
\bibitem{pl_375_301}
J. Bartels {\em et al.},
\newblock  Phys. Lett. {\bf B 375}  (1996) 301\relax
\relax
\bibitem{zfp_68_137}
J.R. Forshaw {\em et al.},
\newblock  Z. Phys. {\bf C 68}  (1995) 137\relax
\relax
\bibitem{pl_299_374}
The H1 Collaboration,
\newblock  Phys. Lett. {\bf B 299}  (1993) 374\relax
\relax
\bibitem{pl_293_465}
The ZEUS Collaboration,
\newblock  Phys. Lett. {\bf B 293}  (1992) 465\relax
\relax
\bibitem{zfp_69_27}
The H1 Collaboration,
\newblock  Z. Phys. {\bf C 69}  (1995) 27\relax
\relax
\bibitem{zfp_63_391}
The ZEUS Collaboration,
\newblock  Z. Phys. {\bf C 63}  (1994) 391\relax
\relax
\bibitem{pl_30_123}
B.L. Ioffe,
\newblock  Phys. Lett. {\bf B 30}  (1969) 123\relax
\relax
\bibitem{hardproc}
B.L. Ioffe, V.A. Khoze, and L.N. Lipatov, {\em Hard Processes} volume~1,
\newblock  (North Holland, 1984)\relax
\relax
\bibitem{pr_120_1834}
L. Hand,
\newblock  Phys. Rev. {\bf 120}  (1963) 1834\relax
\relax
\bibitem{pr_167_1365}
F. Gilman,
\newblock  Phys. Rev. {\bf 167}  (1968) 1365\relax
\relax
\bibitem{zfp_69_607}
The ZEUS Collaboration,
\newblock  Z. Phys. {\bf C 69}  (1996) 607\relax
\relax
\bibitem{zfp_61_139}
A. Donnachie and P. Landshoff,
\newblock  Z. Phys. {\bf C 61}  (1994) 139\relax
\relax
\bibitem{pl_223_485}
The BCDMS Collaboration,
\newblock  Phys. Lett. {\bf B 223}  (1989) 485\relax
\relax
\bibitem{pr_54_3006}
The E665 Collaboration,
\newblock  Phys. Rev. {\bf D 54}  (1996) 3006\relax
\relax
\bibitem{prl_40_1222}
D.O. Caldwell {\em et al.},
\newblock  Phys. Rev. Lett. {\bf 40}  (1978) 1222\relax
\relax
\bibitem{ch_87_1}
S.I. Alekhin {\em et al.},
\newblock  CERN--HERA {\bf 87-001}  (1987)\relax
\relax
\bibitem{pl_269_465}
H. Abramowicz {\em et al.},
\newblock  Phys. Lett. {\bf B 269}  (1991) 465\relax
\relax
\bibitem{marcusthesis}
A. Marcus,
\newblock  Energy Dependence of the $\gamma^* p$ Cross Section,
\newblock  Master's thesis, Tel-Aviv University, 1996\relax
\relax
\bibitem{pa02_25}
The ZEUS Collaboration,
\newblock  PA02--025, XXVIII International Conference on High Energy Physics,
  Warsaw, Poland, July 25--31, 1996\relax
\relax
\bibitem{zfp_48_471}
M. Gl\"uck, E. Reya, and A. Vogt,
\newblock  Z. Phys. {\bf C 48}  (1990) 471\relax
\relax
\bibitem{zfp_53_127}
M. Gl\"uck, E. Reya, and A. Vogt,
\newblock  Z. Phys. {\bf C 53}  (1992) 127\relax
\relax
\bibitem{zfp_67_433}
M. Gl\"uck, E. Reya, and A. Vogt,
\newblock  Z. Phys. {\bf C 67}  (1995) 433\relax
\relax
\bibitem{hep96_08_9}

\newblock  A. Levy, Preprint HEP--PH/96-08-9 (1996)\relax
\relax
\bibitem{pa02_28}
The ZEUS Collaboration,
\newblock  PA02--028, XXVIII International Conference on High Energy Physics,
  Warsaw, Poland, July 25--31, 1996\relax
\relax
\bibitem{pa02_85}
The H1 Collaboration,
\newblock  PA02--085, XXVIII International Conference on High Energy Physics,
  Warsaw, Poland, July 25--31, 1996\relax
\relax
\bibitem{hep97_03_245}

\newblock  N. Cartiglia, Preprint HEP--PH/97-03-245 (1997),
\newblock  publ. in SLAC Summer Inst., 1996\relax
\relax
\bibitem{np_463_3}
The H1 Collaboration,
\newblock  Nucl. Phys. {\bf B 463}  (1996) 3\relax
\relax
\bibitem{pl_356_601}
The ZEUS Collaboration,
\newblock  Phys. Lett. {\bf B 356}  (1995) 601\relax
\relax
\bibitem{np_468_3}
The H1 Collaboration,
\newblock  Nucl. Phys. {\bf B 468}  (1996) 3\relax
\relax
\bibitem{pa02_65}
The H1 Collaboration,
\newblock  PA02--065, XXVIII International Conference on High Energy Physics,
  Warsaw, Poland, July 25--31, 1996\relax
\relax
\bibitem{pa02_50}
The ZEUS Collaboration,
\newblock  PA02--050, XXVIII International Conference on High Energy Physics,
  Warsaw, Poland, July 25--31, 1996\relax
\relax
\bibitem{pa02_53}
The ZEUS Collaboration,
\newblock  PA02--053, XXVIII International Conference on High Energy Physics,
  Warsaw, Poland, July 25--31, 1996\relax
\relax
\bibitem{pa02_51}
The ZEUS Collaboration,
\newblock  PA02--051, XXVIII International Conference on High Energy Physics,
  Warsaw, Poland, July 25--31, 1996\relax
\relax
\bibitem{pa01_88}
The H1 Collaboration,
\newblock  PA01--088, XXVIII International Conference on High Energy Physics,
  Warsaw, Poland, July 25--31, 1996\relax
\relax
\bibitem{zfp_73_73}
The ZEUS Collaboration,
\newblock  Z. Phys. {\bf C 73}  (1996) 73\relax
\relax
\bibitem{pl_377_259}
The ZEUS Collaboration,
\newblock  Phys. Lett. {\bf B 377}  (1996) 259\relax
\relax
\bibitem{pl_380_220}
The ZEUS Collaboration,
\newblock  Phys. Lett. {\bf B 380}  (1996) 220\relax
\relax
\bibitem{pa02_64}
The H1 Collaboration,
\newblock  PA02--064, XXVIII International Conference on High Energy Physics,
  Warsaw, Poland, July 25--31, 1996\relax
\relax
\bibitem{pl_350_120}
The ZEUS Collaboration,
\newblock  Phys. Lett. {\bf B 350}  (1995) 120\relax
\relax
\bibitem{pa02_47}
The ZEUS Collaboration,
\newblock  PA02--047, XXVIII International Conference on High Energy Physics,
  Warsaw, Poland, July 25--31, 1996\relax
\relax
\bibitem{np_472_3}
The H1 Collaboration,
\newblock  Nucl. Phys. {\bf B 472}  (1996) 3\relax
\relax
\bibitem{pa02_86}
The H1 Collaboration,
\newblock  PA02--086, XXVIII International Conference on High Energy Physics,
  Warsaw, Poland, July 25--31, 1996\relax
\relax
\bibitem{sjnp_38_736}
N.N. Achasov and V.A. Karnekov,
\newblock  Sov. J. Nucl. Phys. {\bf 38}  (1983) 736\relax
\relax
\bibitem{damethod}
S. Bentvelsen, J. Engelen, and P. Kooijman, in {\em Proceedings of the Workshop
  on Physics at HERA}, edited by W. Buchm\"uller and G. Ingleman  (DESY,
  Hamburg, Germany, 1992), p.~23\relax
\relax
\bibitem{kpthesis}
K. Piotrzkowski,
\newblock  {\em Experimental Aspects of the Luminosity Measurement in the ZEUS
  Experiment},
\newblock  Ph.D. thesis, University of Hamburg, 1993,
\newblock  DESY F35D--93--06\relax
\relax
\bibitem{pl_19_702}
P. S\"oding,
\newblock  Phys. Lett. {\bf 19}  (1966) 702\relax
\relax
\bibitem{nc_34_1644}
J.D. Jackson,
\newblock  Nuovo Cimento {\bf 34}  (1964) 1644\relax
\relax
\bibitem{prl_5_278}
S. Drell,
\newblock  Phys. Rev. Lett. {\bf 5}  (1960) 278\relax
\relax
\bibitem{pr_149_1172}
M. Ross and L. Stodolsky,
\newblock  Phys. Rev. {\bf 149}  (1966) 1172\relax
\relax
\bibitem{pr_9_126}
R. Spital and D.R. Yennie,
\newblock  Phys. Rev. {\bf D 9}  (1974) 126\relax
\relax
\bibitem{pr_2_1859}
J. Pumplin,
\newblock  Phys. Rev. {\bf D 2}  (1970) 1859\relax
\relax
\bibitem{prl_25_485}
T.H. Bauer,
\newblock  Phys. Rev. Lett. {\bf 25}  (1970) 485\relax
\relax
\bibitem{prl_25_704}
T.H. Bauer,
\newblock  Phys. Rev. Lett. {\bf 25}  (1970) 704,
\newblock  (E)\relax
\relax
\bibitem{pr_3_2671}
T.H. Bauer,
\newblock  Phys. Rev. {\bf D 3}  (1971) 2671\relax
\relax
\bibitem{dr_96_162}
The H1 Collaboration,
\newblock  DESY Report {\bf 96--162}  (1996)\relax
\relax
\bibitem{pa03_48}
The H1 Collaboration,
\newblock  PA03--048, XXVIII International Conference on High Energy Physics,
  Warsaw, Poland, July 25--31, 1996\relax
\relax
\bibitem{pl_338_507}
The H1 Collaboration,
\newblock  Phys. Lett. {\bf B 338}  (1994) 507\relax
\relax
\bibitem{nc_48_541}
P.G.O. Freund,
\newblock  Nuovo Cimento {\bf A 48}  (1967) 541\relax
\relax
\bibitem{pr_175_1669}
R. Erbe {\em et al.},
\newblock  Phys. Rev. {\bf 175}  (1968) 1669\relax
\relax
\bibitem{np_336_1}
J.R. Cudell,
\newblock  Nucl. Phys. {\bf B 336}  (1990) 1\relax
\relax
\bibitem{pl_31_387}
F. Gilman {\em et al.},
\newblock  Phys. Lett. {\bf B 31}  (1970) 387\relax
\relax
\bibitem{pl_65_463}
B. Humbert and A.C.D. Wright,
\newblock  Phys. Lett. {\bf B 65}  (1976) 463\relax
\relax
\bibitem{pr_15_2503}
B. Humbert and A.C.D. Wright,
\newblock  Phys. Rev. {\bf D 15}  (1977) 2503\relax
\relax
\bibitem{np_61_381}
K. Schilling and G. Wolf,
\newblock  Nucl. Phys. {\bf B 61}  (1973) 381\relax
\relax
\bibitem{manybody}
G. Wolf and P. S\"oding, {\em Electromagnetic Interactions of Hadrons}
  volume~1,
\newblock  (New York, 1978)\relax
\relax
\bibitem{futurephysics}
{\em Proceedings of the Workshop on Future Physics at HERA}, edited by G.
  Ingelman, A. De Roeck, R. Klanner  (DESY, Hamburg, Germany, 1996)\relax
\relax
\bibitem{lumiupgrade}
W. Bartel {\em et al.}, in {\em Proceedings of the Workshop on Future Physics
  at HERA}, edited by G. Ingelman, A. De Roeck, R. Klanner  (DESY, Hamburg,
  Germany, 1996), p. 1095\relax
\relax
\bibitem{polprot}
D.P. Barber {\em et al.}, in {\em Proceedings of the Workshop on Future Physics
  at HERA}, edited by G. Ingelman, A. De Roeck, R. Klanner  (DESY, Hamburg,
  Germany, 1996), p. 1205\relax
\relax
\bibitem{heranuc}
M. Arneodo {\em et al.}, in {\em Proceedings of the Workshop on Future Physics
  at HERA}, edited by G. Ingelman, A. De Roeck, R. Klanner  (DESY, Hamburg,
  Germany, 1996), p. 887\relax
\relax
\bibitem{hep96_05_208}

\newblock  J. Nemchik {\em et al.}, Preprint HEP--PH/96-05-208 (1996)\relax
\relax
\bibitem{pl_309_179}
B.Z. Kopeliovich {\em et al.},
\newblock  Phys. Lett. {\bf B 309}  (1993) 179\relax
\relax
\bibitem{pl_382_6}
L.L. Frankfurt and M.I. Strikman,
\newblock  Phys. Lett. {\bf B 382}  (1996) 6\relax
\relax
\bibitem{pl_324_469}
B.Z. Kopeliovich {\em et al.},
\newblock  Phys. Lett. {\bf B 324}  (1994) 469\relax
\relax
\bibitem{pl_383_362}
J. H\"ufner {\em et al.},
\newblock  Phys. Lett. {\bf B 383}  (1996) 362\relax
\relax
\end{mcbibliography}

\end{document}